\documentclass[a4paper,10pt,DIV=12,flalign, abstracton]{scrartcl}

\setkomafont{disposition}{\fontfamily{ptm}}%\fontseries{m}
\setkomafont{title}{\fontfamily{ptm}}%\fontseries{m}

\usepackage[utf8]{inputenc}
\usepackage[T1]{fontenc}
\usepackage{amssymb,amsmath,amsthm}
\usepackage{nicefrac}
\usepackage{graphicx}
\usepackage{hyperref}
\usepackage{csquotes}
\usepackage{color}
\usepackage{subfig}
\usepackage{multirow} % for tables
\usepackage[sort&compress]{natbib}
\usepackage{nomencl}
\usepackage{accents}
\usepackage[normalem]{ulem}
\usepackage{ifthen}

\makenomenclature

\newcommand{\newsym}[4]{%
  \nomenclature[#4]{\ensuremath{#2}}{#3}%
  \newcommand{#1}{#2}%
}

\newcommand{\upmath}[1]{\mathrm{#1}}
\renewcommand{\d}[1]{\matrmh{d}{#1}}

\newcommand{\dutilde}[1]{\underaccent{\approx}{#1}}

\renewcommand*{\dot}[1]{%
  \accentset{\mbox{\large\bfseries .}}{#1}}

\newsym{\ndim}{n}{Dimension of space}{n}
\newcommand{\vnormcomp}{n}
\newsym{\vnorm}{\vect{\vnormcomp}}{Normalenvektor}{n}

\newcommand{\surface}[1]{\partial{#1}}
\renewcommand{\d}{\upmath{d}}

\newcommand{\dV}{\;\d V}
\newcommand{\dsurf}{\;\d S}

\newcommand{\domain}[1][{}]{\varOmega_{#1}}
\newcommand{\domainbound}[1][{}]{\partial\domain[#1]}
\newcommand{\discontsurf}[1][{}]{S^{||}_{#1}}

\newcommand{\Kron}[1]{\delta_{#1}}
\newcommand{\tens}[1]{\undertilde{\mathbf{#1}}}

\newcommand{\Tens}[1]{\dutilde{\mathbf{#1}}}

\newcommand{\vect}[1]{\underline{\mathbf{#1}}}

\newcommand{\partderivf}[3][{}]{\frac{\partial^{#1}{#2}}{\partial{#3}^{#1}}}

\newcommand{\dev}[1]{#1^{\upmath{d}}}

\newcommand{\inv}[1]{{#1}^{-1}}

\newsym{\tident}{\tens{I}}{Einheitstensor}{i}
\newsym{\Tidentfs}{\Tens{I}_{\upmath{FS}}}{vollst. symmetr. Einheitstensor 4. Stufe}{ifs}
\newsym{\Tidents}{\Tens{I}_{\upmath{S}}}{symmetrisierender Einheitstensor 4. Stufe}{ifs}
\newsym{\Tidenttransp}{\Tens{I}_{\upmath{T}}}{transponierender Einheitstensor 4. Stufe}{it}
\newcommand{\ttpermutcomp}{\epsilon}

\newcommand{\placeholder}{(\circ)}
\newcommand{\Lagrangemult}[1][{}]{\lambda^{#1}}

\newsym{\Lagrangefunc}{\mathcal{L}}{Lagrange-Funktion}{L}

\newcommand{\density}{\rho}

\newcommand{\rate}[1]{\dot{#1}}

\newsym{\stress}{\tens{\sigma}}{Innere Spannung}{s}
\newcommand{\stresscomp}{\sigma}
\newsym{\Cauchystress}{\tens{\Sigma}}{Cauchy Spannung}{s}
\newcommand{\tractioncomp}{t}
\newsym{\traction}{\vect{\tractioncomp}}{Schnittspannung}{t}
\newsym{\displ}{\vect{u}}{Verschiebungsvektor}{u}
\newcommand{\displcomp}{u}
\newcommand{\velocitycomp}{v}
\newsym{\velocity}{\vect{\velocitycomp}}{Geschwindigkeitsvektor}{v}
\newsym{\strain}{\tens{\varepsilon}}{Dehnung}{e}

\newcommand{\straincomp}{\varepsilon}

\newsym{\dissipation}{D}{Dissipation}{d}
\newsym{\dissipationpl}{\pi}{plast.\ Dissipation}{p}
\newsym{\defgrad}{\tens{F}}{Deformationsgrad}{F}
\newsym{\velocitygrad}{\tens{L}}{Geschwindigkeitsgradient}{L}
\newsym{\defrate}{\rate{\strain}}{Deformationsgeschwindigkeit}{L}
\newcommand{\defratecomp}{\rate{\straincomp}}
\newsym{\defrateeq}{d_{\upmath{eq}}}{Deformationsgeschwindigkeit}{L}
\newsym{\tstreamfunc}{\tens{\Psi}}{Strömungsfunktion}{P}
\newsym{\vstreamfunc}{\vect{\Psi}}{Strömungsfunktion}{P}
\newsym{\streamfunc}{\Psi}{Strömungsfunktion}{P}
\newsym{\Airyfunc}{F}{Airy stress function}{F}

\newsym{\innerenergy}{\Phi}{inner Energie}{F}
\newcommand{\strainenergy}{W}
\newsym{\entropy}{\eta}{Entropy}{n}

\newsym{\entropyflux}{\vect{h}}{Entropiefluss}{h}
\newsym{\temperature}{\theta}{Temperature}{t}
\newsym{\Helmenergy}{\Psi}{Helmholtz-Energie}{P}
\newsym{\Helmenergyref}{\Helmenergy_0}{Helmholtz-Energie}{P0}
\newsym{\HelmenergyD}{\Helmenergy^{\upmath{D}}}{Defekt-Helmholtz-Energie}{PD}
\newsym{\heatflux}{\vect{q}}{Wärmestrom}{qth}
\newsym{\thermalconduct}{\lambda}{Wärmeleitfähigkeit}{l}

\newsym{\entropyflow}{\vect{h}}{Entropiestrom}{h}
\newsym{\entropysource}{p_{\powerlocextvoltherm}}{Entropieproduction}{pr}
\newsym{\entropysourcespontan}{p_{\upmath{s}}}{spontane Entropieproduktion}{ps}
\newsym{\mechworkflux}{\vect{p}_{\upmath{m}}}{mechanische Arbeit}{pm}
\newsym{\intrvar}{h}{innere Variable}{h}
\newsym{\intrvarset}{\left\{h\right\}}{innere Variablen}{h}

\newsym{\intrvardriveforce}{Y}{Triebkräfte der inneren Variablen}{Y}

\newsym{\dissfunc}{\mathcal{D}}{Dissipationsfunktion}{D}
\newsym{\disspotential}{\Psi}{Dissipationspotential}{P}

\newsym{\powerglobint}{\mathcal{P}_{\upmath{int}}}{Power of internal forces}{Pi}
\newsym{\powerglobext}{\mathcal{P}_{\upmath{ext}}}{Power of external forces}{Pe}
\newsym{\kinenergyglob}{\mathcal{K}}{Kinetic energy}{K}
\newsym{\kinpowerglob}{\rate{\mathcal{K}}}{Change of kinetic energy}{K}
\newsym{\kinenergy}{k}{Kinetic energy density}{K}

\newsym{\powerlocextvolmech}{{p}_{\upmath{m,int}}}{Power of external mechanical volume forces}{pm}
\newsym{\powerlocextvoltherm}{r}{Heat source per volume}{pth}
\newsym{\powerlocintmech}{{p}_{\upmath{m,int}}}{Power of internal mechanical volume forces}{pm}

\newsym{\strainset}{\{\upmath{STRAIN}\}}{set of strain measures}{S}

\newsym{\kinDOFset}{\{\upmath{DOF}\}}{set of primary kinematic quantities}{U}

\newsym{\orderparam}{s}{order parameter}{s}
\newsym{\phasestress}{\vect{\xi}}{micro stress vector}{n}
\newsym{\microforceint}{\pi}{internal microforce}{p}
\newsym{\microforceext}{\gamma}{external microforce}{g}

\newcommand{\microdefcomp}{\chi}
\newsym{\microdef}{\tens{\chi}}{Mikrodeformation}{x}
\newsym{\microdefrate}{\rate{\microdef}}{Mikrodeformation}{x}
\newcommand{\microdefratecomp}{\rate{\microdefcomp}}
\newcommand{\microdefdevcomp}{\microdefcomp^{\upmath{ds}}}

\newcommand{\microdefskwcomp}{\microdefcomp^{\upmath{skw}}}
\newsym{\microdefdiff}{\tens{\microdefdiffcomp}}{Mikrodeformationsdifferenz}{e}
\newcommand{\microdefdiffcomp}{e}
\newsym{\microdefdiffrate}{\rate{\microdefdiff}}{Mikrodeformationsdifferenzgeschwindigkeit}{x}
\newcommand{\microdefdiffratecomp}{\rate{\microdefdiffcomp}}
\newsym{\microdefdiffm}{e^{\upmath{m}}}{Mikroddilatationsdifferenz}{em}
\newsym{\microdefdiffdev}{\microdefdiff^{\upmath{ds}}}{Mikrodeviatordifferenz}{eds}
\newcommand{\microdefdiffdevcomp}{\microdefdiffcomp^{\upmath{ds}}}
\newsym{\microdefdiffskw}{\microdefdiff^{\upmath{skw}}}{Mikroantimetrikdifferenz}{ea}

\newcommand{\microdefgradgradcomp}[1][{}]{K^{#1}}
\newcommand{\microdefgradgradratecomp}[1][{}]{\rate{\microdefgradgradcomp}^{\!#1}}

\newcommand{\microdefgradgradsymcomp}[1][]{\microdefgradgradcomp[\upmath{s}#1]}

\newcommand{\microdefgradgradsymdevcomp}{\microdefgradgradsymcomp[\upmath{d}]}

\newcommand{\microdefgradgradsymhydcomp}{\microdefgradgradcomp[\upmath{h}]}
\newcommand{\microdefgradgradsymhydratecomp}{\microdefgradgradratecomp[\upmath{h}]}

\newcommand{\microdefgradgradsymdevhydcomp}{\microdefgradgradcomp[\upmath{dh}]}
\newcommand{\microdefgradgradsymdevhydratecomp}{\microdefgradgradratecomp[\upmath{dh}]}

\newcommand{\microdefgradgradfulldevsymcomp}{\microdefgradgradcomp[\upmath{dds}]}
\newcommand{\microdefgradgradfulldevsymratecomp}{\microdefgradgradratecomp[\upmath{dds}]}

\newcommand{\microdefgradgradsymdevskewaxmcomp}{\microdefgradgradcomp[\upmath{d\times}]}
\newcommand{\microdefgradgradsymdevskewaxmratecomp}{\microdefgradgradratecomp[\upmath{d\times}]}

\newcommand{\microdefgradgradsymratecomp}[1][]{\microdefgradgradratecomp[\upmath{s}#1]}

\newcommand{\stressdiffcomp}{s}
\newsym{\stressdiff}{\tens{\stressdiffcomp}}{Difference stress}{s}

\newcommand{\stressdiffdevcomp}{\stressdiffcomp^{\upmath{d}}}

\newcommand{\stressdiffskwcomp}{\stressdiffcomp^{\upmath{skw}}}
\newcommand{\hyperstresscomp}[1][{}]{M^{#1}}

\newcommand{\hyperstresshydrocomp}{\hyperstresscomp[\upmath{h}]}

\newcommand{\hyperstressmodhydro}{a_{\upmath{h}}}
\newcommand{\hyperstressmodhydrodevhydro}{a_{\upmath{h-dh}}}
\newcommand{\hyperstressmodhydrocompl}{\tilde{a}_{\upmath{h}}}
\newcommand{\hyperstressmodhydrodevhydrocompl}{\tilde{a}_{\upmath{h-dh}}}

\newcommand{\hyperstressdevhydrocomp}{\hyperstresscomp[\upmath{dh}]}
\newcommand{\hyperstressmoddevhydro}{a_{\upmath{dh}}}
\newcommand{\hyperstressmoddevhydrocompl}{\tilde{a}_{\upmath{dh}}}

\newcommand{\hyperstressfulldevsymcomp}{\hyperstresscomp[\upmath{dds}]}
\newcommand{\hyperstressmodfulldevsym}{a_{\upmath{dds}}}

\newcommand{\hyperstressdevskewcomp}{\hyperstresscomp[\upmath{dskw}]}

\newcommand{\hyperstressdevskewaxmcomp}{\hyperstresscomp[\upmath{d}\times]}
\newcommand{\hyperstressmoddevskewax}{a_{\upmath{dskw}}}

\newsym{\microinert}{\tens{G}^{\density}}{Microinertia}{I}

\newcommand{\shearmodeffstrain}{\bar{\mu}}
\newcommand{\Lamefirsteffstrain}{\bar{\lambda}}
\newcommand{\shearmodeffmicro}{\bar{\mu}_{\microdefcomp}}
\newcommand{\shearmodeffmixed}{g_2}

\newcommand{\microdefgradgradinccomp}[1][]{\microdefgradgradcomp[\upmath{inc}#1]}
\newcommand{\microdefgradgradincratecomp}[1][]{\microdefgradgradratecomp[\upmath{inc}#1]}

\newcommand{\microdefgradgradincaxcomp}[1][{}]{\microdefgradgradinccomp[\times#1]}
\newcommand{\microdefgradgradincaxdevcomp}{\microdefgradgradincaxcomp[\upmath{d}]}
\newcommand{\microdefgradgradincaxhydro}{\microdefgradgradincaxcomp[\upmath{h}]}
\newcommand{\microdefgradgradincaxskwaxcomp}{\microdefgradgradincaxcomp[\times]}
\newcommand{\microdefgradgradincaxratecomp}[1][{}]{\microdefgradgradincratecomp[\times#1]}
\newcommand{\microdefgradgradincaxdevratecomp}{\microdefgradgradincaxratecomp[\upmath{d}]}
\newcommand{\microdefgradgradincaxhydrorate}{\microdefgradgradincaxratecomp[\upmath{h}]}
\newcommand{\microdefgradgradincaxskwaxratecomp}{\microdefgradgradincaxratecomp[\times]}

\newcommand{\hyperstressinccomp}[1][{}]{\hyperstresscomp[\upmath{inc}#1]}
\newcommand{\hyperstressincaxcomp}[1][{}]{\hyperstresscomp[\upmath{inc}\times#1]}
\newcommand{\hyperstressincaxdevcomp}{\hyperstressincaxcomp[\upmath{d}]}
\newcommand{\hyperstressmodincaxdev}{a_{\upmath{inc\times d}}}
\newcommand{\hyperstressmodincaxdevsymskw}{a_{\upmath{inc}-\upmath{dskw}}}
\newcommand{\hyperstressincaxhydro}{\hyperstressincaxcomp[\upmath{h}]}
\newcommand{\hyperstressmodincaxhydro}{a_{\upmath{inc\times h}}}
\newcommand{\hyperstressincaxskwaxcomp}{\hyperstressincaxcomp[\times]}
\newcommand{\hyperstressmodincaxskwax}{a_{\upmath{inc\!\times\!\times}}}
\newcommand{\hyperstressmodhydroincaxskwax}{a_{\upmath{h-inc\!\times\!\times}}}
\newcommand{\hyperstressmoddevhydroincaxskwax}{a_{\upmath{dh-inc\!\times\!\times}}}

\newcommand{\microdilat}{\chi^{\upmath{v}}}

\newcommand{\microdilatdiff}{e^{\upmath{v}}}

\newcommand{\innerstressdilat}{\overline{\sigma}_{\upmath{h}}}
\newcommand{\stressmacrohydro}{\stressmacrocomp^{\upmath{h}}}

\newcommand{\stressdiffdilat}{\stressdiffcomp_{\upmath{h}}}

\newcommand{\compmodeffstrain}{\compmodeff[\strainmacrocomp]\!}
\newcommand{\compmodeffmicrodilat}{\compmodeff[\microdefcomp]}
\newcommand{\compmodeffmixed}{\compmodeff[\strainmacrocomp\microdefcomp]}
\newcommand{\microrotcomp}{\Phi^{\upmath{r}}}
\newcommand{\CossParamSkew}{\kappa}

\newcommand{\CossCharlengthtorsion}{l_{\upmath{tCoss}}}

\newsym{\microstrainrate}{\tens{L}^{\chi\upmath{s}}}{Mikrodeformation}{L}

\newsym{\microstraindiffrate}{\tens{L}^{e}\upmath{s}}{Mikrodeformationsdifferenzgeschwindigkeit}{L}
\newsym{\microstrainratecomp}{L^{\chi\upmath{s}}}{Mikrodeformation}{L}
\newcommand{\microstraingradcomp}[1][{}]{K^{\chi\upmath{s}}}
\newsym{\yieldcond}{\Phi}{Fließbedingung}{f}
\newsym{\yieldstress}{\sigma_0}{yield stress}{s0}
\newsym{\plastmul}{\lambda}{plastischer Multiplikator}{l}
\newsym{\emod}{E^{\upmath{(m)}}}{Elastizitätsmodul}{e}
\newsym{\Poissrat}{\nu}{Querkontraktionszahl}{n}
\newsym{\compmod}{K^{\upmath{(m)}}}{Kompressionsmodul}{k}

\newcommand{\Lamesec}{\mu}
\newcommand{\Lamefirst}{\lambda} %
\newsym{\tstiff}{\Tens{S}}{elastischer Steifigkeitstensor}{c} %^{\upmath{(m)}}
\newsym{\microcompmod}{\tilde{K}}{Mikrokompressionsmodul}{k}
\newsym{\tstiffhigherorder}{\tilde{\vect{C}}}{Mikrosteifigkeit höherer Ordnung}{c}
\newsym{\strainpl}{\strain_{\upmath{pl}}}{plastische Dehnung}{e}
\newsym{\strainratepl}{\dot{\strain}_{\upmath{pl}}}{plastische Dehnungrate}{e}
\newsym{\strainel}{\strain_{\upmath{el}}}{elastische Dehnung}{e}
\newsym{\strainrateplhighorder}{\dot{\vect{K}}^{\upmath{pl}}}{plastische Mikrodehnungsrate 2. Ordnung}{k}

\newsym{\displapprox}{\tilde{\displ}}{lokale Verschiebungsapproximation}{u}
\newcommand{\fluctuation}[1]{\Delta{#1}}
\newsym{\velocityapprox}{\tilde{\velocity}}{lokale Geschwindigkeitssapproximation}{v}
\newcommand{\displmacrocomp}{U}
\newsym{\displmacro}{\vect{\displmacrocomp}}{makroskop.\ Verschiebung}{U}
\newcommand{\vLagrangemultdisplcomp}{\Lagrangemult[\displmacrocomp]} 
 
\newsym{\strainmacrocomp}{E}{makroskop.\ Dehnung}{E}
\newsym{\strainmacro}{\tens{\strainmacrocomp}}{makroskop.\ Dehnung}{E}

\newcommand{\strainmacrodilat}{\strainmacrocomp^{\upmath{v}}}

\newcommand{\strainmacrodevcomp}{\strainmacrocomp^{\upmath{d}}}
\newsym{\defratemacro}{\rate{\strainmacro}}{makroskop.\ Deformationsgeschwindigkeit}{D}
\newcommand{\defratemacrocomp}{\rate{\strainmacrocomp}}

\newsym{\defgradmacro}{\bar{\tens{F}}}{makroskop.\ Deformationsgradient}{F}

\newsym{\densitymacro}{\overline{\density}}{makroskop.\ Dichte}{rho}
\newcommand{\volforcemacrocomp}{\overline{f}}

\newcommand{\displmicrocomp}{\displcomp}

\newsym{\defratemicro}{\defrate}{mikroskop.\ Deformationsgeschwindigkeit}{d}
\newsym{\displerror}{\Delta{\displ}}{lokale Verschiebungsdifferenz}{Du}
\newsym{\velocityerror}{\Delta{\velocity}}{lokale Geschwindigkeitsdifferenz}{Dv}

\newcommand{\locmacrocomp}{X}
\newsym{\locmacro}{\vect{\locmacrocomp}}{makroskop. Ort}{X}
\newcommand{\locmicrocomp}{x}
\newsym{\locmicro}{\vect{\locmicrocomp}}{mikroskop. Ort}{x}
\newcommand{\locmicrorelcomp}{x}
\newsym{\locmicrorel}{\vect{\locmicrorelcomp}}{mikroskop. Ort}{x}

\newsym{\locmacroinit}{\locmacro_0}{makroskop. Ort}{X}

\newsym{\locmicroinit}{\locmicro_0}{mikroskop. Ort}{x}

\newsym{\locmicrorelinit}{\locmicrorel_0}{mikroskop. Ort}{x}
\newcommand{\stressmacrocomp}{\Sigma}

\newcommand{\stressmacrodevcomp}{\dev{\stressmacrocomp}}

\newcommand{\innerstressmacrocomp}{\bar{\sigma}}

\newcommand{\innerstressmacrodevcomp}{\dev{\innerstressmacrocomp}}

\newcommand{\strainenergymacro}[1][{}]{\overline{\strainenergy}_{#1}}

\newcommand{\strainenergymacroint}{\strainenergymacro}

\newcommand{\averop}[2][V]{\left\langle#2\right\rangle_{#1}} %_{\domaincell}
\newcommand{\averopmatrix}[1]{\left\langle#1\right\rangle_{\upmath{M}}}
\newcommand{\weightfuncmatrix}[1][\locmicrorel]{\upmath{H}_{\upmath{M}} \ifthenelse{\equal{#1}{}}{}{\!\left(#1\right)}}
\newcommand{\domainmatrix}{\domain[\upmath{M}]}

\newcommand{\weightfuncmicro}[1][\locmicrorel]{\upmath{H}_{\upmath{M}} \ifthenelse{\equal{#1}{}}{}{\!\left(#1\right)}}

\newcommand{\domaincell}[1][{}]{\Delta V#1}

\newcommand{\domaincellbound}[1][{}]{\surface{\domaincell[]}#1}

\newcommand{\domaincellboundhalf}[1][(\locmacro)]{\surface{\domaincell[]}^+#1}

\newcommand{\ointcell}{\oint\limits_{\domaincellbound[]}\!\!}
\newcommand{\geommomcomp}{G}
\newsym{\geommom}{\tens{\geommomcomp}}{geometr. Moment}{G}
\newsym{\geommommom}{\Tens{G}^{\upmath{M}}}{geometr. Moment 4. Ordnung}{GM}

\newcommand{\geommommicrocomp}{\geommomcomp^{\microdefcomp}}
\newcommand{\geommommicroinvcomp}{\geommomcomp^{\microdefcomp-1}}

\newsym{\Helmenergymicro}{\psi}{lokale Helmholtz-Energie}{P}
\newsym{\Helmenergymacro}{\Psi}{makroskopische Helmholtz-Energie}{P}

\newcommand{\radcell}{R}

\newcommand{\radvoid}{R_{\upmath{v}}}
\newsym{\VVF}{c}{Porenvolumenanteil}{c}
\newcommand{\averopsphereshell}[2][r]{\left\langle#2\right\rangle_{S(#1)}} %_{\domaincell}

\newcommand{\compmodeff}[1][\upmath{eff}]{\bar{K}_{#1}}
\newcommand{\shearmodeff}[1][\upmath{eff}]{\bar{\mu}_{#1}}

\newcommand{\NavierconstA}{C_1}
\newcommand{\NavierconstB}{C_2}

\newcommand{\sphericalharmonicorder}{m}
\newcommand{\sphericalharmonic}[1][\sphericalharmonicorder]{\Phi^{(#1)}}
\newcommand{\spheresolomegacoeff}[1][\sphericalharmonicorder]{\alpha_{(#1)}}
\newcommand{\tractioncomprad}[2][r]{\left.\tractioncomp_{#2}\right|_{#1}}

\newcommand{\intlengthtorsionsym}{\ell_{\mathrm{tsym}}}
\newcommand{\intlengthtorsionskw}{\ell_{\mathrm{tskw}}}
\newcommand{\torque}{T}
\newcommand{\rotangletorsion}{\Phi}
\newcommand{\twist}{\theta}
\newcommand{\torsioncylinderrad}{R_{\circ}}
\newcommand{\torsionmicrostrainfunc}{g}
\newcommand{\torsionmicrorotfunc}{h}
\newcommand{\locmacrocompplane}{\tilde{\locmacrocomp}}
\newcommand{\locmacroradplane}{\tilde{r}}
\newcommand{\crosssec}{A_{\circ}}
\newcommand{\torsionstiffness}{J}
\newcommand{\torsionrelstiffness}{\Omega}

\definecolor{gray}{rgb}{0.8,0.8,0.8}
\newcommand{\zz}{\textcolor{gray}{0}}
\newcommand{\inputsvg}[1]{\includegraphics{#1}}
\newcommand{\inputgnuplot}[1]{\includegraphics{#1}}

\title{Interpretation of micromorphic constitutive relations for porous materials at the microscale via harmonic decomposition}
\subtitle{published in Journal of the Mechanics and Physics of Solids 171 (2023), 105135, DOI:\href{https://www.doi.org/10.1016/j.jmps.2022.105135}{10.1016/j.jmps.2022.105135}}

\author{Geralf Hütter\thanks{Brandenburg University of Technology Cottbus-Senftenberg, Institute of Civil and Structural Engineering, Cottbus, Germany, Geralf.Huetter@b-tu.de} \thanks{Technische Universität Bergakademie Freiberg, Institute of Mechanics and Fluid Dynamics, Germany, Geralf.Huetter@imfd.tu-freiberg.de}}

\begin{document}

\maketitle

\begin{abstract}
Micromorphic theories became an established tool to model size effects in materials like dispersion, localization phenomena or (apparently) size dependent properties. However, the {formulation of adequate constitutive relations with its} large number of constitutive {relations and respective} parameters hinders the usage of the full micromorphic theory, which has 18 constitutive parameters already in the isotropic linear elastic case. Although it is clear that these parameters are related to predicted size effects, the individual meaning of single parameters has been rather unclear.

The present work tries to elucidate the interpretation of {the constitutive relations and their parameters}. For this purpose, a harmonic decomposition is applied to the governing equations of micromorphic theory. The harmonic modes are interpreted at the microscale using a homogenization method for a simple volume element with spherical pore. The resulting boundary-value problem at the microscale is solved analytically {for the linear-elastic case} using spherical harmonics resulting in closed-form expressions for all of the elastic 18 parameters. These values are used to predict the size effect in torsion of slender foam specimens. The predictions are compared with respective experimental results from literature.

%\begin{figure}
%\centering
%\inputsvg{GraphicalAbstract}
%\end{figure}

\end{abstract}

%=====================================================================================
\section{Introduction}
\label{sec:introduction}
%=====================================================================================

Since its initial presentation by \citet{Mindlin1964} and Eringen \citep{Eringen1964}, the micromorphic theory has become a valuable framework for modeling size effects within the mechanical behavior of materials.
Although the potential of this theory is praised widely, mostly only subclasses of the full micromorphic theory like Cosserat theory, microdilatational theory or strain gradient theory are employed. The reason can be found in the large number of constitutive parameters in the full micromorphic theory, amounting to 18 independent parameters even in the linear-elastic, isotropic case.
To the author's knowledge, no attempt has been made yet to determine all of them for a particular (real or virtual) material. {For inelastic material behavior, the construction of yield functions or dissipation potentials has an even greater level of freedom, which has to be filled.}
Shortly after the micromorphic theory had been developed, homogenization techniques \citep{Adomeit1968,Herrmann1968,Sun1968,Achenbach1976,Tauchert1969} have been recognized as a promising tool for providing the required constitutive relations. 
These initial attempts were tailored to particular microstructures and theories. Recently, homogenization methods have been developed \citep{Huetter2017,Biswas2017,Rokos2018} which allow to use any linear or nonlinear material model at the microscale, e.g.\ within micromorphic FE\textsuperscript{2} simulations.
Despite their huge potential, such methods require to use a regularly shaped volume element of the microstructure. Typically a rectangular or cuboidal volume element is used which introduces an artificial anisotropy (even a hexagonal volume element would do so in micromorphic theory \citep{Auffray2010}) and can thus hardly contribute to a fundamental understanding of the micromorphic theory.

The present work tries to shed some light on the interpretation of the {constitutive relations} of the micromorphic theory and the interaction of micropolar, microdilatational and microstrain components therein. For this purpose, a {unique and irreducible decomposition of the individual terms into harmonic tensors} is performed, called shortly \emph{harmonic decomposition} in the following. Each of the harmonic {tensors} is interpreted at the microscale by means of the composite shell model, i.e., analytical homogenization of a spherical volume element, which is geometrically isotropic. 

The manuscript is structured as followed: Section~\ref{sec:micromorphic} briefly recalls the micromorphic theory before the harmonic decomposition is performed in section~\ref{sec:orthogonal_decomposition}. 
The respective microscale problems are solved in closed form and are evaluated in section~\ref{sec:micromorphelastic}. Finally, the resulting predictions for the size effect in torsion specimens of small diameter are evaluated in section~\ref{sec:torsion}. The contribution closes with a summary. 

Aiming at particular {constitutive relations}, the present contribution is written completely in conventional index notation {and focuses on the infinitesimal deformation case}.

%=====================================================================================
\section{Micromorphic theory}
\label{sec:micromorphic}
%=====================================================================================

%------------------------------------------------------
\subsection{General equations}
%------------------------------------------------------

\begin{figure}
	\centering
	\inputsvg{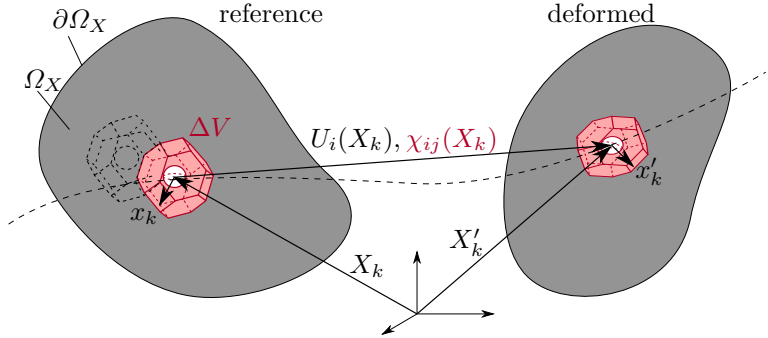}%{micromorphic_continuum_indexnot}
	\caption{Micromorphic continuum}
	\label{fig:micromorphic continuum}
\end{figure}

The micromorphic theory of \citet{Mindlin1964} and Eringen \citep{Eringen1964} introduces the  second-order tensor field $\microdefcomp_{ij}(\locmacrocomp_k)$, called field of \emph{microdeformation}, as additional kinematic degree of freedom besides the conventional displacements $\displmacrocomp_i(\locmacrocomp_k)$ as shown schematically in \figurename~\ref{fig:micromorphic continuum}. 
The microdeformation tensor is intended to allow for a more precise description of deformations of the microstructure of a material. 
The micromorphic theory is a local first order theory in the sense that all objective combinations of these quantities and their first gradients are considered as measures of deformation. 
In addition to the conventional strain tensor $\strainmacrocomp_{ij}=1/2(\partial\displmacrocomp_i/\partial\locmacrocomp_j+\partial\displmacrocomp_j/\partial\locmacrocomp_i)$, these are the relative deformation tensor $\microdefdiffcomp_{ij}=\partial\displmacrocomp_i/\partial\locmacrocomp_j-\microdefcomp_{ij}$ and the gradient of microdeformation 
$\microdefgradgradcomp_{ijk}=\partial\microdefcomp_{ij}/\partial\locmacrocomp_k$.
Note that the order of indices in the non-classical terms is defined partly differently in literature\footnote{Even the employed sets of independent measures of deformations are different in literature. \citet{Eringen1999} uses, equivalently, $\strainmacrocomp^{\microdefcomp}_{ij}=1/2(\microdefcomp_{ij}+\microdefcomp_{ji})$  (denoted as $e_{ij}$ in \citep[Sec.~7.1]{Eringen1999}) instead of $\strainmacrocomp_{ij}$.}. 
Here, we follow the notation of \citet{Forest2019,Forest2013_CISM} which is most widely used nowadays.

The work conjugate stresses to these measures of deformations are derived for hyperelastic material from a strain-energy potential $\strainenergymacroint(\strainmacrocomp_{ij},\microdefdiffcomp_{ij},\microdefgradgradcomp_{ijk})$ as
\begin{align}
 	\innerstressmacrocomp_{ij}&=\partderivf{\strainenergymacroint}{\strainmacrocomp_{ij}}\,, \label{eq:innerstressmacropotentialMindlin} &
% \intertext{and a \enquote{double stress} (???)} 	
 	\stressdiffcomp_{ij}&=\partderivf{\strainenergymacroint}{\microdefdiffcomp_{ij}}\,, &
 	\hyperstresscomp_{ijk}&=\partderivf{\strainenergymacroint}{\microdefgradgradcomp_{ijk}}\,.
\end{align}
The internal stress is symmetric, $\innerstressmacrocomp_{ij}=\innerstressmacrocomp_{ji}$, as its work-conjugate quantity $\strainmacrocomp_{ij}$.
In absence of inertia and higher-{order} body forces, these stresses satisfy the conventional equilibrium condition
\begin{align}
	0&=\partderivf{\left(\innerstressmacrocomp_{ij}+\stressdiffcomp_{ji}\right)}{\locmacrocomp_i}+\densitymacro\volforcemacrocomp_j
	\label{eq:momentumbalancemacro}
\end{align}
and a higher-order balance 
\begin{align}
  0&=\partderivf{\hyperstresscomp_{ijk}}{\locmacrocomp_k}+\stressdiffcomp_{ij}\,.
  \label{eq:momentumbalancehyperstress}
\end{align}
Introducing the external stress as $\stressmacrocomp_{ij}:=\innerstressmacrocomp_{ij}+\stressdiffcomp_{ji}$, equilibrium condition~\eqref{eq:momentumbalancemacro} takes the usual form $\partial\stressmacrocomp_{ij}/\partial\locmacrocomp_i+\densitymacro\volforcemacrocomp_j=0$.
The respective work theorem for a hyperelastic material reads
\begin{align}
  \int_{\domain} \rate{\strainenergymacroint}\dV= \int_{\domainbound} n_i \stressmacrocomp_{ij} \rate{\displmacrocomp}_j + n_k \hyperstresscomp_{ijk} \microdefratecomp_{ij} \dsurf+\int_{\domain} \densitymacro\volforcemacrocomp_i \rate{\displmacrocomp}_i \dV\,.
	\label{eq:worktheorem}
\end{align}

%------------------------------------------------------
\subsection{Isotropic, linear-elastic material}
\label{sec:isotropiclaw}
%------------------------------------------------------

The general form of the strain-energy density for centro-symmetric, isotropic and linear-elastic material has been derived by \citet{Mindlin1964} as
\begin{equation}
	\begin{split}
	\strainenergymacro=&
	 \shearmodeffstrain \, \strainmacrocomp_{ij} \, \strainmacrocomp_{ij} 
		+ \frac{\Lamefirsteffstrain}{2} \, \strainmacrocomp_{ii} \, \strainmacrocomp_{jj}
		+ \frac{b_1}{2} \, \microdefdiffcomp_{ii} \, \microdefdiffcomp_{jj} 
		+ \frac{b_2}{2} \, \microdefdiffcomp_{ij} \, \microdefdiffcomp_{ij} 
		+ \frac{b_3}{2} \, \microdefdiffcomp_{ij} \, \microdefdiffcomp_{ji}
		+ g_1 \, \microdefdiffcomp_{ii} \, \strainmacrocomp_{jj}  
		+ g_2 \, \left(\microdefdiffcomp_{ij} + \microdefdiffcomp_{ji} \, \right)\strainmacrocomp_{ij} \\
%------------------------------------------		
		&+ a_1 \, \microdefgradgradcomp_{kii} \, \microdefgradgradcomp_{jjk} 
		+ a_2 \, \microdefgradgradcomp_{kii} \, \microdefgradgradcomp_{jkj}
		+ \frac{1}{2} \, a_3 \, \microdefgradgradcomp_{kii} \, \microdefgradgradcomp_{kjj} 
		+ \frac{1}{2} \, a_4 \, \microdefgradgradcomp_{jji} \, \microdefgradgradcomp_{kki} 
		+ a_5 \, \microdefgradgradcomp_{jji} \, \microdefgradgradcomp_{kik}
		+ \frac{1}{2} \, a_8 \, \microdefgradgradcomp_{iji} \, \microdefgradgradcomp_{kjk} \\
		&+ \frac{1}{2} \, a_{10} \, \microdefgradgradcomp_{ijk} \, \microdefgradgradcomp_{ijk} 
		+ a_{11} \, \microdefgradgradcomp_{kji} \, \microdefgradgradcomp_{ikj}
		+ \frac{1}{2} \, a_{13} \, \microdefgradgradcomp_{kji} \, \microdefgradgradcomp_{jki} 
		+ \frac{1}{2} \, a_{14} \, \microdefgradgradcomp_{kji} \, \microdefgradgradcomp_{kij}
		+ \frac{1}{2} \, a_{15} \, \microdefgradgradcomp_{kji} \, \microdefgradgradcomp_{ijk}
  \end{split}
	\label{eq:strainenergymacroisotropic}
\end{equation}
in terms of 18 Lamé-type constants. Note that {the conventional Lamé parameters $\shearmodeffstrain$ and $\Lamefirsteffstrain$} are introduced here with an overbar to distinguish them from their microscale counterparts.
The corresponding stresses according to \eqref{eq:innerstressmacropotentialMindlin} read
\begin{align}
	\innerstressmacrocomp_{ij}=&2\shearmodeffstrain \, \strainmacrocomp_{ij} +\Lamefirsteffstrain \, \strainmacrocomp_{kk} \, \Kron{ij} + g_1 \, \microdefdiffcomp_{kk} \, \Kron{ij} 
		+ g_2 \, \left(\microdefdiffcomp_{ij} + \microdefdiffcomp_{ji} \, \right)
		\label{eq:innerstressisotropic} \\
	\stressdiffcomp_{ij}=&b_1 \, \microdefdiffcomp_{kk} \, \Kron{ij}
		+ b_2 \, \microdefdiffcomp_{ij} + b_3 \, \microdefdiffcomp_{ji} 
		+ g_1 \,  \strainmacrocomp_{kk}\Kron{ij} + 2g_2 \, \strainmacrocomp_{ij} \label{eq:stressdiffisotropic}\\
%	\stressmacrocomp_{ij}&=...\\
	\hyperstresscomp_{ijk}=&a_1\left(\Kron{jk}\microdefgradgradcomp_{mmi}\!+\!\Kron{ij}\microdefgradgradcomp_{kmm}\right)
	+ a_2 \, \left(\Kron{jk} \microdefgradgradcomp_{mim}\!+\!\Kron{ik}\microdefgradgradcomp_{jmm}  \right)
	 + a_3 \microdefgradgradcomp_{imm} \Kron{jk}
		+ a_4 \microdefgradgradcomp_{mmk} \Kron{ij} \nonumber \\
		&+ a_5 \left(\Kron{ij} \microdefgradgradcomp_{mkm}\!+\!\microdefgradgradcomp_{mmj}\Kron{ik}\right)
		+ a_8 \Kron{ik} \microdefgradgradcomp_{mjm} 
	 + a_{10} \microdefgradgradcomp_{ijk}  
		+ a_{11} \left( \microdefgradgradcomp_{kij}\!+\!\microdefgradgradcomp_{jki}\right)
		+ a_{13} \microdefgradgradcomp_{jik} \nonumber \\
		&+ a_{14} \microdefgradgradcomp_{ikj}
		+ a_{15} \microdefgradgradcomp_{kji}
	\label{eq:hyperstressisotropic}
\end{align}

%=====================================================================================
\subsection{Homogenization}
\label{sec:homogenization}
%=====================================================================================

In this section, the theory from \citep{Huetter2019,Huetter2017} for the homogenization of a heterogeneous Cauchy material at the microscale towards a macroscopic micromorphic continuum is recalled briefly. This theory is a synthesis of the homogenization approaches of \citet{Eringen1968} and \citet{Forest2002} that contains the second-order homogenization of \citet{Kouznetsova2002} as a special case. Generally, lower-case symbols refer to microscopic quantities whereas macroscopic quantities are denoted by capital letters or by an overbar. 
Firstly, the micro-macro relations for the kinetic quantities read
\begin{align}
 	%\stressmacro&=\frac{1}{\domaincell[]}\ointcell \locmicrorel\tprod \left(\vnorm \sprod \stress\right)\dsurf\,, &
	%\innerstressmacro&=\averop{\stress}\,, &
	%\hyperstress&=\frac{1}{\domaincell[]}\ointcell  \left(\vnorm \sprod \stress\right) \tprod \locmicrorel\locmicrorel\dsurf\, \nonumber\\
	\stressmacrocomp_{ij}&=\frac{1}{\domaincell[]}\ointcell \locmicrorelcomp_i \vnormcomp_k \stresscomp_{kj}\dsurf\,, &
	\innerstressmacrocomp_{ij}&=\averop{\stresscomp_{ij}}\,, &% \label{eq:averopinnerstress} &
	\hyperstresscomp_{ijk}&=\frac{1}{\domaincell[]}\ointcell  \vnormcomp_k \stresscomp_{ki} \locmicrorelcomp_j\locmicrorelcomp_k\dsurf\,.
	\label{eq:kineticmicromacro}
\end{align}
%$\locmicrorel=\locmicro-\locmacro$
Therein, it is assumed without loss of generality that the origin of the coordinate system corresponds to the geometric center of the microstructural volume element $\domaincell[]$ (having a boundary $\domaincellbound[]$ with normal $\vnormcomp_i$). Note that the hyperstress $\hyperstresscomp_{ijk}$ is symmetric a priori with respect to $j$ and $k$. The volume average over the volume element is denoted as usual as
$\averop{\placeholder}=1/\domaincell[] \int_{\domaincell[]}\placeholder \dV$.
%$\microinert=\averop{\density\locmicrorel\,\locmicrorel}$ 
%$\volforcehighorder=\averop{\density\volforce\,\locmicrorel}/\densitymacro$ 
%$\densitymacro=\averop{\density}$.

The micro-macro {relation} for the displacement is $\displmacrocomp_i=\averopmatrix{\displmicrocomp_i}$. Thereby, $\averopmatrix{\placeholder}$ denotes the volume average over the matrix (since a displacement field cannot be defined uniquely in a pore, cf.~\citep{Huetter2017a}).
The micro-macro relation for the displacement gradient
\begin{align}
	%\gradmacro{\displmacro}=&\frac{1}{\domaincell[]}\ointcell \vnorm\tprod\displmicro \dsurf
	\partderivf{\displmacrocomp_i}{\locmacrocomp_j}=&\averop{\partderivf{\displcomp_i}{\locmicrocomp_j}}=\frac{1}{\domaincell[]}\ointcell \displmicrocomp_i\vnormcomp_j \dsurf
	\label{eq:velocitygradmacrovoids}
\end{align}
corresponds to the classical theory of \citet{Hill1963}.
The tensor of microdeformation 
\begin{align}
	%\displmacro=&\averopmatrix{\displmicro},& %\label{eq:displmacro} &
	%\microdef=&\frac{1}{|\discontsurf|} \int\limits_{\discontsurf} \displmicro\tprod\locmicrorel \dsurf \sprod\geommommicroinv
 		%& \text{with }
 		%\geommommicro&=\frac{1}{|\discontsurf|} \int\limits_{\discontsurf} \locmicrorel\tprod\locmicrorel \dsurf\,.
	\microdefcomp_{ij}=&\frac{1}{|\discontsurf|} \int\limits_{\discontsurf} \displmicrocomp_i\locmicrorelcomp_{k} \dsurf\, \geommommicroinvcomp_{kj}
 		& \text{with }
 		\geommommicrocomp_{ij}&=\frac{1}{|\discontsurf|} \int\limits_{\discontsurf} \locmicrorelcomp_{i}\locmicrorelcomp_{j} \dsurf\,.	
 		\label{eq:defmacrodeformationratessurface}
\end{align}
is introduced as first moment of the displacements at a surface $\discontsurf$ of internal heterogeneity (the void surface in the following application). The second geometric moment $\geommommicrocomp_{ij}$ of $\discontsurf$ ensures that $\microdefcomp_{ij}$ is identical to the macroscopic state in the case of homogeneous deformations. 
The micro-macro relation for the symmetric part of the gradient of microdeformation (as explained below in Section~\ref{sec:orthogonal_decomposition})
\begin{align}
\microdefgradgradsymcomp_{ijk}
=&\frac{1}{4\domaincell[]}\left[\ointcell \displcomp_{i} n_k \locmicrorelcomp_m \dsurf \inv{\geommomcomp}_{jm}+ \ointcell \displcomp_{i} n_j \locmicrorelcomp_m \dsurf \inv{\geommomcomp}_{km} -\frac{2}{2+\ndim} \ointcell  \displcomp_{i} n_m\locmicrorelcomp_m \dsurf \inv{\geommomcomp}_{jk} \right] -\frac{1}{2+n}\displmacrocomp_i \inv{\geommomcomp}_{jk}\,.
\label{eq:microdefgradgradmicrosym}
\end{align}
comprises a surface integral of microscopic displacements and a contribution of the macroscopic displacement in the spherical part (which is necessary for objectivity reasons and appears also in strain gradient theory \citep{Luscher2010}), both terms being weighted with the geometric moment $\geommomcomp_{ij}=\averop{\locmicrorelcomp_i \locmicrorelcomp_j}$ of the volume element.
Therein, $\ndim=\Kron{kk}$ refers to the dimension of space (either $\ndim=2$ or $\ndim=3$).
The generalized Hill-Mandel lemma
\begin{equation}
	\averop{\stresscomp_{ij}\defratecomp_{ij}}
=\innerstressmacrocomp_{ij}\defratemacrocomp_{ij}+\stressdiffcomp_{ij}\microdefdiffratecomp_{ij}+\hyperstresscomp_{ijk}\microdefgradgradsymratecomp_{ijk}
	\label{eq:energybalanceaveragedmomentbalanceColemanNollmech}
\end{equation}
imposes restrictions on the choice of boundary conditions. In the following, periodic boundary conditions
\begin{align}
\displmicrocomp_i&=\displmacrocomp_i+\partderivf{\displmacrocomp_i}{\locmacrocomp_j}\locmicrorelcomp_j +\microdefgradgradsymcomp_{ijk}\locmicrorelcomp_j\locmicrorelcomp_k+ \fluctuation{\displmicrocomp_i}& \text{on }& \domaincellbound
	\label{eq:periodicBCmicromorphic} \text{ with }
	%\fluctuation{\displmicro}(\locmicrorel^+)=\fluctuation{\displmicro}(\locmicrorel^-)
	\fluctuation{\displmicrocomp_i}(\locmicrorelcomp_k^+)=\fluctuation{\displmicrocomp_i}(\locmicrorelcomp_k^-)
%	\label{eq:periodicfluctuations}
\end{align}
are employed with a fluctuation field $\fluctuation{\displmicrocomp_i}$. Note that \eqref{eq:periodicBCmicromorphic} does not satisfy the kinematic micro-macro relations for the displacement and for the gradient of microdeformation~\eqref{eq:microdefgradgradmicrosym} {a priori}. 
Rather, these conditions and \eqref{eq:periodicBCmicromorphic} are enforced by Lagrange multipliers leading to a functional (after eliminating $\fluctuation{\displmicrocomp_i}$)
\begin{equation}
		\begin{split}
		\Lagrangefunc=&\averop{\strainenergy}		
		-\frac{1}{\domaincell[]}\!\!\!\!\int\limits_{\domaincellboundhalf[]}\!\!\!\!\!\Lagrangemult[]_i(\locmicrorel^+)\left[\displcomp_i(\locmicrorel^+)-\displcomp_i(\locmicrorel^-) -\displmacrocomp_{i,j}\left(\locmicrorelcomp_j^+\!\!-\locmicrorelcomp_j^-\right)+\microdefgradgradsymcomp_{ijk}(\locmicrorelcomp_j^+\locmicrorelcomp_k^+ - \locmicrorelcomp_j^-\locmicrorelcomp_k^-)\right]\dsurf \\		
		&+\Lagrangemult_{kij}\left\{\microdefgradgradsymcomp_{ijk}\!\!\!-\frac{1}{4\domaincell[]}\left[\ointcell \displcomp_i   \locmicrorelcomp_m n_k \dsurf\, \inv{\geommomcomp}_{mj}\!+ \ointcell   \displcomp_i n_j \locmicrorelcomp_m \dsurf\,\inv{\geommomcomp}_{km}  - \!\frac{2}{2+\ndim}\!\!\ointcell  \displcomp_i n_m  \locmicrorelcomp_m \dsurf\, \inv{\geommomcomp}_{jk}\right]\right. \\		
			&\left.+\frac{1}{2+\ndim}\averopmatrix{\displcomp_i}\inv{\geommomcomp}_{jk} \right\}
			+ \vLagrangemultdisplcomp_j\left(\displmacrocomp_j-\averopmatrix{\displcomp_j} \right)
		-\stressdiffcomp_{ij}\left(\microdefcomp_{ij}-
		\frac{1}{|\discontsurf|} \int\limits_{\discontsurf} \displmicrocomp_i\locmicrorelcomp_k \dsurf\, \geommommicroinvcomp_{kj}\right)\,.
		\end{split}
		\label{eq:LagrangefuncminconstrperiodicBC}
\end{equation}
whose stationarity conditions are
\begin{align}
		n_i\sigma_{ij}&=\pm {\Lagrangemult[]}_j(\locmicrocomp_k) +\frac12 {\Lagrangemult[]}_{jkm} n_m  \inv{\geommomcomp}_{kp} \locmicrorelcomp_p- \frac{1}{2(2+\ndim)} n_i \locmicrorelcomp_i {\Lagrangemult[]}_{jmn}\inv{\geommomcomp}_{mn} & \forall \locmicrocomp_j\in\domaincellbound[] 
	\label{eq:natBCminconstrperiodic}\\
	 n_i\sigma_{ij}&=-\frac{\domaincell[]}{|\discontsurf|} \stressdiffcomp_{jk} \geommommicroinvcomp_{kl}\locmicrorelcomp_{l} & \forall \locmicrocomp_j\in\discontsurf 
	\label{eq:natBCminconstrdiscont}\\
	\stresscomp_{ij,i}&=- \frac{|\domaincell|}{|\domainmatrix|}  \vLagrangemultdisplcomp_j
		& \forall \locmicrocomp_j\in\domainmatrix
	\label{eq:momentumbalancelocalminconstrvoid}
\end{align}
The Hill-Mandel condition~\eqref{eq:energybalanceaveragedmomentbalanceColemanNollmech} shows the relation
\begin{equation}
	\hyperstresscomp_{ijk}=\Lagrangemult[]_{ijk}+\frac{1}{\domaincell[]}\!\!\!\int\limits_{\domaincellbound[]^+} \vLagrangemultdisplcomp_i\left(\locmicrorelcomp_j^+\locmicrorelcomp_k^+ -\locmicrorelcomp_j^-\locmicrorelcomp_k^-\right) \dsurf
\end{equation}
between hyperstress $\hyperstresscomp_{ijk}$ and Lagrange multiplier $\Lagrangemult[]_{ijk}$ and, due to global equilibrium of the volume element, that the Lagrange multiplier for the macroscopic displacements is related to the spherical part of the hyperstress $\vLagrangemultdisplcomp_i=-\frac{1}{\ndim+2}\hyperstresscomp_{ijk}\inv{\geommomcomp}_{jk}$. In addition to those hyperstress terms, anti-periodic tractions $\pm {\Lagrangemult[]}_j$ appear in \eqref{eq:natBCminconstrperiodic} (as part of the solution), compare \citep{Miehe2007}.

%------------------------------------------------------
\section{{Irreducible decomposition into harmonic parts}}
\label{sec:orthogonal_decomposition}
%------------------------------------------------------

{The aim of the present investigation is to gain further insight into the micromorphic theory and its constitutive equations, in particular the non-classical ones, by decomposing each part of the theory into {unique and irreducible} modes which can be characterized by harmonic tensors, i.e.\ totally symmetric and totally deviatoric tensors \citep{Applequist1989,Gluege2016,Lazar2016}, in order to be able to interpret these modes at the microscale using the aforementioned homogenization approach.}

\subsection{{Measures of stress and deformation}}

{For any kind of material behavior, the measures of stress ($\innerstressmacrocomp_{ij}$, $\stressdiffcomp_{ij}$, $\hyperstresscomp_{ijk}$) and deformation ($\strainmacrocomp_{ij}$, $\microdefcomp_{ij}$, $\microdefdiffcomp_{ij}$, $\microdefgradgradcomp_{ijk}$) can be decomposed uniquely into their irreducible, harmonic parts. For second order tensors, this is quite common in mechanics of materials. Symmetric second order tensors like internal stress $\innerstressmacrocomp_{ij}$ and conventional strain $\strainmacrocomp_{ij}$ are split into a spherical (hydrostatic/dilatational)	and a deviatoric part.} Non-symmetric second order tensors like 
stress difference $\stressdiffcomp_{ij}=\stressdiffdilat\Kron{ij}+\stressdiffdevcomp_{ij}+\stressdiffskwcomp_{ij}$ 
incorporate additionally a skew symmetric part $\stressdiffskwcomp_{ij}=\stressdiffcomp_{[ij]}$, which can be represented by an axial vector. 
{For the microdeformation, this axial vector $\microrotcomp_k=-\microdefskwcomp_{ij}\ttpermutcomp_{ijk}=-\microdefcomp_{ij}\ttpermutcomp_{ijk}$ is known as microrotation.}

More elaborate than these standard relations is the split of the hyperstress $\hyperstresscomp_{ijk}$	and of the gradient of microdeformation $\microdefgradgradcomp_{ijk}$.
\citet{Forest2017} interpreted the microdeformation $\microdefcomp_{ij}$ as \enquote{relaxed \dots deformation gradient}. 
If $\microdefcomp_{ij}$ corresponded to the deformation gradient, then $\microdefgradgradcomp_{ijk}$ would be the second gradient and would thus have a symmetry. Consequently, we split $\microdefgradgradcomp_{ijk}$ up 
\begin{align}
	\microdefgradgradcomp_{ijk}=\microdefgradgradsymcomp_{ijk}+\microdefgradgradinccomp_{ijk}
\end{align}
into this symmetric part $\microdefgradgradsymcomp_{ijk}=(\microdefgradgradcomp_{ijk}+\microdefgradgradcomp_{ikj})/2$ and the incompatible, skew part $\microdefgradgradinccomp_{ijk}=(\microdefgradgradcomp_{ijk}-\microdefgradgradcomp_{ikj})/2$. 
The latter can be represented by a second-order axial tensor $\microdefgradgradincaxcomp_{ij}= \microdefgradgradcomp_{imn}\ttpermutcomp_{mnj}=\microdefcomp_{im,n}\ttpermutcomp_{mnj}$. This tensor corresponds to the curl of the field of micro-deformations and is employed in the  {reduced micromorphic theories of \citet{Neff2014} (\enquote{relaxed micromorphic theory}) and \citet{Teisseyre1974}}. If $\microdefcomp_{ij}$ were the plastic {distortion} field, $\microdefgradgradincaxcomp_{ij}$ would be the Nye-Kröner dislocation density tensor \citep{Nye1953}. {This second tensor order can again be split into spherical, deviatoric and skew parts, 
\begin{align}
\microdefgradgradincaxhydro&=\microdefgradgradcomp_{ijk}\ttpermutcomp_{ijk} \label{eq:microdefgradgradincaxhydro}\\
\microdefgradgradincaxdevcomp_{ij}&=\microdefgradgradincaxcomp_{(ij)}-\frac13 \Kron{ij}\microdefgradgradincaxhydro\\
\microdefgradgradincaxskwaxcomp_i&=\frac12(\microdefgradgradcomp_{mmi}-\microdefgradgradcomp_{mim})
\end{align}
with their well-known interpretations as dislocation related lattice curvatures.}

{The third-order tensor with a single sub-symmetry like $\microdefgradgradsymcomp_{ijk}$ can be decomposed as shown by \citet{Lazar2016}, and similarly by \citet{Gluege2016}, for the strain-gradient theory. Firstly, it is split
up into a spherical part $\microdefgradgradsymhydcomp_i=\microdefgradgradcomp_{imm}$ and a deviatoric part $\microdefgradgradsymdevcomp_{ijk}=\microdefgradgradsymcomp_{ijk}-1/\ndim\microdefgradgradsymhydcomp_i\Kron{jk}$ with respect to the symmetric pair of indices. 
Subsequently, the symmetric deviatoric part $\microdefgradgradsymdevcomp_{ijk}$ can again be split up into a vector
$\microdefgradgradsymdevhydcomp_i$ as trace with respect to the remaining indices, a totally symmetric and deviatoric part  $\microdefgradgradfulldevsymcomp_{ijk}$ and a skew part with respect to the other two indices, represented by a deviatoric and symmetric axial tensor $\microdefgradgradsymdevskewaxmcomp_{ij}$:}
\begin{align}
  \microdefgradgradsymdevhydcomp_i=&\microdefgradgradsymdevcomp_{kik} 
	{=\microdefgradgradcomp_{k(ik)}-\frac{1}{\ndim}\microdefgradgradcomp_{imm}}
	\label{eq:defhyperstressdevhydro}\\
	\microdefgradgradfulldevsymcomp_{ijk}=&\frac{1}{3}\left[\microdefgradgradsymdevcomp_{ijk}+\microdefgradgradsymdevcomp_{jik}+\microdefgradgradsymdevcomp_{kij}-\frac{2}{\ndim+2} \left(\microdefgradgradsymdevhydcomp_i \Kron{jk}+ \microdefgradgradsymdevhydcomp_j \Kron{ik}+ \microdefgradgradsymdevhydcomp_k \Kron{ij}\right)\right] \nonumber\\
	=&{\frac{1}{3}\!\left\lbrace\microdefgradgradcomp_{i(jk)}\!\!+\!\microdefgradgradcomp_{j(ik)}\!\!+\!\microdefgradgradcomp_{k(ij)}\right.} \nonumber \\
	&{\left.\quad-\frac{1}{\ndim\!+\!2} \left[\left(\microdefgradgradcomp_{imm}\!\!+\!2\microdefgradgradcomp_{m(im)}\!\right) \Kron{jk}\!+\!\left(\microdefgradgradcomp_{jmm}\!+\!2\microdefgradgradcomp_{m(jm)}\!\right) \Kron{ik}\!+\!\left(\microdefgradgradcomp_{kmm}\!+\!2\microdefgradgradcomp_{m(km)}\!\right) \Kron{ij}\right]\right\rbrace}\\
\microdefgradgradsymdevskewaxmcomp_{ij}=&\frac{1}{2}\left(\microdefgradgradsymdevcomp_{kil} \ttpermutcomp_{jlk}+\microdefgradgradsymdevcomp_{kjl} \ttpermutcomp_{ilk} \right)
{=\frac{1}{2}\left(\microdefgradgradcomp_{k(il)} \ttpermutcomp_{jlk}+\microdefgradgradcomp_{k(jl)} \ttpermutcomp_{ilk} \right)}
\label{eq:defhyperstressdevskewax}
\end{align}
From these contributions, {the gradient of microdeformation tensor can be (re-)composed} as
\begin{equation}
\begin{split}
\microdefgradgradcomp_{ijk}=&\microdefgradgradfulldevsymcomp_{ijk}\!+\frac{1}{\ndim}
\frac{\ndim}{(\ndim\!-\!1)(\ndim\!+\!2)}\left(\microdefgradgradsymdevhydcomp_j\Kron{ik}\!+\!\microdefgradgradsymdevhydcomp_k\Kron{ij}\right)\!-\frac{2}{(\ndim\!-\!1)(\ndim\!+\!2)} \microdefgradgradsymdevhydcomp_i \Kron{jk}
\!+\!\frac{\ndim\!-\!2}{\ndim} \left(\ttpermutcomp_{imj}\microdefgradgradsymdevskewaxmcomp_{km}\!+\!\ttpermutcomp_{kim}\microdefgradgradsymdevskewaxmcomp_{jm}  \right)\\
&+\microdefgradgradsymhydcomp_i\Kron{jk}
+\frac{\ndim\!-\!2}{2} \microdefgradgradincaxdevcomp_{im} \ttpermutcomp_{jkm}
+\frac{\ndim\!-\!2}{2\ndim} \microdefgradgradincaxhydro \ttpermutcomp_{ijk}+\frac12\left(\microdefgradgradincaxskwaxcomp_k \Kron{ij}\!-\!\microdefgradgradincaxskwaxcomp_j \Kron{ik}\right)
\end{split}
\label{eq:hyperstresssplit}
\end{equation}
Note that the skew part $\microdefgradgradsymdevskewaxmcomp_{ij}$ is relevant only for the spatial case $\ndim=3$ \citep{Auffray2021,Lazar2016}, {and so are the axial quantities $\microdefgradgradincaxdevcomp_{ij}$ and $\microdefgradgradincaxhydro$,} as can be found already by counting the independent components of the respective tensors as listed in \tablename~\ref{tab:numbercomponents}. {The table contains also the microscale interpretation of the individual parts as discussed above for $\microdefgradgradincaxcomp_{ij}$ and derived in Section~\ref{sec:micromorphelastichyperstress} below for the parts of $\microdefgradgradsymcomp_{ijk}$.}
\begin{table}%
\caption{{Irreducible parts} of the gradient of microdeformation for spatial case ($\ndim=3$) and the planar case ($\ndim=2$), {and their microscopic interpretation}}
\label{tab:numbercomponents}
\centering
\begin{tabular}{l|c|c|l|l}
                               & $\ndim=2$& $\ndim=3$ & {type} & {interpretation} \\ \hline
	$\microdefgradgradsymhydcomp_i$	               & 2 & 3 & polar & relative displacement\\
  $\microdefgradgradsymdevhydcomp_i$             & 2 & 3 & polar & \multirow{2}{*}{{\large{$\rbrace$}} microbending}\\
	$\microdefgradgradfulldevsymcomp_{ijk}$        & 2 & 7 & polar\\
  $\microdefgradgradsymdevskewaxmcomp_{ij}$      & - & 5 & axial & microtorsion\\ %\hline \hline	
	%%$\microdefgradgradsym_{ijk}$                 & 6 & 18	
	{$\microdefgradgradincaxhydro$}        & - & 1 & axial & screw dislocation-type lattice curvature\\
	{$\microdefgradgradincaxdevcomp_{ij}$} & - & 5 & axial & \multirow{2}{*}{{\large{$\rbrace$}} edge dislocation-type lattice curvature}\\
	{$\microdefgradgradincaxskwaxcomp_i$}  & 2 & 3 & polar\\ \hline \hline
	{$\microdefgradgradcomp_{ijk}$}            & 8 & 27 & polar & gradient of microdeformation
\end{tabular}
\end{table}
{This decomposition~\eqref{eq:hyperstresssplit} can be applied to the hyperstress tensor $\hyperstresscomp_{ijk}$ in the same way so that the internal power contribution of these terms becomes}
\begin{equation}
	\begin{split}
 	\hyperstresscomp_{ijk}\microdefgradgradratecomp_{ijk}
	=\hyperstressfulldevsymcomp_{ijk}\microdefgradgradfulldevsymratecomp_{ijk}
	+\frac{1}{\ndim}\hyperstresshydrocomp_i\microdefgradgradsymhydratecomp_i
	+\frac{2\ndim}{(\ndim\!-\!1)(\ndim\!+\!2)}\hyperstressdevhydrocomp_i\microdefgradgradsymdevhydratecomp_i
	+2\frac{\ndim\!-\!2}{\ndim}\hyperstressdevskewaxmcomp_{ij} \microdefgradgradsymdevskewaxmratecomp_{ij}\\
	%+\frac{1}{2}\hyperstressincaxcomp_{ij}\microdefgradgradincaxratecomp_{ij}
	{+\frac{\ndim-2}{2} \hyperstressincaxdevcomp_{ij}\microdefgradgradincaxdevratecomp_{ij}
	+\frac{\ndim-2}{6} \hyperstressincaxhydro\microdefgradgradincaxhydrorate
	+\hyperstressincaxskwaxcomp_i\microdefgradgradincaxskwaxratecomp_i}
	\end{split}
  \label{eq:hyperstresssplitwork}
\end{equation}
{It shows that the individual parts are orthogonal in an energetic sense. Though, they are not necessarily decoupled in the constitutive equations. Even isotropic constitutive laws allow a coupling between irreducible parts of the same tensorial order and type.}

\subsection[Kinematic relations]{{Kinematic relations}}

Splitting up the microdeformation in the same way
$\microdefcomp_{ij}=\frac{1}{\ndim}\microdilat\Kron{ij}+\microdefdevcomp_{ij}-\ttpermutcomp_{ijk}\microrotcomp_k$
into microdilatation $\microdilat$, deviatoric microstrain $\microdefdevcomp_{ij}$ and microrotation $\microrotcomp_k$, {the kinematic relation $\microdefgradgradcomp_{ijk}=\partial\microdefcomp_{ij}/\partial\locmacrocomp_k$ is decomposed according to Eqs.~\eqref{eq:microdefgradgradincaxhydro}--\eqref{eq:defhyperstressdevskewax} into}
\begin{align}
  \microdefgradgradsymhydcomp_i&=\microdefdevcomp_{ij,j} +\ttpermutcomp_{ijk} \microrotcomp_{j,k}+\frac{1}{\ndim}  \microdilat_{,i}, \\
	\microdefgradgradsymdevhydcomp_i&=\frac{\ndim-2}{2\ndim}\microdefdevcomp_{ij,j}+\frac{(\ndim+2)(\ndim-1)}{2\ndim^2}  \microdilat_{,i}-\frac{\ndim+2}{2\ndim} \ttpermutcomp_{ijk} \microrotcomp_{j,k},\\
	 \microdefgradgradsymdevskewaxmcomp_{ij}
		&=	-\frac{1}{2} \left[\microdefdevcomp_{(ik,l} \ttpermutcomp_{j)kl}+3\microrotcomp_{(i,j)}-\Kron{ij} \microrotcomp_{m,m} \right],\\
	\microdefgradgradfulldevsymcomp_{ijk}&=\frac13 \left[\microdefdevcomp_{jk,i}+\microdefdevcomp_{ij,k} +\microdefdevcomp_{ik,j}-\frac{2}{\ndim+2} \left(\microdefdevcomp_{jm,m}\Kron{ik}+\microdefdevcomp_{km,m}\Kron{ij} + \microdefdevcomp_{im,m}\Kron{jk}\right) \right],\\
{\microdefgradgradincaxhydro}&{=-2\microrotcomp_{m,m},}\\
{\microdefgradgradincaxskwaxcomp_i}&{=\frac{1}{2} \left( \frac{\ndim-1}{\ndim} \microdilat_{,i}-\microdefdevcomp_{ij,j}+\ttpermutcomp_{ijm} \microrotcomp_{j,m} \right),}\\
{\microdefgradgradincaxdevcomp_{ij}}&{=\microdefdevcomp_{(ik,l}\ttpermutcomp_{j)kl}+\microrotcomp_{(i,j)}-\frac13 \Kron{ij}\microrotcomp_{m,m}}
	\,.
\end{align}
Obviously, these harmonic modes of the gradient of microdeformation do \emph{not} decouple into the gradients of the harmonic components of microdeformation (microdilatation, deviatoric microstrain and microrotation). {Solely $\microdefgradgradfulldevsymcomp_{ijk}$ contains only certain components of the gradient of the deviator of microdeformation and $\microdefgradgradincaxhydro$ is directly related to the divergence of microrotation field.
But the vectorial irreducible components  $\microdefgradgradsymhydcomp_i$, $\microdefgradgradsymdevhydcomp_i$ and $\microdefgradgradincaxskwaxcomp_i$ are related to gradient of microdilatation, curl of microrotation and divergence of deviatoric microstrain in a coupled way. And the symmetric second order components $\microdefgradgradsymdevskewaxmcomp_{ij}$ and $\microdefgradgradincaxdevcomp_{ij}$ exhibit a coupling between devatoric parts of gradient of microrotation and curl of microdeviatoric strain.}
{Consequently, these kinematic relations relations introduce a coupling into the global balances \eqref{eq:momentumbalancehyperstress}, even if the constitutive equations between respective irreducible parts of hyperstress and gradient of microdeformation were decoupled.}

\subsection{Isotropic, linear-elastic material}

{In the following, the general isotropic law from section~\ref{sec:isotropiclaw} is taken as simplest example to demonstrate the harmonic decomposition.   
Firstly, the hydrostatic parts of \eqref{eq:innerstressisotropic}--\eqref{eq:stressdiffisotropic} become}
\begin{align}
  \innerstressdilat
	&=	\underbrace{\left(\frac{2\shearmodeffstrain}{\ndim} +\Lamefirsteffstrain\right)}_{=:\compmodeffstrain} \, \strainmacrodilat  
	  + \underbrace{\left(g_1 + \frac{2g_2}{\ndim}\right)}_{=:\compmodeffmixed} \microdilatdiff , &
	\stressdiffdilat
	&=\underbrace{\left(b_1  + \frac{b_2+b_3}{\ndim}\right)}_{=:\compmodeffmicrodilat} \, \microdilatdiff
		+ \left(g_1+\frac{2g_2}{\ndim}\right) \, \strainmacrodilat
	\label{eq:stresshydrostaticisotropic}
\end{align}
in terms of the dilatational parts $\strainmacrodilat=\strainmacrocomp_{kk}$ and $\microdilatdiff=\microdefdiffcomp_{kk}$ of strain and microdeformation difference, respectively. 
Therein, the abbreviations 
$\compmodeffstrain$, $\compmodeffmixed$ and $\compmodeffmicrodilat$ 
have been introduced as bulk moduli for dilatational loadings.
Evaluating the respective terms for the deviatoric, symmetric parts yields
\begin{align}
	\innerstressmacrodevcomp_{ij}=&2\shearmodeffstrain \, \strainmacrodevcomp_{ij}  + 2g_2 \, \microdefdiffdevcomp_{ij}
		\label{eq:innerstressdevisotropic} \\
		\stressdiffdevcomp_{ij}=&2g_2 \, \strainmacrodevcomp_{ij}+\underbrace{(b_2+b_3)}_{=:2\shearmodeffmicro}\,\microdefdiffdevcomp_{ij}\,,
	\label{eq:stressdiffdevisotropic}
\end{align} 	
showing that $g_2$ has the role of a coupling shear modulus. 
Furthermore, $\shearmodeffmicro$ was introduced as shear modulus for the microdeformations.
Finally, the skew part of Eq.~\eqref{eq:stressdiffisotropic} reads
\begin{equation}
	\stressdiffcomp_{[ij]}=\stressmacrocomp_{[ji]}=\underbrace{(b_2-b_3)}_{=:\CossParamSkew}\microdefdiffcomp_{[ij]}
		 \label{eq:stressdiffskewisotropic}
\end{equation}
showing that $b_2-b_3$ corresponds to the Cosserat coupling modulus $\CossParamSkew$. 
{
Decomposing the the isotropic hyperstress law \eqref{eq:hyperstressisotropic} 
according to \eqref{eq:defhyperstressdevhydro}--\eqref{eq:defhyperstressdevskewax} and 
inserting Eq.~\eqref{eq:hyperstresssplit} to the respective right-hand side terms yields, after some manipulations
\begin{align}
\hyperstressfulldevsymcomp_{ijk}=&\underbrace{(a_{10}+2a_{11}+a_{13}+a_{14}+a_{15})}_{=:\hyperstressmodfulldevsym}\microdefgradgradfulldevsymcomp_{ijk}
\label{eq:hyperstressisotropicfulldevsym}\\	
\begin{pmatrix}
		\frac{2}{3}\hyperstressdevskewaxmcomp_{ij} \\
		\frac{1}{2}\hyperstressincaxdevcomp_{ij}
	\end{pmatrix}
	=&
	\begin{pmatrix}
		 \frac{2}{3}\hyperstressmoddevskewax & \frac{1}{2}\hyperstressmodincaxdevsymskw \\
		 \frac{1}{2}\hyperstressmodincaxdevsymskw & \frac{1}{2}\hyperstressmodincaxdev 
	\end{pmatrix}
	\cdot
	\begin{pmatrix}
		 \microdefgradgradsymdevskewaxmcomp_{ij}\\
		 \microdefgradgradincaxdevcomp_{ij}
	\end{pmatrix}
	\label{eq:hyperstressisotropicsecondorderpartsmatrix}
	\\
	\begin{pmatrix}
		\frac{1}{\ndim}\hyperstresshydrocomp_i \\
		\frac{2\ndim}{(\ndim\!-\!1)(\ndim\!+\!2)}\hyperstressdevhydrocomp_i\\	
		\hyperstressincaxskwaxcomp_i
	\end{pmatrix}
	=&
	\begin{pmatrix}
		\frac{1}{\ndim}\hyperstressmodhydro & \frac{1}{\ndim}\hyperstressmodhydrodevhydro & \frac{1}{\ndim}\hyperstressmodhydroincaxskwax \\
			& \frac{2\ndim}{(\ndim\!-\!1)(\ndim\!+\!1)}    \hyperstressmoddevhydro & \frac{2\ndim}{(\ndim\!-\!1)(\ndim\!+\!1)}\hyperstressmoddevhydroincaxskwax \\
		  \text{  sym.}               &								& \hyperstressmodincaxskwax
	\end{pmatrix}
	\cdot
	\begin{pmatrix}
		 \microdefgradgradsymhydcomp_i\\
		 \microdefgradgradsymdevhydcomp_i \\
		 \microdefgradgradincaxskwaxcomp_i
	\end{pmatrix}
	\label{eq:hyperstressisotropicvectorpartsmatrix} \\
%%%	
	\hyperstressincaxhydro=&\underbrace{\frac{\ndim-1}{2} (2a_{11}+2\ndim a_{10}-a_{13}-2\ndim a_{14}-a_{15} )}_{=:\hyperstressmodincaxhydro} \microdefgradgradincaxhydro
	\label{eq:hyperstressisotropicinchydro}\,.
\end{align}
Therein, \emph{eleven modal hyperstress moduli} ($\hyperstressmodfulldevsym$, $\hyperstressmoddevskewax$, $\hyperstressmodincaxdev$, $\hyperstressmodincaxdevsymskw$, $\hyperstressmodhydro$, $\hyperstressmoddevhydro$, $\hyperstressmodincaxskwax$, $\hyperstressmodhydrodevhydro$, $\hyperstressmodhydroincaxskwax$, $\hyperstressmoddevhydroincaxskwax$, $\hyperstressmodincaxhydro$) have been introduced as indicated partly in Eqs.~\eqref{eq:hyperstressisotropicfulldevsym} and \eqref{eq:hyperstressisotropicinchydro}. These eleven modal hyperstress moduli are related to the \emph{eleven Lamé-type} hyperstress moduli of Mindlin 
%($a_{1}$, $a_{2}$, $a_{3}$, $a_{4}$, $a_{5}$, $a_{8}$, $a_{10}$, $a_{11}$, $a_{13}$, $a_{14}$, $a_{15}$) 
via
\begin{equation}
  \begin{pmatrix}
		\hyperstressmodfulldevsym\\
		\hyperstressmoddevskewax\\
		\hyperstressmodincaxdev\\
		\hyperstressmodincaxdevsymskw\\
		\hyperstressmodhydro\\
		\hyperstressmoddevhydro\\
		\hyperstressmodincaxskwax\\
		\hyperstressmodhydrodevhydro\\
		\hyperstressmodhydroincaxskwax\\
		\hyperstressmoddevhydroincaxskwax\\
		\hyperstressmodincaxhydro
	\end{pmatrix}
	=
	\begin{pmatrix}
%	a_1	&a_2&a_3&a_4&a_5&a_8&a_10&a_11&a_13&a_14|a_15\\	
%\hyperstressmodfulldevsym=a_{10}+2a_{11}+a_{13}+a_{14}+a_{15}
	\zz&\zz&\zz&\zz&\zz&\zz& 1 & 2 & 1 & 1 &1\\
%\hyperstressmoddevskewax=a_10−a_11+a_14−(a_13+a_15)/2\\
	\zz&\zz&\zz&\zz&\zz&\zz& 1 &-1 &\hspace{-0.8em}-\frac12&1& \hspace{-0.8em}-\frac12\\
%\hyperstressmodincaxdev=(n−2)/2 (2a_10−2a_11+a_13−2a_14+a_15 )= \\
	\zz&\zz&\zz&\zz&\zz&\zz& 1 & -1&\frac12& -1&\frac12\\
%\hyperstressmodincaxdevsymskw=−(a_13−a_15 )\\			
	\zz&\zz&\zz&\zz&\zz&\zz&\zz&\zz&-1 & \zz& 1 \\
%\hyperstressmodhydro=(2na_1+2na_2+n^2 a_3+a_4+2a_5+a_8+na_10+2a_11+a_13+na_14+a_15)/n\\
		2 & 2 &\ndim&\frac1{\ndim}&\frac2{\ndim}&\frac1{\ndim}&1&\frac2{\ndim}&\frac1{\ndim}&1&\frac1{\ndim}\\
%\hyperstressmoddevhydro=((n^2+n−2)(a_4+2a_5+a_8 )+2n(a_10+a_14)+(n−2)(2a_11+a_13+a_15))/(2n)\\
	\zz&\zz&\zz&\frac{\ndim^2\!+\!\ndim\!-\!2}{2\ndim}&\frac{\ndim^2\!+\!\ndim\!-\!2}{\ndim}&\frac{\ndim^2\!+\!\ndim\!-\!2}{2\ndim}&1&\frac{\ndim\!-\!2}{\ndim}& \frac{\ndim\!-\!2}{2\ndim}& 1 &\frac{\ndim\!-\!2}{2\ndim} \\
%\hyperstressmodincaxskwax=na_1−na_2+a_4−a_8−a_13+a_15\\
\ndim&\hspace{-0.5em}-\ndim&\zz&1&-1&\zz&\zz&\zz& -1  &\zz& 1\\
%\hyperstressmodhydrodevhydro=na_1+na_2+a_4+2a_5+a_8+2a_11+a_13+a_15 \\
\ndim&\hspace{-0.5em}\phantom{-}\ndim&\zz& 1& 2 & 1 &\zz& 2 & \phantom{-}1 & \zz& 1\\
%\hyperstressmodhydroincaxskwax=na_1−na_2+a_4−a_8−a_13+a_15\\
\ndim&\hspace{-0.5em}-\ndim&\zz& 1&\zz&-1 &\zz&\zz&-1 &\zz& 1\\
%\hyperstressmoddevhydroincaxskwax=((n^2+n−2)(a_4−a_8 )+(n+2)(a_13−a_15))/2n
%\zz&\zz&\zz&\zz&\zz&\zz&{\scriptstyle\ndim(\ndim\!-\!1)}&{\scriptstyle\ndim\!-\!1}&\frac{1\!-\!\ndim}2&{\scriptstyle\!\ndim(1\!-\!\ndim)}&\frac{1\!-\!\ndim}2 \\
\zz&\zz&\zz&\frac{\ndim^2\!+\ndim\!-\!2}{2\ndim}&\zz&-\frac{\ndim^2\!+\ndim\!-\!2}{2\ndim}&\zz&\zz&\frac{\ndim\!+\!2}{2\ndim}&\zz&-\frac{\ndim\!+\!2}{2\ndim}\\
%\hyperstressmodincaxhydro=(n−1)/2 (2na_10+2a_11−a_13−2na_14−a_15 )=\frac{\ndim-1}{2} (2\ndim a_{10}+2a_{11}-a_{13}-2\ndim a_{14}-a_{15} )
\zz&\zz&\zz&\zz&\zz&\zz&{\scriptstyle\ndim(\ndim\!-\!1)}&{\scriptstyle \ndim\!-\!1}&\frac{1\!-\!\ndim}{2}&{\scriptstyle\ndim(1\!-\!\ndim)}&\frac{1\!-\!\ndim}{2}
%\zz&\zz&\zz&\zz&\zz&\zz&\zz&{\scriptstyle\ndim\!-\!1}&{\scriptstyle\ndim\!-\!1}&{\scriptstyle\!\ndim(1\!-\!\ndim)}&\frac{1\!-\!\ndim}{2}
	\end{pmatrix}
	\cdot
	\begin{pmatrix}
		 a_{1}\\
		 a_{2}\\
		 a_{3}\\
		 a_{4}\\
		 a_{5}\\
		 a_{8}\\
		 \smash{a_{10}}\\
		 \smash{a_{11}}\\
		 \smash{a_{13}}\\
		 \smash{a_{14}}\\
		 \smash{a_{15}}
	\end{pmatrix}
	\label{eq:hyperstressmoduliLametomodal}
\end{equation} 
The modal constitutive equations \eqref{eq:hyperstressisotropicfulldevsym}--\eqref{eq:hyperstressisotropicinchydro} show that the irreducible parts of same tensorial order can be coupled component wise. 
For the parts of third order in Eq.~\eqref{eq:hyperstressisotropicfulldevsym} and zeroth order (scalar) in Eq.~\eqref{eq:hyperstressisotropicinchydro}, there is only one irreducible part for each order so that respective parts of hyperstress, and $\hyperstressfulldevsymcomp_{ijk}$ and $\hyperstressincaxhydro$, are coaxial to respective parts of the gradient of microdeformation, $\microdefgradgradfulldevsymcomp_{ijk}$ and $\microdefgradgradincaxhydro$. 
For the irreducible parts of order one and two in Eqs.~\eqref{eq:hyperstressisotropicsecondorderpartsmatrix} and \eqref{eq:hyperstressisotropicvectorpartsmatrix} there will, in general, be a coupling between components of same tensorial order. 
The prefactors in the column vectors at the left-hand sides in Eqs.~\eqref{eq:hyperstressisotropicsecondorderpartsmatrix}--\eqref{eq:hyperstressisotropicvectorpartsmatrix} incorporate the prefactors from the power expression in Eq.~\eqref{eq:hyperstresssplitwork}. Consequently, the coupling matrices in Eqs.~\eqref{eq:hyperstressisotropicsecondorderpartsmatrix}--\eqref{eq:hyperstressisotropicvectorpartsmatrix} are symmetric.
The inverse relation of Eq.~\eqref{eq:hyperstressmoduliLametomodal} is given in Appendix~\ref{sec:hypertressmodulimodaltoLame}.
}

{Note that the homogenization theory, section~\ref{sec:homogenization}, predicts that the hyperstress has one sub-symmetry. In modal decomposition, this means that $\hyperstressincaxcomp_{ij}$ vanishes, corresponding to $\hyperstressmodincaxdev=\hyperstressmodincaxdevsymskw=\hyperstressmodincaxskwax=\hyperstressmodhydroincaxskwax=\hyperstressmoddevhydroincaxskwax=\hyperstressmodincaxhydro=0$, or $a_1=a_2$, $a_4=a_5=a_8$, $a_{10}=a_{14}$, $a_{11}=a_{13}=a_{15}$, for linear isotropic material.}

%=====================================================================
\section[Microscopic interpretation for a material with voids]{{Microscopic interpretation} for a material with voids}
\label{sec:micromorphelastic}
%=====================================================================

\subsection{Composite shell approach}

\begin{figure}
	\centering
	\inputsvg{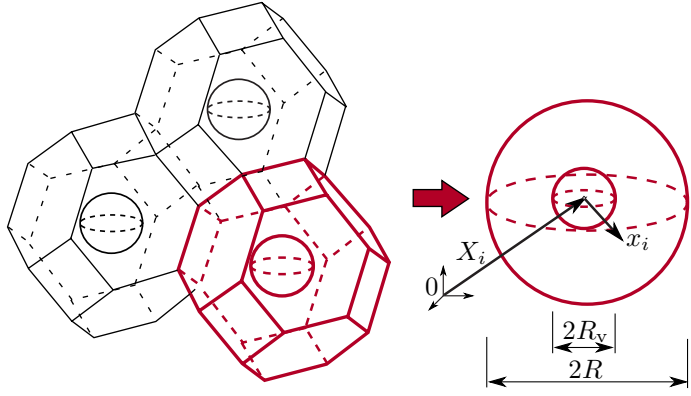}
	\caption{Unit cell of a porous material}
	\label{fig:sphericalcell}
\end{figure}

In Hashin's composite shell approach \citep{Hashin1962}, a spherical volume element with radius $\radcell$ with spherical pore (or inclusion) with radius $\radvoid$ is {employed at the microscale} as sketched in \figurename~\ref{fig:sphericalcell}. 
The analogon for the plane case $\ndim=2$ is a circular volume element with circular pore. Both cases will be termed as \enquote{sphere} in a generalized sense in the following and treated by a common approach for which the spatial dimension $\ndim$ has to be set to $\ndim=2$ or $\ndim=3$, respectively. In this sense, the porosity is $\VVF=(\radvoid/\radcell)^{\ndim}$ and the required second geometric moments amount to
\begin{align}
\geommomcomp_{ij}&=\frac{\radcell^2}{\ndim+2} \Kron{ij}, &
\geommommicrocomp_{ij}=\frac{\radvoid^2}{\ndim} \Kron{ij}\,.
\label{eq:geommomciricle}
\end{align}
Hence, the relevant Eqs.~\eqref{eq:defmacrodeformationratessurface}, \eqref{eq:natBCminconstrperiodic}, \eqref{eq:natBCminconstrdiscont}, \eqref{eq:momentumbalancelocalminconstrvoid} become
\begin{align}
 	\microdefcomp_{ij}=&
	\frac{\ndim}{\radvoid^2} \averopsphereshell[\radvoid]{\displmicrocomp_i\locmicrorelcomp_j}
	\label{eq:defmicrodefsphere}\\
			\microdefgradgradsymdevcomp_{ijk}
=&\frac{\ndim+2}{2\radcell^2}\left[\frac{\ndim}{\radcell^2}\averopsphereshell[\radcell]{\displcomp_{i} \locmicrorelcomp_j \locmicrorelcomp_k } - \averopsphereshell[\radcell]{  \displcomp_{i} } \Kron{jk} \right]
\label{eq:microdefgradgradmicrocomppartintdevsphere}\\
\microdefgradgradsymhydcomp_i=&\microdefgradgradsymcomp_{ijj}
=\frac{\ndim}{\radcell^2}\left[\averopsphereshell[\radcell]{\displcomp_{i}}-\averopmatrix{\displcomp_{i}}\right]
\label{eq:microdefgradgradmicrocomppartinthydsphere}\\	
		n_i\sigma_{ij}&=\pm {\Lagrangemult[]}_j(\locmicrorelcomp_q) +\frac{\ndim+2}{2\radcell^2}  {\hyperstresscomp[]}_{jkm}   \locmicrorelcomp_k n_m- \frac{1}{2\radcell}  \hyperstresscomp_{jmm}   & \text{for }r&=\radcell 
	\label{eq:natBCminconstrperiodicsphere}\\
	 n_i\sigma_{ij}%=&-\frac{\ndim}{\radvoid^2}\frac{\domaincell[]}{|\discontsurf|} \stressdiffcomp_{jk} \locmicrorelcomp_{k}
	=&-\frac{1}{\VVF\radvoid} \stressdiffcomp_{jk} \locmicrorelcomp_{k} & \text{for }r&=\radvoid 
	\label{eq:natBCminconstrdiscontcircle}\\
	\stresscomp_{ji,j}&= \frac{1}{1-\VVF}\, \frac{1}{\radcell^2}\hyperstresshydrocomp_i
		& \forall \locmicrocomp_i&\in\text{ matrix}
	\label{eq:momentumbalancelocalminconstrsphere}
\end{align}
Therein, it was taken into account that the normal $\vnormcomp_i$ and the location vector $\locmicrorelcomp_i$ are coaxial at spherical surfaces. The operator $\averopsphereshell[r]{\placeholder}$ computes the average over a spherical surface of radius $r=\sqrt{\locmicrorelcomp_i \locmicrorelcomp_i}$ \citep{Leblond1995}.
Equation~\eqref{eq:microdefgradgradmicrocomppartinthydsphere} shows that the spherical part $\microdefgradgradsymhydcomp_i$ of the gradient of microdeformation corresponds to a relative translation between the surface and the bulk of the volume element.
{Furthermore, it should be noted that eqs.~\eqref{eq:natBCminconstrperiodicsphere}--\eqref{eq:momentumbalancelocalminconstrsphere} correspond to} periodic boundary conditions for microdeformations $\microdefcomp_{ij}$ and strains $\strainmacrocomp_{ij}$ with superimposed static boundary conditions for the hyperstresses $\hyperstresscomp_{ijk}$.

{In the following, the microscopic loading modes to the spherical volume, which result from the harmonic decomposition of the macroscopic measures of stress and strain, are inspected individually in order to provide a qualitative interpretation of their physical meaning. Furthermore, analytical solutions of the respective micro-scale boundary value problems are derived for the linear-elastic case to gain a quantitative insight as well.}

\subsection{Elastic solution in terms of spherical harmonics}
\label{sec:spherical harmonics}

{The isotropic elastic behavior of the matrix material is characterized by the Lamé parameters $\Lamefirst$ and $\Lamesec$.
The solutions of the resulting Navier-Lamé equations $(\Lamefirst+\Lamesec) \displcomp_{j,ji}+\Lamesec \displcomp_{i,jj}=0$} for a sphere in terms of solid spherical harmonics from \citep[§172]{Love1944} are generalized here to include also the plane case $\ndim=2$.  
A spherical harmonic $\sphericalharmonic(\locmicrocomp_j)$ of degree $\sphericalharmonicorder$ satisfies Laplace's equation $\sphericalharmonic_{,kk}=0$ and is homogeneous of degree $\sphericalharmonicorder$, so that $\sphericalharmonic(\beta x_j)=\beta^{\sphericalharmonicorder}\sphericalharmonic(x_j)$ holds for any scalar $\beta$.
\citet{Love1944} provided three types of solutions (\enquote{Type $\omega$}, \enquote{Type $\phi$}, and \enquote{Type $\chi$}), the latter not to be confused with the microdeformation.
The respective ansatz is listed in appendix~\ref{sec:generalsolutions} together with the corresponding traction vector $\tractioncomprad{i}=\stresscomp_{ij}\locmicrocomp_j/r$ at a surface $r=\mathrm{const}$. These tractions are required to satisfy the boundary conditions {for the individual loading modes} and to evaluate the respective micro-macro relations. The solutions of type $\omega$ contain a dimensionless coefficient $\spheresolomegacoeff$ which depends on Poisson's ratio as listed in the appendix. 
According to \citet{Applequist1989}, the spherical harmonics {$\sphericalharmonic(\locmicrocomp_j)$ themselves} can be represented by harmonic tensors, i.e., totally traceless and totally symmetric tensors as outlined in appendix~\ref{sec:generalsolutions}. This representation is used in the following instead of the individual choice of spherical harmonics of \citet{Hashin1962}.

%-------------------------------------------------------
\subsection{Strains and microdeformations}
\label{sec:micromorphelasticstrains}
%-------------------------------------------------------

In the following, {the loading modes for the individual irreducible components are considered}, starting with the modes for microdeformations $\microdefcomp_{ij}$ and strains $\strainmacrocomp_{ij}$.

%-------------------------------------------------------
\subsubsection{Dilatation}
\label{sec:microscaledilatation}
%-------------------------------------------------------

\begin{figure}
	\centering
	\subfloat[]{\inputsvg{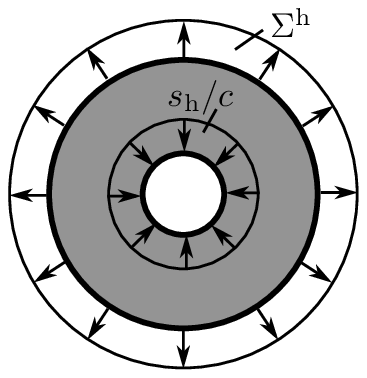} \label{fig:dilat_sketch}}
	\hspace{1cm}
	\subfloat[]{\inputgnuplot{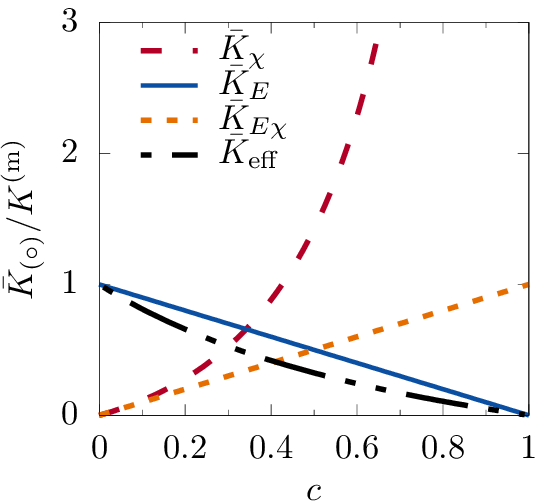} \label{fig:dilat_plot}}
	\caption{Dilatation: a) mode of loading for plane case and b) derived moduli ($\ndim=3$)} 
	\label{fig:dilat}
\end{figure}

Dilatation corresponds to a macroscopic stress state $\stressdiffcomp_{ij}=\stressdiffdilat\Kron{ij}$ and $\stressmacrocomp_{ij}=\stressmacrohydro\Kron{ij}$. It leads to Navier's problem of a hollow sphere under sphero-symmetrical loading. 
The respective micro-macro relations and boundary conditions from eqs.~\eqref{eq:kineticmicromacro}, \eqref{eq:velocitygradmacrovoids}, \eqref{eq:defmicrodefsphere}, \eqref{eq:natBCminconstrdiscontcircle} in spherical coordinates read
\begin{align}
	\strainmacrodilat=&\frac{\ndim}{\radcell}\displcomp(\radcell), &
	\microdilat=&\ndim \frac{\displcomp_r(\radvoid)}{\radvoid}, & 
	\label{eq:micromacroNavier} \\
	\stresscomp_{rr}(\radcell)=&\stressmacrohydro, &
	\stresscomp_{rr}(\radvoid)=&\frac{\stressdiffdilat}{\VVF}
\end{align}
as visualized in \figurename~\ref{fig:dilat_sketch}.
{It can be seen that it corresponds to the classical case if $\stressdiffdilat=0$ as expected. Furthermore, the aforementioned boundary conditions allow for a homogeneous stress state in the matrix if $\stressdiffdilat=\VVF\stressmacrohydro$.}

The {elastic solution of this boundary-value problem by} \citet{Hashin1962} was extended towards microdilatational contributions in \citep{Huetter2017a} and is generalized here to incorporate the plane case $\ndim=2$, too.
The well-known solution for the (purely radial) displacements $\displcomp(r)=\displcomp_i\locmicrorelcomp_i/r$ and respective radial stresses \citep{Kachanov2003} is 
\begin{align}
	\displcomp(r)&= \NavierconstA r + \NavierconstB \frac{\radcell^{\ndim}}{r^{\ndim-1}}, &
	\label{eq:solutionNavier} 
	\stresscomp_{rr}&= \ndim \compmod \NavierconstA - 2\Lamefirst(\ndim-1) \NavierconstB \frac{\radcell^{\ndim}}{r^{\ndim}}
\end{align}
with two constants of integration $\NavierconstA$ and $\NavierconstB$, and $\compmod=2\Lamefirst/\ndim+\Lamesec$ being the bulk modulus of the matrix material.
Inserting solution \eqref{eq:solutionNavier} (irrelevant whether static or kinematic boundary conditions are employed for this sphero-symmetric case) leads to the macroscopic constitutive relation
\begin{align}
  \innerstressdilat=&\stressmacrohydro-\stressdiffdilat
			=\underbrace{(1-\VVF)\compmod}_{=\compmodeffstrain}\strainmacrodilat
		+\underbrace{\vphantom{(1)}\VVF\compmod}_{=\compmodeffmixed}\microdilatdiff,&
		\stressdiffdilat%\VVF\stresscomp_{rr}(r=\radvoid)
		=&\underbrace{\vphantom{(1)}\VVF \compmod}_{=\compmodeffmixed} \strainmacrodilat+\underbrace{\left(\VVF \compmod + 2\Lamefirst \frac{\ndim-1}{\ndim}\right)\frac{\VVF}{1-\VVF}}_{=\compmodeffmicrodilat}\microdilatdiff
	\label{eq:compressionmodulimicro}
\end{align}
Comparing this solution with the general isotropic law~\eqref{eq:stresshydrostaticisotropic} allows to identify the {resulting macroscopic moduli $\compmodeffstrain$, $\compmodeffmixed$ and $\compmodeffmicrodilat$ as indicated in the equation.}
They are plotted in \figurename~\ref{fig:dilat_plot}. The strain modulus $\compmodeffstrain$ corresponds to the Voigt-Taylor estimate and the coupling modulus $\compmodeffmixed$ corresponds to the difference of $\compmodeffstrain$ to the bulk modulus $\compmod$ of the matrix material. 
The micromodulus $\compmodeffmicrodilat$ has a slightly more complicated form which forms the \enquote{difference} to the effective bulk modulus $\compmodeff=\compmodeffstrain-(\compmodeffmixed)^2/\compmodeffmicrodilat$. The latter is also plotted in \figurename~\ref{fig:dilat_plot} and coincides with the result of \citet{Hashin1962}.
Both non-classical moduli, $\compmodeffmixed$ and $\compmodeffmicrodilat$  vanish for homogenous material $\VVF=0$, so that the governing macroscopic equilibrium condition~\eqref{eq:momentumbalancemacro} becomes independent of $\microdilatdiff$ or $\microdilat$, resp.

%-------------------------------------------------------
\subsubsection{Deviatoric Strains}
\label{sec:microscaledevstrain}
\begin{figure}
	\centering
	\subfloat[]{\inputsvg{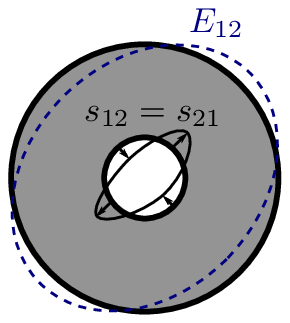} \label{fig:deviatoric_sketch}}
	\hspace{1cm}
	\subfloat[]{\inputgnuplot{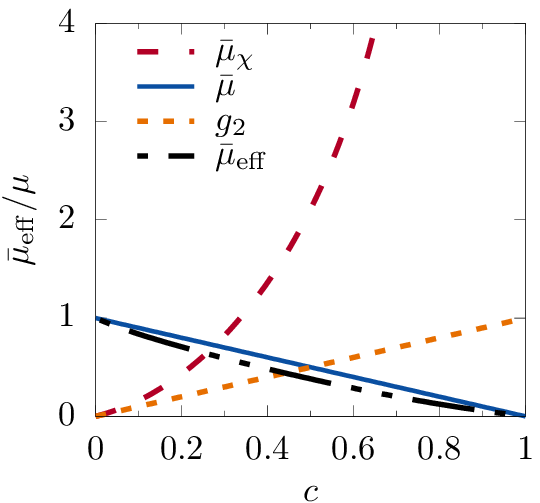} \label{fig:deviatoric_plot}}
	\caption{Deviatoric strains: a) mode of loading for plane case and b) derived moduli ($\ndim=3$)} 
	\label{fig:deviatoric}
\end{figure}

%-------------------------------------------------------
For the loading mode of deviatoric and symmetric parts of strain and microdeformation, we start with kinematic boundary conditions at $r=\radcell$, i.e.\ Eq.~\eqref{eq:periodicBCmicromorphic} with $\fluctuation{\displmicrocomp_i}=0$, together with condition \eqref{eq:natBCminconstrdiscontcircle} at the void surface
\begin{align}
	\displcomp_i(r=\radcell) =& \strainmacrodevcomp_{ij} \locmicrocomp_j\,, &
	\tractioncomprad[r=\radvoid]{i}=&\frac{1}{\VVF\radvoid} \stressdiffdevcomp_{ij} \locmicrorelcomp_{j}\,.
	\label{eq:BCsdeviatoric}
\end{align}
as illustrated in \figurename~\ref{fig:deviatoric_sketch}. 
And we will see that the result satisfies also the periodic boundary conditions for the spherical (circular) volume element under consideration. 
{The pictures is similar to the dilatational case of the previous section: The classical Hill-Mandel problem is obtained for a vanishing difference stress $\stressdiffdevcomp_{ij}=0$, whereas a homogeneous state of stress and strain becomes a possible solution if $\VVF\stressdiffdevcomp_{ij}$ corresponds to the constitutive response of the matrix material for the particular $\strainmacrodevcomp_{ij}$.}

For the {elastic case, the displacement ansatz} of \citet{Hashin1962} is adopted with a combination of terms of type $\omega$ and of type $\phi$ in terms of spherical harmonics of degree 2 and $-\ndim$ and extended here to arbitrary loading directions:
\begin{align}
	\displcomp_i=&2r^2 A_{ij} \locmicrocomp_j+\spheresolomegacoeff[2] \locmicrocomp_i A_{jk} \locmicrocomp_j \locmicrocomp_k+\frac{2}{r^\ndim}  B_{ij} \locmicrocomp_j+\frac{\spheresolomegacoeff[\!-\!\ndim]\!-\!n\!-\!2}{r^{\ndim+2}}  \locmicrocomp_i  B_{jk} \locmicrocomp_j \locmicrocomp_k 
	+C_{ij} \locmicrocomp_j \nonumber \\
	&+\frac{2}{r^{\ndim+2}}  D_{ij} \locmicrocomp_j-\frac{\ndim\!+\!2}{r^{\ndim+4}}  \locmicrocomp_i D_{jk} \locmicrocomp_j \locmicrocomp_k
	\label{eq:ansatzdispldeviatoric}
	\intertext{corresponding to tractions at a surface $r=\mathrm{const}$ of}
	%tractions
	\tractioncomprad{i}=&2\Lamesec(4\!+\!\spheresolomegacoeff[2])r A_{ij} \locmicrocomp_j+(4(\Lamefirst\!+\!\Lamesec)+((2\!+\!\ndim)\Lamefirst\!+\!4\Lamesec) \spheresolomegacoeff[2] ) \frac{\locmicrocomp_i}{r}A_{jk} \locmicrocomp_j \locmicrocomp_k\nonumber\\
	&+2\Lamesec\frac{\spheresolomegacoeff[\!-\!\ndim]\!-\!2\ndim}{r^{\ndim+1}}  B_{ij} \locmicrocomp_j 
	+\left[\Lamesec(\ndim\!+\!1)\!-\!\Lamefirst\!-\!\Lamesec\spheresolomegacoeff[\!-\!\ndim]\right]  \frac{2\ndim}{r^{\ndim+3}}  \locmicrocomp_i B_{jk} \locmicrocomp_j \locmicrocomp_k
	+\frac{2\Lamesec}{r} C_{ij} \locmicrocomp_j \nonumber \\
	&+2\Lamesec\frac{(\ndim\!+\!1)(\ndim\!+\!2)}{r^{\ndim+5}}  \locmicrocomp_i D_{jk} \locmicrocomp_j \locmicrocomp_k-4\Lamesec\frac{\ndim\!+\!1}{r^{\ndim+3}}  D_{ij} \locmicrocomp_j
	\label{eq:ansatztractdeviatoric}
\end{align}
Inserting these expressions into the left-hand side of the boundary conditions~\eqref{eq:BCsdeviatoric} and comparing the coefficients of linear and cubic terms yields four linear equations (decoupled for each pair of indices $ij$) for the four coefficient tensors $A_{ij}$, $B_{ij}$, $C_{ij}$ and $D_{ij}$.
Furthermore, the obtained displacements and tractions can be inserted to the micro-macro relations~\eqref{eq:kineticmicromacro}$_1$ and \eqref{eq:defmicrodefsphere} obtaining
\begin{align}
	\microdefdevcomp_{ij}=&2\frac{\ndim+2+\spheresolomegacoeff[2]}{\ndim+2}\VVF^{\frac{2}{\ndim}}\radcell^2A_{ij}+\frac{\spheresolomegacoeff[\!-\!\ndim]}{\ndim+2}\frac2{\VVF\radcell^\ndim}B_{ij}+C_{ij} \\
	\stressmacrodevcomp_{ij}=&\frac{2}{\ndim\!+\!2}\left[\Lamesec(12\!+\!4\ndim\!+\!(6\!+\!\ndim)\spheresolomegacoeff[2])+(4\!+\!(2\!+\!\ndim)\spheresolomegacoeff[2] )\Lamefirst\right] R^2 A_{ij} \nonumber\\
	&-\frac{2}{\ndim\!+\!2}\left[\Lamesec(2\ndim\!+\!\spheresolomegacoeff[\!-\!\ndim] (\ndim\!-\!2))\!+\!2\ndim\Lamefirst\right]  \frac{B_{ij}}{\radcell^{\ndim}} +2\Lamesec C_{ij}\,.
\end{align}
Expressing $\stressmacrodevcomp_{ij}$ in terms of the internal stress $\innerstressmacrocomp_{ij}$ and the difference stress $\stressdiffcomp_{ij}$ as well as $\microdefcomp_{ij}$ in terms of the difference deformation $\microdefdiffcomp_{ij}$ and conventional strain $\strainmacrocomp_{ij}$ finally yields the desired constitutive equations
\begin{align}
  \innerstressmacrodevcomp_{ij}=&2\underbrace{(1-\VVF)\Lamesec}_{=\shearmodeffstrain}\strainmacrodevcomp_{ij}+2\underbrace{\VVF\Lamesec}_{\shearmodeffmixed}\microdefdiffdevcomp_{ij}\,,
	\label{eq:innerstressdevmicro}\\
	\stressdiffdevcomp_{ij}=& 2\underbrace{\vphantom{(1)}\VVF\Lamesec}_{=\shearmodeffmixed}\strainmacrodevcomp_{ij}+2\shearmodeffmicro\microdefdiffdevcomp_{ij}
	\label{eq:stressdiffdevmicro}
\end{align}
with 
\begin{align}
\shearmodeffmicro=&\frac{\Lamesec\VVF}{1\!-\!\VVF^{\frac{1}3}}\,\frac{\frac{98}{25}\!-\!\frac{42}{5}\Poissrat\!+\!4\Poissrat^2\!+(\frac{28}{25}-\frac35\Poissrat-\Poissrat^2)\VVF^{\frac{10}3}+(\frac72-\frac12 \Poissrat^2)\VVF^{\frac73}-\frac{126}{25}\VVF^{\frac53}+(7\!-\!12\Poissrat\!+\!8\Poissrat^2)\VVF}{
(\frac{28}{25}-\frac35\Poissrat\!-\!\Poissrat^2)(1\!+\!\VVF^{\frac13}\!+\!\VVF^{\frac23})\VVF^{\frac73}\!-\frac{63}{25}(1\!+\!\VVF^{\frac13}\!-\!\VVF^{\frac23}\!-\!\VVF)\VVF+(\frac{112}{25}-\!12\Poissrat\!+\!8\Poissrat^2)(1\!+\!\VVF^{\frac13}\!+\!\VVF^{\frac23})}
\label{eq:shearmodeffmicroplane}
\intertext{for the 3D case ($\ndim=3$) and}
\shearmodeffmicro=&\frac{\Lamesec\VVF}{1-\VVF}\,\frac{(3-4\Poissrat)(1+\VVF^4)+4\VVF(3-6\Poissrat+4\Poissrat^2)-6\VVF^2+4\VVF^3}{(3-4\Poissrat)^2-3\VVF(1-\VVF)+\VVF^3(3-4\Poissrat)} 
\label{eq:shearmodeffmicrospatial}
\end{align}
for the {plane} strain case $\ndim=2$.

Note that the obtained displacement field~\eqref{eq:ansatzdispldeviatoric} and tractions~\eqref{eq:ansatztractdeviatoric} contain only linear and cubic terms with respect to the location vector $\locmicrocomp_i$. Thus, they are antiperiodic with respect to homologous points $\locmicrorelcomp_k^-=-\locmicrorelcomp_k^+$ at the boundary so that the present solution satisfies also periodic boundary conditions $\displcomp_i(\locmicrorelcomp_k^+)-\displcomp_i(\locmicrorelcomp_k^-)=\strainmacrodevcomp_{ij}(\locmicrocomp_j^+-\locmicrocomp_j^-$) and $\tractioncomprad[\radcell]{i}(\locmicrorelcomp_k^+)=-\tractioncomprad[\radcell]{i}(\locmicrorelcomp_k^-)$.

The resulting moduli are plotted in \figurename~\ref{fig:deviatoric_plot}. The picture is very similar to the dilatational case in the previous section: the strain modulus $\shearmodeffstrain$ corresponds to the Taylor-Voigt bound and the coupling modulus $\shearmodeffmixed$ to the difference of the former to the value of the matrix material. 
And again the micromodulus $\shearmodeffmicro$ forms the link to the effective bulk modulus $\shearmodeff=\shearmodeffstrain-(\shearmodeffmixed)^2/\shearmodeffmicro$. The latter is also plotted in \figurename~\ref{fig:deviatoric_plot} and coincides with the one given by \citet{Hashin1962} (when letting the stiffness of his inclusion going to zero).

%-------------------------------------------------------
\subsubsection{Relative Rotation}
\label{sec:microscalerelrotation}
%-------------------------------------------------------

\begin{figure}
	\centering
	\subfloat[]{\inputgnuplot{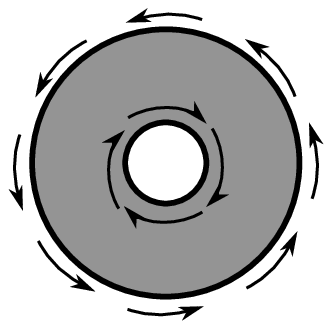}\label{fig:stressskew_sketch}}
	\subfloat[]{\inputgnuplot{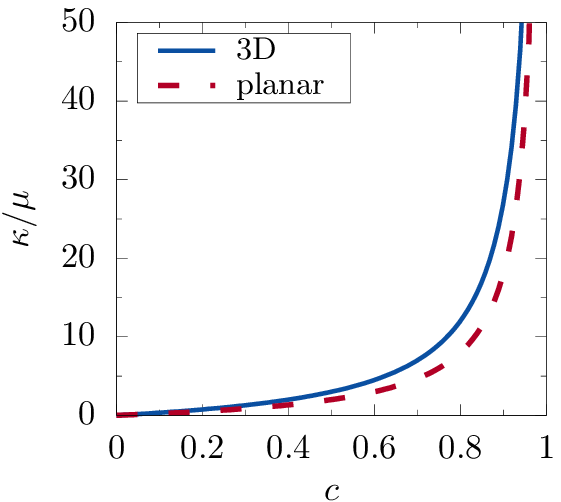}\label{fig:stressskew_plot}}
	\caption[]{Skew-symmetric part of stress: \subref{fig:stressskew_sketch} mode of loading and \subref{fig:stressskew_plot} resulting modulus}
	\label{fig:stressskew}
\end{figure}

For a purely skew loading $\stressdiffcomp_{ij}=\stressdiffcomp_{[ij]}=\stressmacrocomp_{[ji]}$, the boundary condition~\eqref{eq:natBCminconstrdiscontcircle} at the void surface reads
\begin{align}
%	\displcomp_i(\radcell)=&0, &
	%---------------------------------------
%	\stresscomp_{ij}(\radvoid)n_j=&-\frac{1}{\VVF\radvoid}\ttpermutcomp_{ijk}\stressmacroskewcomp_k
	\stresscomp_{ij}(\radvoid)n_j=&-\frac{1}{\VVF\radvoid}\stressdiffcomp_{[ij]}\locmicrorelcomp_{j}
	\label{eq:BCsskewsphere}
\end{align}
as illustrated in \figurename~\ref{fig:stressskew_sketch}. {It drives a relative rotation between the boundary of the volume element and the surface of the void. Such a \enquote{relative rotations of the microstructural elements} \citep{Ehlers2018} is typically associated qualitatively with the skew part $\stressmacrocomp_{[ij]}=\stressdiffcomp_{[ji]}$ of the (force) stress tensor. But the employed homogenization theory allows to capture this effect more precisely and also quantitatively.}

{For the elastic case, the respective}
microscale problem can be solved by an ansatz of $\omega$-type with spherical harmonics of degree $1$ and $1-\ndim$:
\begin{equation}
	\displcomp_{i} = \ttpermutcomp_{ijk} \locmicrocomp_j  \frac{1}{r^{\ndim}}  A_k  +\ttpermutcomp_{ijk} \locmicrocomp_j B_k
	\label{eq:displmicrotwist}
\end{equation}
resulting in tractions $\tractioncomprad{i}=-\ndim\Lamesec/r^{\ndim+1} \ttpermutcomp_{ijk} \locmicrocomp_j A_k$. 
Inserting this relation into the boundary condition~\eqref{eq:BCsskewsphere} allows to identify %$A_k=-\radvoid^{\ndim}/(\ndim\Lamesec\VVF)\stressmacroskewcomp_k$. 
$A_k=-\radvoid^{\ndim}/(2\ndim\Lamesec\VVF)\stressdiffcomp_{[ij]}\ttpermutcomp_{ijk}$. 
The term with $B_k$ corresponds to a rigid rotation and could be computed from a boundary condition at $r=\radcell$, if required. Finally, the displacement field~\eqref{eq:displmicrotwist} can be inserted into the kinematic micro-macro relations~\eqref{eq:defmicrodefsphere} to obtain the final macroscopic constitutive relation
\begin{equation}
		\stressdiffcomp_{[ij]}=\underbrace{\ndim\Lamesec\frac{\VVF}{1-\VVF}}_{=\CossParamSkew}\microdefdiffcomp_{[ij]}\,.
		\label{eq:couplmod}
\end{equation}
A comparison with the general isotropic law~\eqref{eq:stressdiffskewisotropic} allows to identify the micropolar coupling modulus  $\CossParamSkew=b_2-b_3=\ndim\Lamesec\VVF/(1-\VVF)$. This solution is independent of whether static, periodic or kinematic boundary conditions are applied.
Equation~\eqref{eq:couplmod} contains the solution for a circular void $\ndim=2$ from \citep{Huetter2019} as a special case.
The resulting coupling modulus is plotted in \figurename~\ref{fig:stressskew_plot}, showing again that the coupling modulus vanishes for homogeneous material {$\VVF=0$}. For high porosities $\VVF\rightarrow1$ it diverges so that the microrotation $\microrotcomp_k=-\microdefcomp_{ij}\ttpermutcomp_{ijk}$ is forced to follow the macroscopic rotation, {as it is to be expected from the microscale when the matrix becomes infinitesimally thin}.

%========================================================
\subsection{Hyperstresses}
\label{sec:micromorphelastichyperstress}
%=======================================================

{The boundary condition for the hyperstress terms follows from inserting the decomposition~\eqref{eq:hyperstresssplit} to the boundary condition~\eqref{eq:natBCminconstrperiodicsphere} as}
\begin{align}
\tractioncomprad[\radcell]{j} &=\frac{\ndim\!+\!2}{2\radcell^3}\hyperstressfulldevsymcomp_{ijk} \locmicrorelcomp_i \locmicrorelcomp_k 
+\frac{1}{\ndim\radcell} \hyperstresshydrocomp_j
		+ \frac{1}{\radcell^3}\,\frac{\ndim}{\ndim\!-\!1}\hyperstressdevhydrocomp_k \locmicrorelcomp_k \locmicrorelcomp_j-\frac{1}{(\ndim-1)\radcell} \hyperstressdevhydrocomp_j
+\frac{n^2\!-\!4}{\ndim\radcell^3} \hyperstressdevskewaxmcomp_{im} \ttpermutcomp_{kjm} \locmicrorelcomp_i \locmicrorelcomp_k \,.
\label{eq:natBChyperstressspheresplit}
\end{align}
%{in the matrix $\locmicrorelcomp_k\in\domainmatrix$. 
{The loading modes which result at the microscale from the individual irreducible components are sketched in \figurename~\ref{fig:hyperstresses_loading_modes}. 
\begin{figure}
  \centering
	\subfloat[$\hyperstressfulldevsymcomp_{111}=-\hyperstressfulldevsymcomp_{212}$]{\inputsvg{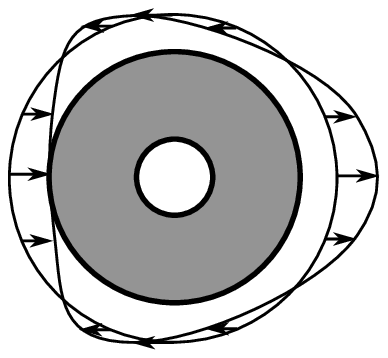} \label{fig:hyperstress_Mdds_sketch}}
	\hspace{2cm}
	\subfloat[$\hyperstresshydrocomp_1$]{\inputsvg{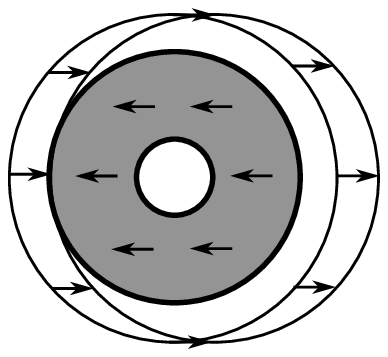} \label{fig:hyperstress_Mh_sketch}}
	\hspace{2cm}
	\subfloat[$\hyperstressdevhydrocomp_1$]{\inputsvg{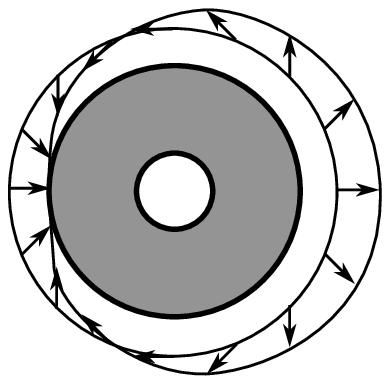} \label{fig:hyperstress_Mdh_sketch}}
	\hspace{2cm}
	\subfloat[$\hyperstressdevskewaxmcomp_{22}=\hyperstressdevskewaxmcomp_{33}=-\frac12\hyperstressdevskewaxmcomp_{11}$]{\scalebox{0.75}{\inputsvg{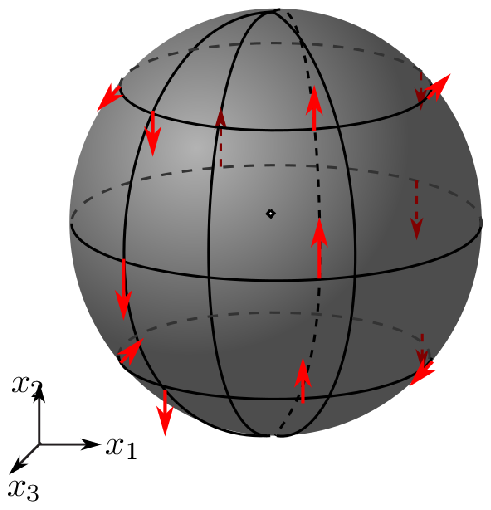}} \label{fig:hyperstress_Mdskew_sketch}}
	\caption[]{Loading modes of the volume element due to hyperstresses: \subref{fig:hyperstress_Mdds_sketch} totally symmetric part $\hyperstressfulldevsymcomp_{ijk}$, \subref{fig:hyperstress_Mh_sketch} spherical part $\hyperstresshydrocomp_i$, \subref{fig:hyperstress_Mdh_sketch} trace of deviator $\hyperstressdevhydrocomp_i$ and \subref{fig:hyperstress_Mdskew_sketch} skew part of deviator of hyperstress $\hyperstressdevskewcomp_{ij}$ }
	\label{fig:hyperstresses_loading_modes}
\end{figure}}
{Figure~\ref{fig:hyperstress_Mdds_sketch} illustrates, that the totally symmetric and deviatoric part $\hyperstressfulldevsymcomp_{ijk}$ drives a bending-like deformation mode with coaxial tractions. A bending-type behavior is driven also by $\hyperstressdevhydrocomp_i$ according to \figurename~\ref{fig:hyperstress_Mdh_sketch}, but with additional dilatational contributions. In contrast, the spherical part $\hyperstresshydrocomp_i$ corresponds to constant tractions at the boundary which are balanced by opposing volume forces in the matrix, thus implying a relative translation between both.} 
Note that this mode would be a zero-energy mode if the micro-macro relation~\eqref{eq:momentumbalancelocalminconstrvoid} for the displacement were not enforced \citep{Auffray2010, Gologanu1997}, compare also \citep{Lopes2021}. 
{Finally, the part $\hyperstressdevskewaxmcomp_{ij}$, which appears only in the spatial case $\ndim=3$, induces a twisting of the microscopic volume element.}

{In addition to this qualitative interpretation, quantitative results shall be provided again in form of the analytical solution of the micro-scale boundary-value problem for the linear-elastic case. In this case, the solutions of the modes can be derived separately and then superimposed. 
The solution procedure is analogous to sections~\ref{sec:microscaledevstrain} and \ref{sec:microscalerelrotation}. Details of the derivation can be found in Appendix~\ref{sec:hyperstressdetails}.
Finally, the obtained hyperstress moduli are obtained as}
\begin{align}
	\hyperstressmodfulldevsym=2\Lamesec\radcell^2{\frac { \left(1+4\,\VVF+\VVF^{2} \right)  \left(1-\VVF\right)^3}{1+\VVF-4(1+\nu)\VVF^2+4(3-\nu)\VVF^3+(3-4\nu)(\VVF^4+\VVF^5)}}
	\label{eq:hyperstressmodfulldevsymplane}
\end{align}
for the planar case $\ndim=2$ ({plane} strain) and
\begin{align}
\hyperstressmodfulldevsym=\frac{8\Lamesec\radcell^2}{5}\frac{
(1\!-\!\rho)
\left\lbrace
(13\!-\!7\nu)  (13\!+\!7\nu)  
\left[\sum\limits_{k=0}^4(k\!\!+\!\!1)(\rho^k\!\!\!+\!\!\rho^{12\!\!-\!\!k})\right]
\!-\!5 (101\!+\!49{\nu}^{2}) 
(\!\sum\limits_{k=5}^{7}\!\rho^k\!\!)
\!-\!1350\rho^6
\right\rbrace
}{
2 (13\!+\!7\nu )  (11\!-\!14\nu)  
%( \rho^{13}\!+\!\rho^{12}\!+\!\rho^{11}\!+\!\rho^{10}\!+\!\rho^{9} )
(\!\sum\limits_{k=9}^{13}\!\rho^k\!\!)
 \!+\!5 (361\!-\!42\nu\!-\!49{\nu}^{2}) 
%( \rho^{8}\!+\!\rho^{7}\!+\!\rho^{6}\!+\!\rho^5 )
(\!\sum\limits_{k=5}^{13}\!\rho^k\!\!)
\!-\!2700(\rho^5\!+\!\rho^6)
\!+\! (169\!-\!49{\nu}^{2})  
(\!\sum\limits_{k=0}^{4}\!\rho^k\!\!)
%(\rho^4\!+\!\rho^3\!+\!\rho^2\!+\!\rho\!+\!1 )
}
\label{eq:hyperstressmodfulldevsymspatial}
\end{align}
for the spherical case $\ndim=3$.  
{Only for the latter case, the hyperstress modulus for the skew part of the deviator gets}
\begin{equation}
  \hyperstressmoddevskewax=\Lamesec\radcell^2  \frac{8}{5}\,\frac{1-\VVF^{5/3}}{4+\VVF^{5/3}}\,.
	\label{eq:hyperstressmoddevskewaxmicro}
\end{equation}
{The moduli for the vectorial irreducible parts are computed as}
\begin{align}
	\hyperstressmodhydro=&\frac{\ndim\hyperstressmoddevhydrocompl}{\hyperstressmodhydrocompl\hyperstressmoddevhydrocompl+\hyperstressmodhydrodevhydrocompl^2}\,, &
	\hyperstressmoddevhydro=&\frac{(\ndim\!-\!1)(\ndim\!+\!2)}{2\ndim}\,\frac{\hyperstressmodhydrocompl}{\hyperstressmodhydrocompl\hyperstressmoddevhydrocompl+\hyperstressmodhydrodevhydrocompl^2}\,, &
	\hyperstressmodhydrodevhydro=&\frac{\ndim\hyperstressmodhydrodevhydrocompl}{\hyperstressmodhydrocompl\hyperstressmoddevhydrocompl+\hyperstressmodhydrodevhydrocompl^2}\,.
	\label{eq:hyperstressmodhydrodev}
\end{align}
{from inversion\footnote{{Taking into account that the homogenization theory, section~\ref{sec:homogenization}, predicts that $\hyperstressincaxcomp_{ij}$ vanishes, corresponding to $\hyperstressmodincaxdev=\hyperstressmodincaxdevsymskw=\hyperstressmodincaxskwax=\hyperstressmodhydroincaxskwax=\hyperstressmoddevhydroincaxskwax=\hyperstressmodincaxhydro=0$.}} of Eq.~\eqref{eq:hyperstressisotropicvectorpartsmatrix} using the compliances $\hyperstressmodhydrocompl$, $\hyperstressmoddevhydrocompl$ and $\hyperstressmoddevhydrocompl$.
These compliances are obtained from the homogenization procedure (since stresses $\hyperstresshydrocomp_i$ and $\hyperstressdevhydrocomp_i$ are prescribed) as}
\begin{align}
\hyperstressmoddevhydrocompl=&\frac{1}{8\Lamesec\radcell^2} \frac{(\ndim\!-\!1)(\ndim\!+\!2)^2}{\ndim^2}  \frac{\frac{1+\ndim(1-2\Poissrat)}{1+\Poissrat(\ndim-2)}+\VVF^{\frac{\ndim+2}{\ndim}}}{1-\VVF^{\frac{\ndim+2}{\ndim}} }
\label{eq:hyperstressmodcompldevhydro} \\
\hyperstressmodhydrodevhydrocompl=&\frac{1}{8\Lamesec\radcell^2} \frac{(\ndim\!-\!1)(\ndim\!+\!2)^2}{\ndim} \frac{1}{1+\Poissrat(\ndim-2)}  \frac{1-\frac{\ndim}{\ndim+2}  \frac{1-\VVF^{\frac{\ndim+2}{\ndim}}}{1-\VVF}}{1-\VVF^{\frac{\ndim+2}{\ndim}}}
\label{eq:hyperstressmodcompldevhydrohydro}
\end{align}
{together with}
\begin{align}
	\hyperstressmodhydrocompl&=\frac{2(1\!-\!\VVF)\left[1\!-\!\Poissrat\!-\!2(1\!-\!\Poissrat)\VVF\!-\!(2\!-\!3\Poissrat)\VVF^2\right]\!-\!(3\!-\!4\Poissrat)\VVF^2(1\!+\!\VVF)\ln(\VVF)}{4\Lamesec\radcell^2(1\!+\!\VVF)(1\!-\!\VVF)^2(1\!-\!\Poissrat)}
	\label{eq:hyperstressmodcomplhydroplane}
\end{align}
for the circular volume element $\ndim=2$ and
\begin{align}
	\hyperstressmodhydrocompl&=
	\frac{2(5\rho^{11}\!-\!9\rho^{10}\!+\!5\rho^8\!+\!8\rho^5\!-\!5\rho^6\!-\!5\rho^3\!+\!1)(2\!-\!\Poissrat\!-\!6\Poissrat^2)\!+\!5(6\rho^{11}\!-\!9\rho^{10}\!+\!12\rho^{5}\!-\!\rho^{6}\!-\!10\rho^3\!+\!2)}{20\Lamesec\radcell^2(1\!-\!\Poissrat^2)(1\!-\!\rho^3\!-\!\rho^{5}\!+\!\rho^{8})(1\!-\!\rho^3)}
	\label{eq:hyperstressmodcomplhydrospatial}
\end{align}
for the spherical volume element. In these equations, the abbreviation $\rho=\sqrt[3]{\VVF}$ was introduced for a shorter writing.

{Figure~\ref{fig:hyperstressmoduli} presents plots of these modal hyperstress moduli (normalized to matrix shear modulus $\Lamesec$ and volume element radius $\radcell$) against porosity $\VVF$.
\begin{figure}
	\centering
	\subfloat[$\hyperstressmodfulldevsym$]{\inputsvg{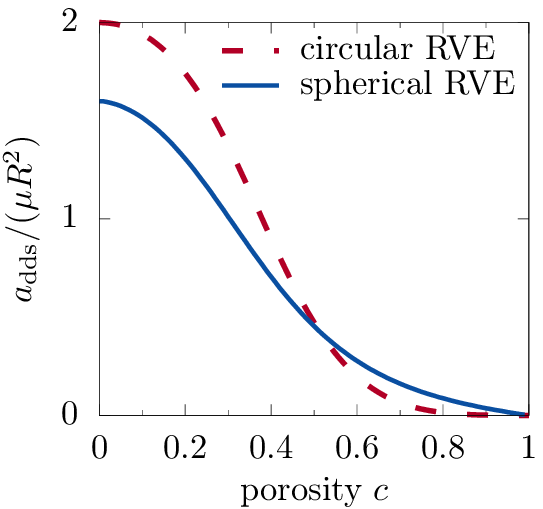} \label{fig:hyperstress_Mdds_plot}} \hspace{2cm}
	\subfloat[$\hyperstressmoddevskewax$]{\inputsvg{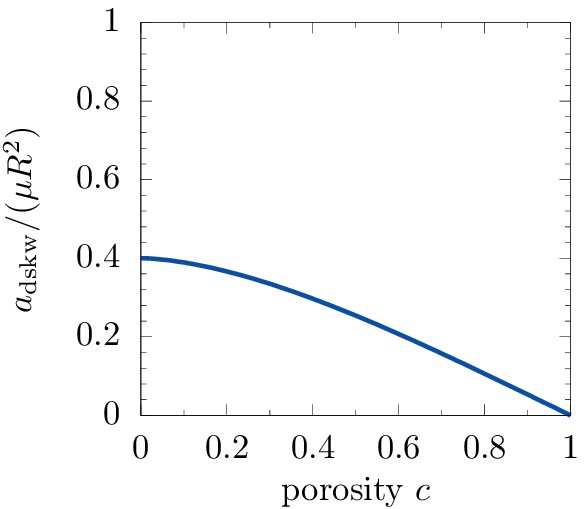} \label{fig:hyperstress_Mdskw_plot}} \\
	\subfloat[$\hyperstressmoddevhydro$ and $\hyperstressmodhydrodevhydro$]{\inputgnuplot{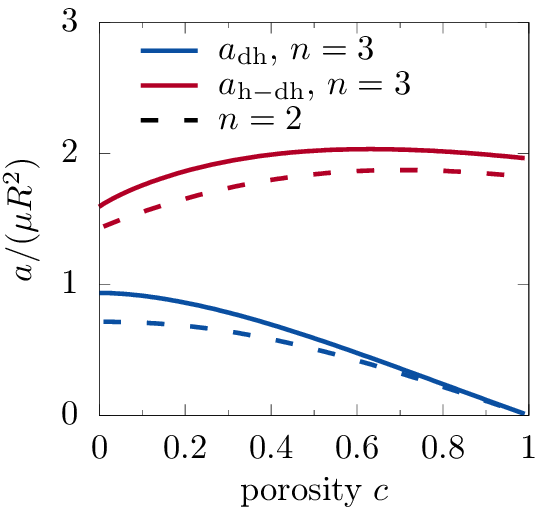}\label{fig:hyperstress_vectors_dev_plot}} \hspace{2cm}
	\subfloat[$\hyperstressmodhydro$]{\inputgnuplot{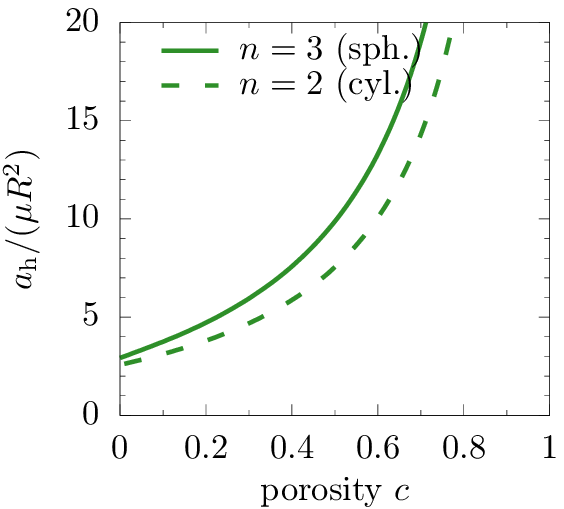}\label{fig:hyperstress_vectors_hydro_plot}}
	\caption{Results for modal hyperstress moduli ($\Poissrat=0.3$)}
	\label{fig:hyperstressmoduli}
\end{figure}
The moduli $\hyperstressmodfulldevsym$ and $\hyperstressmoddevhydro$, which are related to microscopic bending modes, in Figures~\ref{fig:hyperstress_Mdds_plot} and \ref{fig:hyperstress_vectors_dev_plot} show a similar trend. Starting from a finite value for the compact material $\VVF=0$, they decrease to zero for a highly porous material $\VVF\rightarrow 1$. Qualitatively, the same trend is observed for the stiffness $\hyperstressmoddevskewax$ against the torsional micro-mode, though quantitatively at a lower magnitude.
Such a monotonic decrease with increasing porosity is known from effective stiffnesses in classical homogenization (e.g., cmp.\ $\compmodeff$ and $\shearmodeff$ in figures~\ref{fig:dilat_plot} and \ref{fig:deviatoric_plot}, resp.), where the loading is applied via the boundary of the microscopic volume element.
In contrast, \figurename~\ref{fig:hyperstress_vectors_hydro_plot} shows that the modulus $\hyperstressmodhydro$ against a relative translation of volume element (cmp. \figurename~\ref{fig:hyperstress_Mh_sketch}) even diverges for $\VVF\rightarrow1$. This behavior is plausible as such a relative translation becomes nearly infeasible for highly porous structures, the same way as a relative rotation being penalized by the Cosserat coupling modulus as plotted in \figurename~\ref{fig:stressskew_plot}.
The coupling modulus $\hyperstressmodhydrodevhydro$ between both modes in \figurename~\ref{fig:hyperstress_vectors_dev_plot} is predicted to behave, in some sense, in an intermediate way between $\hyperstressmodhydro$ and $\hyperstressmoddevhydro$ and does depend only moderately on $\VVF$ at all.
All the plots in \figurename~\ref{fig:hyperstressmoduli} indicate that, within the given normalization, the difference between a spherical void $\ndim=3$ and a circular one $\ndim=2$ is below 30\%.}

\subsection{Relation to Mindlin's Lamé-type parameters}

{Although the modal moduli, which have been determined in the previous section, have a clear physical meaning at the microscopic scale, the Lamé-type parameters of Mindlin (Section~\ref{sec:isotropiclaw}) are sometimes more convenient for macroscopic computations  with the micromorphic continuum. 
For this reason, all the relations between both types of (equivalent) elastic moduli and respective results of the performed analytical homogenization are summarized in \tablename~\ref{tab:summarymoduli}.}
\begin{table}[hb]
\centering
{
\caption{Summary of relations between modal and Mindlin's Lamé-type elastic moduli}
\label{tab:summarymoduli}
\begin{tabular}{p{1.4cm}|p{2.7cm}|p{2.0cm}||p{2.0cm}|p{2.1cm}}
modal modulus & \raggedright relation to Mindlin's moduli & microscopic relation(s) & Mindlin modulus & in terms of modal moduli \\
\hline
	$\shearmodeffstrain$ & $=\shearmodeffstrain$ & \multirow{2}{*}{$\left\rbrace \begin{array}{l} \\ \end{array}\right.$\hspace{-1.5em} \eqref{eq:innerstressdevmicro}} & $\shearmodeffstrain$ & $=\shearmodeffstrain$ \\
	$\shearmodeffmixed$ & $=g_2$&  & $\Lamefirsteffstrain$ & $=\compmodeffstrain\,-\frac{2\shearmodeffstrain}{\ndim}$\\
	$\shearmodeffmicro$ & $=b_2+b_3$ &\eqref{eq:shearmodeffmicroplane}, \eqref{eq:shearmodeffmicrospatial} & $g_1$ & $=\compmodeffmixed + \frac{2g_2}{\ndim}$ \\
	$\compmodeffstrain$ & $=\frac{2\shearmodeffstrain}{\ndim} +\Lamefirsteffstrain$ &\multirow{3}{*}{$\left\rbrace \begin{array}{l} \\ \\ \end{array}\right.$\hspace{-1.5em} \eqref{eq:compressionmodulimicro}} & $g_2$ & $=\shearmodeffmixed$  \\
	$\compmodeffmixed$ & $=g_1 + \frac{2g_2}{\ndim}$ & & $b_1$ & $=\compmodeffmicrodilat-\frac{\shearmodeffmicro}{\ndim}$ \\
	$\compmodeffmicrodilat$	& $=b_1  + \frac{b_2+b_3}{\ndim}$ & & $b_2$ & $=(\shearmodeffmicro+\CossParamSkew)/2$ \\
	$\CossParamSkew$ & $=b_2-b_3$ &\eqref{eq:couplmod} & $b_3$ & $=(\shearmodeffmicro-\CossParamSkew)/2$ \\
\hline
$\hyperstressmodfulldevsym$ & \multirow{5}{*}{$\left\rbrace \begin{array}{l} \\ \\ \\ \\ \end{array}\right.$\hspace{-1.5em} Eq.~\eqref{eq:hyperstressisotropicvectorpartsmatrix}} & \eqref{eq:hyperstressmodfulldevsymplane}, \eqref{eq:hyperstressmodfulldevsymspatial} & $a_{1}=a_2$ & \multirow{5}{*}{$\left\rbrace \begin{array}{l} \\ \\ \\ \\ \end{array}\right.$\hspace{-1.5em} Eqs.~\eqref{eq:hypertressmodulimodaltoLamecompspatial},\eqref{eq:hypertressmodulimodaltoLamecompplane}} \\
$\hyperstressmoddevskewax$ & &\eqref{eq:hyperstressmoddevskewaxmicro} & $a_{3}$ \\
$\hyperstressmodhydro$ & &\multirow{3}{*}{$\left\rbrace \begin{array}{l} \\ \\ \\ \end{array}\right.$ \hspace{-1.5em}\begin{minipage}{2.5cm}
	\raggedright
  \eqref{eq:hyperstressmodhydrodev} using 
	\eqref{eq:hyperstressmodcompldevhydro}--\eqref{eq:hyperstressmodcomplhydrospatial}
	%\eqref{eq:hyperstressmodcompldevhydro},\eqref{eq:hyperstressmodcompldevhydrohydro}, \eqref{eq:hyperstressmodcomplhydrodevhydro}--\eqref{eq:hyperstressmodcomplhydrospatial}
\end{minipage}}
& $a_{4}\!\!=a_5\!\!=a_8\!$\\
$\hyperstressmoddevhydro$ & & & $a_{10}=a_{14}$ \\
$\hyperstressmodhydrodevhydro$ & & &$a_{11}\!\!=\!\!a_{13}\!\!=\!a_{15}\!$
\end{tabular}
}
\end{table}
%
%Inverting Eqs.~\eqref{eq:stresshydrostaticisotropic} as well as \eqref{eq:stressdiffdevisotropic} and \eqref{eq:stressdiffskewisotropic}
%\begin{align}
	%\compmodeffstrain&=	\frac{2\shearmodeffstrain}{\ndim} +\Lamefirsteffstrain,\\ 
	%\compmodeffmixed&=g_1 + \frac{2g_2}{\ndim} \\
	%\compmodeffmicrodilat&=b_1  + \frac{b_2+b_3}{\ndim}\, \microdilatdiff\\
%\end{align}
%\begin{align}
  %\shearmodeffmicro&=b_2+b_3 \\
	%\CossParamSkew=b_2-b_3
%\end{align}
{Regarding the hyperstress moduli $a_{*}$, it is recalled that the employed homogenization theory (Section~\ref{sec:homogenization}) predicts $\hyperstressinccomp_{ij}=0$ (or does not account for this part, depending on the point of view). 
For the linear-elastic case, this implies $\hyperstressmodincaxdev=\hyperstressmodincaxdevsymskw=\hyperstressmodincaxskwax=\hyperstressmodincaxhydro=0$. 
Taking this fact into account, the modal hyperstress moduli $\hyperstressmodfulldevsym$, 		$\hyperstressmoddevskewax$, $\hyperstressmodhydro$, $\hyperstressmoddevhydro$ and 		$\hyperstressmodhydrodevhydro$ from the elastic homogenization can be converted to Lamé-type counterparts via 
%from \eqref{eq:hypertressmodulimodaltoLamespatial} 
\begin{equation}
	\begin{pmatrix}
		 a_{1}=a_2\\
		 a_{3}\\
		 a_{4}=a_5=a_8\\
		 a_{10}=a_{14}\\
		 a_{11}=a_{13}=a_{15}
	\end{pmatrix}
 =
 \left( \begin {array}{rrrrr} 
-\frac{1}{15}&\frac{1}{6}&0&-{\frac{1}{10}}&{\frac{1}{6}}\\ 
-\frac{1}{15}&-\frac{1}{3}&\frac{1}{3}&\frac{1}{15}&-\frac{2}{9}\\ 
-\frac{1}{15}&-\frac{1}{12}&0&\frac{3}{20}&0\\
\frac{1}{6}&\frac{1}{3}&0&0&0\\ 
\frac{1}{6}&-{\frac{1}{6}}&0&0&0
\end {array} \right) 
  \begin{pmatrix}
		\hyperstressmodfulldevsym\\
		\hyperstressmoddevskewax\\
		\hyperstressmodhydro\\
		\hyperstressmoddevhydro\\
		\hyperstressmodhydrodevhydro
	\end{pmatrix}
	\label{eq:hypertressmodulimodaltoLamecompspatial}
\end{equation} 
for the spatial case $\ndim=3$ and by 
\begin{equation}
	\begin{pmatrix}
		 a_{1}=a_2\\
		 a_{3}\\
		 a_{4}=a_5=a_8\\
		 %a_{10}=a_{14}\\
		 %a_{11}=a_{13}=a_{15}
		 a_{10}=a_{11}=a_{13}=a_{14}=a_{15}
	\end{pmatrix}
 =
\left(\begin {array}{cccc} 
-\frac{1}{12}&0&-\frac{1}{4}&\frac{1}{4}\\
-\frac{1}{12}&\frac{1}{2}&\frac{1}{4}&-\frac{1}{2}\\
-\frac{1}{12}&0&\frac{1}{4}&0\\
\frac{1}{6}&0&0&0\\
%\frac{1}{6}&0&0&0
\end{array} \right)
  \begin{pmatrix}
		\hyperstressmodfulldevsym\\
%		\hyperstressmoddevskewax\\
		\hyperstressmodhydro\\
		\hyperstressmoddevhydro\\
		\hyperstressmodhydrodevhydro
	\end{pmatrix}
	\label{eq:hypertressmodulimodaltoLamecompplane}
\end{equation} 
for the plane case $\ndim=2$.} %from \eqref{eq:hypertressmodulimodaltoLamespatial}

%========================================================================================
\section[Benchmark Problem: Torsion of an elastic cylinder]{Benchmark Problem: Torsion of {an elastic} cylinder}
\label{sec:torsion}
%========================================================================================

%---------------------------------------------------------------------
\subsection{Analytical solution}
\label{sec:torsionsolution}
%---------------------------------------------------------------------

\begin{figure}[hb]
	\begin{minipage}{0.45\textwidth}
	\centering
	\inputsvg{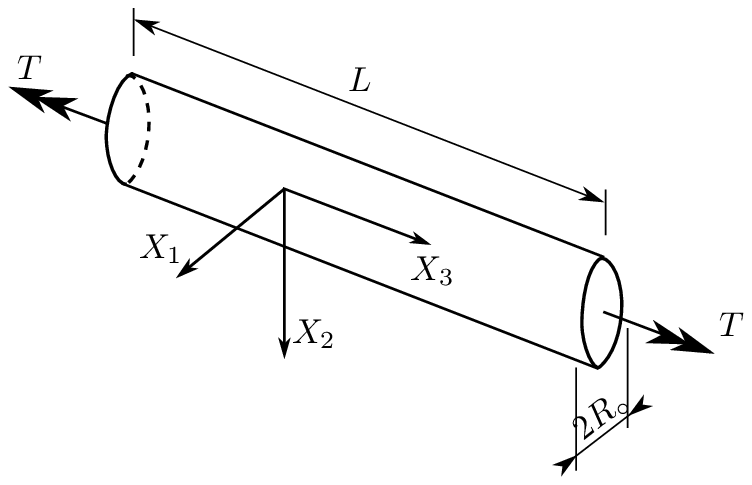}
	\caption{Torsion of a cylinder}
	\label{fig:torsion_sketch}
	\end{minipage}\hfill
  \begin{minipage}{0.45\textwidth}
		\centering
		\inputgnuplot{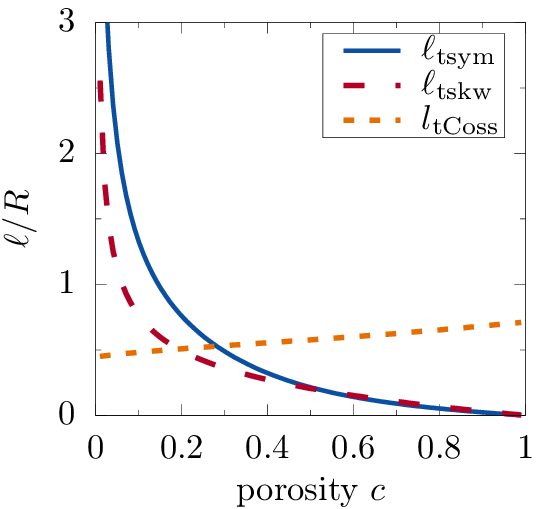}
		\caption{Internal length scales and predicted size effect ($\Poissrat=0.3$)}
		\label{fig:torsion_intlength}
	\end{minipage}
\end{figure}

The size effect in torsion of a slender cylinder with radius $\torsioncylinderrad$ as sketched in \figurename~\ref{fig:torsion_sketch} is a well-studied benchmark problem, starting from the pioneering work of \citet{Gauthier1975} who derived the analytical solution of this problem for the {elastic} Cosserat continuum. Recently, the ansatz for the micromorphic continuum has been found by \citet{Rizzi2021a} for certain special cases of the hyperstress moduli. This ansatz is shown in the following to work also for the isotropic micromorphic continuum in general.
For the displacement field, the classical St.~Venant ansatz of rigidly rotating cross sections is retained. With respect to the chosen coordinate system as sketched in \figurename~\ref{fig:torsion_sketch}, the ansatz reads 
\begin{equation}
	\displmacrocomp_i=-\rotangletorsion(\locmacrocomp_3) \ttpermutcomp_{ik3}\locmacrocomp_k
	\label{eq:torsionansatzdisplacement}
\end{equation}
and leads to a {distortion} field
\begin{equation}
	\displmacrocomp_{i,j}=-\rotangletorsion(\locmacrocomp_3) \ttpermutcomp_{ij3} -\twist\ttpermutcomp_{ik3}\locmacrocomp_k\Kron{j3} 
\end{equation}
in terms of the twist $\twist:=\rotangletorsion'(\locmacrocomp_3)=\mathrm{const.}$
The ansatz for microdeformation field 
\begin{equation}
	\microdefcomp_{ij}=-\rotangletorsion(\locmacrocomp_3) \ttpermutcomp_{ij3} 
	-\frac12 \twist\torsionmicrostrainfunc(\locmacroradplane)\left(\ttpermutcomp_{ik3}\Kron{j3}+\ttpermutcomp_{jk3}\Kron{i3}\right)\locmacrocomp_k
	-\frac12 \twist\torsionmicrorotfunc(\locmacroradplane)\left(\ttpermutcomp_{ik3}\Kron{j3}-\ttpermutcomp_{jk3}\Kron{i3}\right)\locmacrocomp_k \label{eq:ansatzmicrodef}
\end{equation}
incorporates the radial distributions $\torsionmicrostrainfunc(\locmacroradplane)$ and $\torsionmicrorotfunc(\locmacroradplane)$ of the microstrain and microrotation, respectively, with respect to the distance $\locmacroradplane:=\sqrt{\locmacrocomp^2_1+\locmacrocomp^2_2}$ from the central $\locmacrocomp_3$-axis. The ansatz~\eqref{eq:ansatzmicrodef} is a superposition of the ansatzes from \citep{Gauthier1975,Huetter2016} for the micropolar and microstrain continuum, respectively.
{Using the constitutive relations \eqref{eq:innerstressisotropic}--\eqref{eq:stressdiffisotropic}, it}
can be verified that the ansatz fulfills the balance of linear momentum~\eqref{eq:momentumbalancemacro} ad hoc.

Inserting this ansatz to the kinematic relations yields vanishing vectorial harmonic components $\microdefgradgradsymdevhydcomp_{i}=
\microdefgradgradsymhydcomp_{i}=0$ of the gradient of microdeformation and
\begin{align}
	\microdefgradgradfulldevsymcomp_{ijk}=&-\frac{\twist}6  \frac{\torsionmicrostrainfunc'(\locmacroradplane)}{\locmacroradplane} \left[(\ttpermutcomp_{jm3} \Kron{k3}\!+\!\ttpermutcomp_{km3}\Kron{j3} ) \locmacrocompplane_i\!+(\ttpermutcomp_{jm3}\Kron{i3}\!+\!\ttpermutcomp_{im3} \Kron{j3}) \locmacrocompplane_k\!+\!(\ttpermutcomp_{im3}\Kron{k3}\!+\!\ttpermutcomp_{km3}\Kron{i3} ) \locmacrocompplane_j 
	\right] \locmacrocompplane_m  \\
	\microdefgradgradsymdevskewaxmcomp_{ij}=&\frac{\twist}{4} \left[\frac{\torsionmicrostrainfunc'\!\!+\!3\torsionmicrorotfunc'}{\locmacroradplane} \locmacrocompplane_i\locmacrocompplane_j\!-\left(\torsionmicrostrainfunc(\locmacroradplane)\!+\!\locmacroradplane\torsionmicrostrainfunc'\!-\!8\!+\!5\torsionmicrorotfunc(\locmacroradplane)\!+\!4\locmacroradplane\torsionmicrorotfunc'(\locmacroradplane)\right)\Kron{ij}\!+\left(3\torsionmicrostrainfunc\!+\!2\locmacroradplane\torsionmicrostrainfunc'\!-\!6\!-\!3\torsionmicrorotfunc(\locmacroradplane)\right)\Kron{i3}\Kron{j3}\right].
\end{align}
Therein, the abbreviation $\locmacrocompplane_i:=\locmacrocomp_i-\Kron{i3}\locmacrocomp_3$ has been introduced for the in-plane location vector.
Inserting these relations into the constitutive law~\eqref{eq:hyperstressisotropicfulldevsym}--\eqref{eq:hyperstressisotropicsecondorderpartsmatrix} 
{(with $\hyperstressmodincaxdev=\hyperstressmodincaxdevsymskw=0$ according to Section~\ref{sec:micromorphelastichyperstress})}
allows to compute the harmonic modes of the hyperstress and with them  the full hyperstress tensor according to Eq.~\eqref{eq:hyperstresssplit}. Subsequently, the higher-order balance~\eqref{eq:momentumbalancehyperstress} 
%(or directly the respective equation of motion from Appendix~\ref{sec:harmonicdecompositionbalances}) 
is evaluated, in terms of its symmetric and skew parts, as
\begin{align}
0&=\partderivf{\hyperstresscomp_{(ij)k}}{\locmacrocomp_k}\!+\!\stressdiffcomp_{(ij)}=\shearmodeffmicro \twist\left[\frac{2\hyperstressmodfulldevsym\!+\!\hyperstressmoddevskewax}{12\shearmodeffmicro} \left(\torsionmicrostrainfunc''(\locmacroradplane)\!+\!3 \frac{\torsionmicrostrainfunc'(\locmacroradplane)}{\locmacroradplane}\!\right)-\torsionmicrostrainfunc(\locmacroradplane)\!+\frac{\shearmodeffmixed\!+\!\shearmodeffmicro}{\shearmodeffmicro}\right](\ttpermutcomp_{jm3} \Kron{k3}\!+\!\ttpermutcomp_{km3} \Kron{j3} ) \locmacrocomp_k\\
0&=\left[\partderivf{\hyperstresscomp_{ijk}}{\locmacrocomp_k}+\stressdiffcomp_{ij}\right]\ttpermutcomp_{ijp}=\CossParamSkew\twist\left[\frac{\hyperstressmoddevskewax}{2\CossParamSkew} \left(\torsionmicrorotfunc''(\locmacroradplane)+3 \frac{\torsionmicrorotfunc'(\locmacroradplane)}{\locmacroradplane}\right)-\torsionmicrorotfunc(\locmacroradplane)+1\right] \locmacrocompplane_p\,.
	\label{eq:torsionhigherorderquilibrium}
\end{align}
Obviously, the square bracket terms therein have to vanish. These uncoupled ODEs have the solution
\begin{align}
	\torsionmicrostrainfunc(\locmacroradplane)&=\frac{\shearmodeffmixed+\shearmodeffmicro}{\shearmodeffmicro} +C_1  \frac{I_1\left(\locmacroradplane/\intlengthtorsionsym\right)}{\locmacroradplane/\intlengthtorsionsym}+C_2\frac{K_1\left(\locmacroradplane/\intlengthtorsionsym\right)}{\locmacroradplane/\intlengthtorsionsym}\\   
	\torsionmicrorotfunc(\locmacroradplane)&=1+C_3  \frac{I_1\left(\locmacroradplane/\intlengthtorsionskw\right)}{\locmacroradplane/\intlengthtorsionskw}+C_4\frac{K_1\left(\locmacroradplane/\intlengthtorsionskw\right)}{\locmacroradplane/\intlengthtorsionskw} 
	\label{eq:torsionsolutionfunc}
\end{align}
Therein, $I_1(x)$ and $K_1(x)$ are the modified Bessel functions of first order and the characteristic length scales have been defined as
\begin{align}
\intlengthtorsionsym&=\sqrt{\frac{2\hyperstressmodfulldevsym+\hyperstressmoddevskewax}{12\shearmodeffmicro}}, &
\intlengthtorsionskw&=\sqrt{\frac{\hyperstressmoddevskewax}{2\CossParamSkew}}.
\label{eq:torsionintlengths}
\end{align}
The modified Bessel function of second kind $K_1(\locmacroradplane)$ is unbounded at $\locmacroradplane\rightarrow0$, so that the respective coefficients $C_2=C_4=0$ vanish for the compact cylinder under consideration.
The higher-order trivial Neumann boundary {condition} $\hyperstresscomp_{ijk}n_k=0$ at the barrel $\locmacroradplane=\torsioncylinderrad$ of the cylinder (with normal $n_i=\locmacrocompplane_i/\locmacroradplane$) yields two independent equations\footnote{Remarkably, a coupling between $\torsionmicrostrainfunc(\locmacroradplane)$ and $\torsionmicrorotfunc(\locmacroradplane)$, i.e., between microstrain and microrotation, arises due to this boundary condition~\eqref{eq:BCtorsion} (in contrast to the {simplified} hyperstress term in \citep{Rizzi2021a}), even though the respective ODEs \eqref{eq:torsionhigherorderquilibrium} and thus their solution~\eqref{eq:torsionsolutionfunc} are uncoupled.} from its symmetric and skew part 
\begin{align}
  2+\torsionmicrostrainfunc(\torsioncylinderrad)+\torsionmicrorotfunc(\torsioncylinderrad)+\frac{4\shearmodeffmicro}{\CossParamSkew} \frac{\intlengthtorsionsym^2}{\intlengthtorsionskw^2} \torsioncylinderrad\torsionmicrostrainfunc'(\torsioncylinderrad)&=0\, ,  &
	2+\torsionmicrostrainfunc(\torsioncylinderrad)+\torsionmicrorotfunc(\torsioncylinderrad)+2\torsioncylinderrad\torsionmicrorotfunc'(\torsioncylinderrad)=&0 
	\label{eq:BCtorsion}
\end{align}
which allow to determine the coefficients as
\begin{align}
  C_1&=2 \frac{\shearmodeffmixed+4\shearmodeffmicro}{\shearmodeffmicro}\,\frac{I_2\left(\torsioncylinderrad/\intlengthtorsionskw\right)}{det}, &
	C_3&=4 \frac{\shearmodeffmixed+4\shearmodeffmicro}{\CossParamSkew}\,\frac{\intlengthtorsionsym^2}{\intlengthtorsionskw^2}\,\frac{I_2\left(\torsioncylinderrad/\intlengthtorsionsym\right)}{det}
\end{align}
with	
\begin{equation*}
	det=\left(1\!-\frac{6\shearmodeffmicro}{\CossParamSkew} \frac{\intlengthtorsionsym^2}{\intlengthtorsionskw^2}\right) I_2\!\left(\!\frac{\torsioncylinderrad}{\intlengthtorsionsym}\!\right) I_2\!\left(\!\frac{\torsioncylinderrad}{\intlengthtorsionskw}\!\right)-\frac{2\shearmodeffmicro}{\CossParamSkew}  \frac{\intlengthtorsionsym^2}{\intlengthtorsionskw^2} I_2\!\left(\!\frac{\torsioncylinderrad}{\intlengthtorsionsym}\!\right) I_0\!\left(\!\frac{\torsioncylinderrad}{\intlengthtorsionskw}\!\right)-I_2\!\left(\!\frac{\torsioncylinderrad}{\intlengthtorsionskw}\!\right) I_0\!\left(\!\frac{\torsioncylinderrad}{\intlengthtorsionsym}\!\right)\,.
\end{equation*}
The conventional boundary condition $n_i\stressmacrocomp_{ij}=0$ at the barrel is satisfied identically by the employed ansatz \eqref{eq:torsionansatzdisplacement}--\eqref{eq:ansatzmicrodef}.

Next, the work theorem \eqref{eq:worktheorem} is applied to a cut-free cylinder of length $L$ at any position. Thus, the surface $\domainbound$ comprises the cylinder barrel and the two cross sections $\crosssec(\locmacrocomp_3)$ and $\crosssec(\locmacrocomp_3+L)$, the latter with normals $n_i=-\Kron{i3}$ and $n_i=\Kron{i3}$, respectively. 
The barrel is traction-free and does thus not contribute to the external work term, but only $\crosssec(\locmacrocomp_3)$ and $\crosssec(\locmacrocomp_3+L)$ do.
Inserting furthermore the ansatz \eqref{eq:torsionansatzdisplacement}, \eqref{eq:ansatzmicrodef} into \eqref{eq:worktheorem} and taking into account that the terms with $\torsionmicrostrainfunc(\locmacroradplane)$ and $\torsionmicrorotfunc(\locmacroradplane)$ do not depend on $\locmacrocomp_3$, and thus cancel out from $\crosssec(\locmacrocomp_3)$ and $\crosssec(\locmacrocomp_3+L)$, yields
\begin{align}
  \int_{\domain} \rate{\strainenergymacroint}\dV
		&= \left[\rate{\rotangletorsion}(\locmacrocomp_3+L)-\rate{\rotangletorsion}(\locmacrocomp_3)\right] \underbrace{\int_{\crosssec}- \stressmacrocomp_{3j} \ttpermutcomp_{ik3}\locmacrocomp_k  - \hyperstresscomp_{ij3} \ttpermutcomp_{ij3}  \dsurf}_{:=\torque}
		\label{eq:deftorque}
\end{align}
and allows to introduce the torque $\torque$ as work-conjugate quantity to the relative rotation $\rotangletorsion(\locmacrocomp_3+L)-\rotangletorsion(\locmacrocomp_3)$ of the two cross sections. In terms of the harmonic decomposition~\eqref{eq:hyperstresssplit}, the hyperstress term therein can be written as $\hyperstresscomp_{ij3} \ttpermutcomp_{ij3}=-\hyperstressdevskewaxmcomp_{33}$, so that the expression for the torque becomes
\begin{equation}
	\torque=\int_{\crosssec} \stressmacrocomp_{3j} \ttpermutcomp_{i3k}\locmacrocomp_k  + \hyperstressdevskewaxmcomp_{33} \dsurf\,,
\end{equation}
substantiating the microscale interpretation from Section~\ref{sec:micromorphelastic} of $\hyperstressdevskewaxmcomp_{ij}$ as a micro-torque stress.
For the present solution, this expression becomes
\begin{equation}
	\begin{split}
	\torque=\twist \frac{\pi\shearmodeff\torsioncylinderrad^4}{2} 
		\left\lbrace 1\!+\!\frac{2\CossParamSkew}{\shearmodeff} \frac{\intlengthtorsionskw^2}{\torsioncylinderrad^2}\,\frac{\shearmodeffmixed\!+\!4\shearmodeffmicro}{\shearmodeffmicro}
		+2C_1\!\left[\frac{\CossParamSkew}{\shearmodeff}\,\frac{\intlengthtorsionskw^2}{\torsioncylinderrad^2}\,\frac{I_1\left(\frac{\torsioncylinderrad}{\intlengthtorsionsym}\right)}{\frac{\torsioncylinderrad}{\intlengthtorsionsym}}\!-\!2 \frac{\shearmodeffmixed\!+\!\shearmodeffmicro}{\shearmodeff}\frac{\intlengthtorsionsym^2}{\torsioncylinderrad^2}I_2\left(\frac{\torsioncylinderrad}{\intlengthtorsionsym}\right)\right]\right.\\
		\left.\!+\!C_3  \frac{\CossParamSkew}{\shearmodeff}\frac{\intlengthtorsionskw^2}{\torsioncylinderrad^2}\left[I_0\left(\frac{\torsioncylinderrad}{\intlengthtorsionskw}\right)\!-\!3I_2\left(\frac{\torsioncylinderrad}{\intlengthtorsionskw}\right)\right]\right\rbrace
		\end{split}
		\label{eq:torsionsolution}
\end{equation}
For the following ease of notation, the symbol $\torsionstiffness=\torque/\twist$ is introduced for the torsion stiffness and $\torsionrelstiffness=2 \torsionstiffness/(\pi\shearmodeff\torsioncylinderrad^4)$ as ratio to its classical, and limiting, value \citep{Gauthier1975}. Obviously, $\torsionrelstiffness$ corresponds to the curly bracket in Eq.~\eqref{eq:torsionsolution}.

\subsection{Predictions from micromorphic homogenisation}

In the following section, the homogenized micromorphic parameters from section~\ref{sec:micromorphelastic} are used to evaluate the torsion solution. 
Firstly, the length parameters from Eq.~\eqref{eq:torsionintlengths} are plotted in \figurename~\ref{fig:torsion_intlength} against the porosity. 
%\begin{figure}
	%\centering
	%\inputgnuplot{TorsionIntlengthHomogen}
	%\caption{Internal length scales and predicted size effect ($\Poissrat=0.3$)}
	%\label{fig:torsion_intlength}
%\end{figure}
At the first glimpse, it looks surprising that both lengths, $\intlengthtorsionsym$ and $\intlengthtorsionskw$, \emph{decrease} with increasing porosity $\VVF$, and do even diverge for $\VVF\rightarrow0$. However, the observable size effect arises only in interaction with the coupling moduli which vanish for $\VVF\rightarrow0$, compare section~\ref{sec:micromorphelasticstrains}. 
In this context, it shall be remarked that $\intlengthtorsionskw$ does \emph{not} correspond to the internal length $\CossCharlengthtorsion$ defined by \citet{Gauthier1975} for the Cosserat solution via the ratio\footnote{Exactly speaking, $\intlengthtorsionskw$ corresponds to $1/p$ in the work of \citet{Gauthier1975}.} of the hyperstress modulus and effective shear modulus $\shearmodeff$. 
For comparison, the respective {value} $\CossCharlengthtorsion=\sqrt{\hyperstressmoddevskewax/(2\shearmodeff)}$ is plotted as well in \figurename~\ref{fig:torsion_intlength}. It can be observed that this quantity does hardly depend on porosity $\VVF$.

\begin{figure}
	\centering
	\subfloat[]{\inputgnuplot{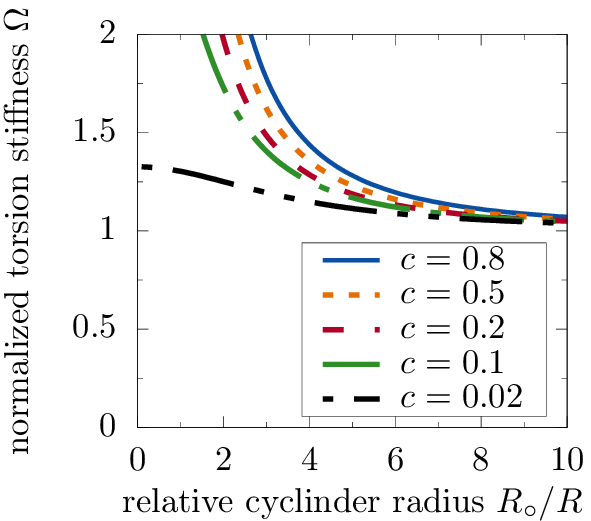}	\label{fig:sizeffecttorsionRelSizeEffect}}
	\hspace*{1cm}
	\subfloat[]{\inputgnuplot{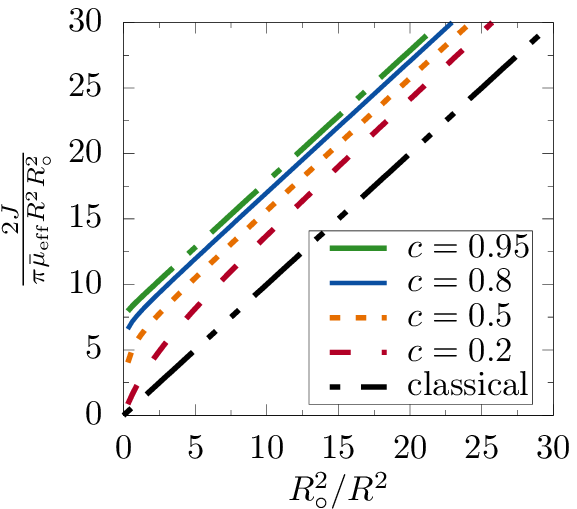} \label{fig:sizeffecttorsionSizeEffectPlot}}
	\caption{Predicted size effect in torsion in two equivalent representations ($\Poissrat=0.3$)}
	\label{fig:sizeffecttorsion}
\end{figure}

Figure~\ref{fig:sizeffecttorsion} shows the predicted size effect within the torsion stiffness in two different representations for different values of the porosity $\VVF$.
Figure~\ref{fig:sizeffecttorsionRelSizeEffect} shows the {ratio} $\torsionrelstiffness$ of torsion stiffness to its classical counterpart as employed by \citet{Gauthier1975} versus the scale separation ratio, whereas \figurename~\ref{fig:sizeffecttorsionSizeEffectPlot} shows a representation $\torsionstiffness/\torsioncylinderrad^2$ versus $\torsioncylinderrad$ as proposed by \citet{Lakes1983} for calibrating material parameters to measured data. The latter representation is furthermore normalized here in such a way that the classical solution corresponds to a line of unit slope through the origin and with respect to the intrinsic length $\radcell$.
Both representations show that the present theory predicts that the classical theory is recovered for microhomogeneous material $\VVF\rightarrow0$. The reason is that the coupling moduli tend to zero in this case (even though the hyperstress moduli remain finite!). In the opposite limiting case of highly porous material $\VVF\rightarrow 1$, the size effect is predicted to saturate (in the normalization with respect to the effective shear modulus $\shearmodeff$). {Figure~\ref{fig:sizeffecttorsionRelSizeEffect} shows the region where the present theory is assumed to yield reasonable results for the size effect. For a scale separation ratio (the relative cylinder radius) below about 2, the continuum assumption becomes questionable. For a scale separation ratio higher than 10, the predictions become virtually indistinguishable from the classical theory. This predictions complies with the general rule of thumb regarding the applicability limit of classical homogenization approaches.}

\begin{figure}
	\centering
	\subfloat[]{\inputgnuplot{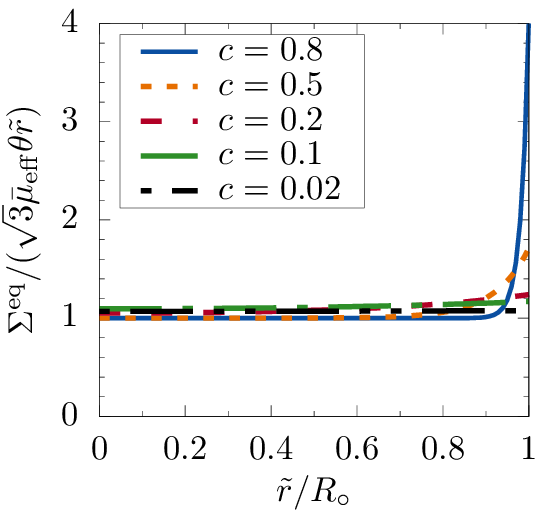}\label{fig:torsionstresssym}}
	\hspace{1cm}
	\subfloat[]{\inputgnuplot{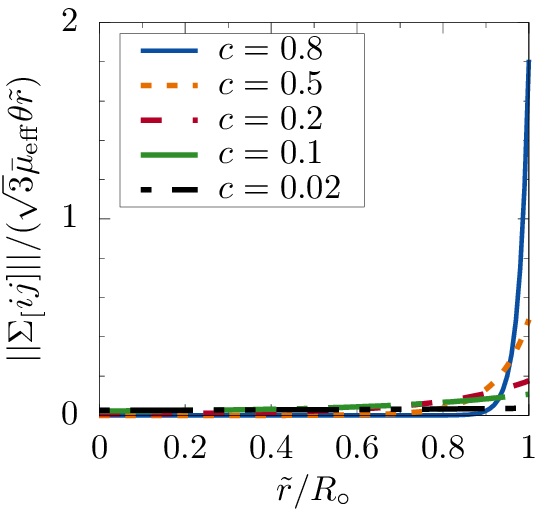}\label{fig:torsionstressskw}}
	\caption[]{Stresses in torsion, normalized to classical solution ($\Poissrat=0.3$, $\torsioncylinderrad/\radcell=3$): \subref{fig:torsionstresssym} Mises stress for symmetric part, \subref{fig:torsionstressskw} skew part of stress}
	\label{fig:torsionstress}
\end{figure}

The resulting symmetric and skew parts of the stresses are plotted in \figurename~\ref{fig:torsionstress} for a specimen of particular size, normalized to the classical solution. 
Again, it shows that the classical solution is recovered for microhomogeneous material $\VVF\rightarrow0$. In presence of voids $\VVF>0$, a boundary layer is predicted to form, whose amplitude increases with porosity $\VVF$. {The formation of such boundary layers is a well-known feature from experiments and fully resolved simulations.}

\subsection{Comparison to experiments}

Qualitatively, the predictions agree with the numerous experiments of Lakes and co-workers on porous materials as reviewed in \citep{Lakes2020}. Firstly, the coupling moduli increase with porosity as found in \citep{Rueger-Lakes2019}, compare figures~\ref{fig:deviatoric_plot} and \ref{fig:stressskew_plot}, and so does the maximum attainable size effect in torsion, compare \figurename~\ref{fig:sizeffecttorsion}. And of course the prediction that any size effect vanishes for microhomogeneous material is plausible and required, although it is not ensured by certain higher-order homogenization theories, e.g.~\citep{Gologanu1997,Huetter2016,Zybell2009}. Also the fact that a full micromorphic theory with micropolar and microstrain contributions like the present one is required to explain the nonclassic effects in foams (size dependent apparent moduli in bending and torsion, dispersion of longitudinal and transversal waves, skew part of stress tensor) is generally accepted, compare \citep{Lakes2020} and references therein. 

In the following, the predictions of the present theory are compared \emph{quantitatively} with respective experimental results of Lakes et al. %\citep{Anderson1994a,Rueger-Lakes2019,Lakes1986}. 
\citet{Anderson1994a}, studied a polymethacrylimide (PMI) closed-cell foam Rohacell WF300 with porosity $\VVF\approx0.7$ and cell size 0.65~mm whereas a high-density polyurethane (PU) closed-cell foam FR3720 with $\VVF\approx0.3$ and (largest) cell size of 0.15~mm was investigated in \citep{Anderson1994a,Rueger-Lakes2019,Lakes1986}.

\begin{figure}
	\centering
	\inputgnuplot{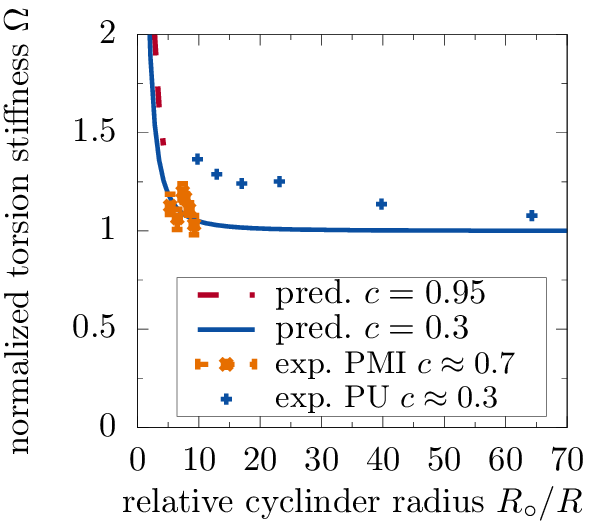}
	\caption{Internal length scales and predicted size effect ($\Poissrat=0.3$) in comparison to experimental data of PMI foam \citep{Anderson1994a} and dense PU foam \citep{Rueger-Lakes2019,Lakes1986} }
	\label{fig:torsion_cmp_exp}
\end{figure}

Figure~\ref{fig:torsion_cmp_exp} shows the normalized size effects from theory and experiments, whereby $\radcell=\text{cell size}/2$ was assumed.
The present theory predicts a negligible effect of porosity $\VVF$ on the normalized size effect within the range of $\VVF$ {and scale separation} under consideration {(i.e., where a continuum theory can reasonably be applied)}. The behavior of the PMI foam is predicted quite well by the present theory, but the size effect of the dense PU foam is underestimated. Remarkably, the latter exhibits a stronger size effect ({regarding magnitude $\torsionrelstiffness$ of stiffness increase as well maximum scale separation ratio at which a deviation from classical theory is observed}) even though its porosity is lower. 
Presumably, the size effect, even in the employed normalization, is not only determined by the porosity $\VVF$ (or, equivalently, relative density $1-\VVF$) of a foam, but also by its topology as indicated by recent numerical experiments \citep{Pham2021}. The present composite shell is a strong geometric simplification of the topology and can thus not reproduce this effect. The effect of the foam topology on classical effective properties is well-known, see e.g. \citep{GibsonAshby,Yoder2019,Soyarslan2019,Jang2008,Storm2019,Tekoglu2011,Redenbach2012}.

%---------------------------------------
\section{Summary and conclusions}
\label{sec:summary}
%---------------------------------------

The scope of the present contribution was to contribute to the understanding {and interpretation of the constitutive relations in micromorphic theory and the large number of constitutive parameters contained therein}. For this purpose, a harmonic decomposition was applied to the constitutive equations as well as to the governing equations. Besides the well-known decomposition of the second order tensors into trace, deviator and skew parts, it turns out that the constitutive relations for the hyperstress tensor decompose into independent modes as well. 

The harmonic modes have been interpreted at the micro-scale for a porous medium using Hashin's composite shell model, i.e., a spherical microscopic volume element with spherical pore. The microdeformation corresponds to the deformation of the pore surface. In particular, the skew part of the conventional stress is found to drive a relative rotation between the surfaces of pore and volume element. The harmonic modes of hyperstress and gradient of microdeformation, respectively, have different characters. There are modes which are of bending type and one of torsion type (which is active only in 3D problems). Another one, which is driven by the trace of the hyperstresses, corresponds to a relative translation between pore and matrix. {The three modes being related to the incompatible part of the gradient of microdeformation can be related to dislocation-induced lattice curvature.}

Quantitatively, the resulting elastic boundary value problem at the microscale has been solved for periodic boundary conditions using spherical harmonics. The known elastic basic solutions in terms of spherical harmonics have been written in index notation using harmonic tensors and can thus be used (relatively) conveniently, not only for the present problem. 
Furthermore, the known solutions have been generalized to incorporate the 2D case as well.

In consequence, all 18 constitutive parameters of the {linear-elastic} micromorphic continuum have been derived in closed form. In particular, the resulting non-classical moduli for the microdeformations take remarkably simple forms: The (bulk and shear) strain moduli are identical to the Taylor-Voigt estimates, the coupling moduli corresponds to the virtual stiffness of the pore if it were filled with material and the micromodulus is the \enquote{difference} to the effective modulus under periodic boundary conditions. 
The strain moduli correspond to the maximum stiffness of very, theoretically infinitesimally, small specimens \citep{Neff2019}. 
Heuristically, \citet{Neff2019} proposed to use the RVE solution with kinematic boundary conditions for this limit. But for the present composite shell model, kinematic boundary conditions yield the same result as periodic boundary conditions, so that this heuristic approach would predict a vanishing size effect.
More plausible, the present theory predicts that this (theoretical) limit corresponds to the Taylor-Voigt bound.
Nevertheless, the predicted size effect vanishes for homogeneous material since Taylor-Voigt bound and effective value by periodic boundary conditions do coincide. Such a prediction has been recognized as a necessary requirement for micromorphic homogenization procedures, which has now been proven to be satisfied by the employed homogenization method of the author.    

As an application, the torsion of {an elastic} slender cylinder has been investigated. It has been demonstrated that a previously published ansatz for the microdeformation field, comprising microrotations and deviatoric microstrains, works also for the full micromorphic theory, though with an additional coupling of microstrain and microrotation components via the higher-order boundary conditions. The microstrain and microrotation parts give rise to two internal length scales. The obtained expressions for these two length scales comprise the hyperstress moduli of two harmonic modes which are activated under torsion. This finding substantiates the interpretation of the respective {harmonic} hyperstress {modes}.

Finally, a quantitative prediction for the size effect under torsion is generated using the obtained constitutive parameters from homogenization. The predictions complies qualitatively and partly quantitatively with respective experiments from literature. 

Despite these partly quantitative deviations due to the highly simplified microstructure in the employed composite shell model, the derived closed form expressions for all of the 18 parameters can serve as a good starting point for future applications of the linear micromorphic theory to porous materials. 
The found simple relations between the non-classical moduli for the microdeformations and classical bounds of homogenization theory have the potential for an even wider usage and may thus reduce obstacles in using the micromorphic theory.

\section*{Acknowledgments}

The authors wishes to thank Nicolas Auffray, Rainer Glüge, Roderic Lakes, Markus Lazar, Patrizio Neff and Gianlucca Rizzi for stimulating discussions.

\bibliographystyle{elsarticle-harv}
\bibliography{MicromorphicPropsFoam3D}

\appendix

\section{Obtaining general Mindlin parameters from their modal counterparts}
\label{sec:hypertressmodulimodaltoLame}

{
For the spatial case $\ndim=3$, the inversion of Eq.~\eqref{eq:hyperstressisotropicvectorpartsmatrix} reads
\begin{align}
	\begin{pmatrix}
		 a_{1}\\
		 a_{2}\\
		 a_{3}\\
		 a_{4}\\
		 a_{5}\\
		 a_{8}\\
		 \smash{a_{10}}\\
		 \smash{a_{11}}\\
		 \smash{a_{13}}\\
		 \smash{a_{14}}\\
		 \smash{a_{15}}
	\end{pmatrix}
	=
\left( \begin {array}{ccccccccccc} -{\frac{1}{15}}&{\frac{1}{6}}&\zz&-{\frac{1}{4}}&\zz&-{\frac{1}{10}}&\zz&{\frac{1}{6}}&{\frac{1}{6}}&-{\frac{1}{10}}&\zz\\ 
-{\frac{1}{15}}&{\frac{1}{6}}&\zz&{\frac{1}{4}}&\zz&-{\frac{1}{10}}&\zz&{\frac{1}{6}}&-{\frac{1}{6}}&{\frac{1}{10}}&\zz\\
-{\frac{1}{15}}&-{\frac{1}{3}}&\zz&\zz&{\frac{1}{3}}&{\frac{1}{15}}&\zz&-{\frac{2}{9}}&\zz&0&\zz\\
-{\frac{1}{15}}&-{\frac{1}{12}}&\zz&{\frac{3}{8}}&\zz&{\frac{3}{20}}&{\frac{1}{2}}&\zz&
-{\frac{1}{2}}&{\frac{9}{20}}&\zz\\
-{\frac{1}{15}}&-{\frac{1}{12}}&\zz&-{\frac{1}{8}}&\zz&{\frac{3}{20}}&-{\frac{1}{2}}&\zz&{\frac{1}{2}}&-{\frac{3}{20}}&\zz\\
-{\frac{1}{15}}&-{\frac{1}{12}}&\zz&-{\frac{1}{8}}&\zz&{\frac{3}{20}}&{\frac{1}{2}}&\zz&-{\frac{1}{2}}&-{\frac{3}{20}}&\zz\\
{\frac{1}{6}}&{\frac{1}{3}}&{\frac{1}{8}}&\zz&\zz&\zz&\zz&\zz&\zz&\zz&{\frac{1}{16}}\\ 
{\frac{1}{6}}&-{\frac{1}{6}}&-{\frac{3}{8}}&\zz&\zz&\zz&\zz&\zz&\zz&\zz&{\frac{1}{16}}\\ 
{\frac{1}{6}}&-{\frac{1}{6}}&{\frac{3}{8}}&-{\frac{1}{2}}&\zz&\zz&\zz&\zz&\zz&\zz&-{\frac{1}{16}}\\
{\frac{1}{6}}&{\frac{1}{3}}&-{\frac{1}{8}}&\zz&\zz&\zz&\zz&\zz&\zz&\zz&-{\frac{1}{16}}\\
{\frac{1}{6}}&-{\frac{1}{6}}&{\frac{3}{8}}&{\frac{1}{2}}&\zz&\zz&\zz&\zz&\zz&\zz&-{\frac{1}{16}}
\end {array}
 \right)
\cdot
  \begin{pmatrix}
		\hyperstressmodfulldevsym\\
		\hyperstressmoddevskewax\\
		\hyperstressmodincaxdev\\
		\hyperstressmodincaxdevsymskw\\
		\hyperstressmodhydro\\
		\hyperstressmoddevhydro\\
		\hyperstressmodincaxskwax\\
		\hyperstressmodhydrodevhydro\\
		\hyperstressmodhydroincaxskwax\\
		\hyperstressmoddevhydroincaxskwax\\
		\hyperstressmodincaxhydro
	\end{pmatrix}
	\label{eq:hypertressmodulimodaltoLamespatial}
\end{align}
For the plane case $\ndim=2$ and under usage of $\hyperstressmoddevskewax=\hyperstressmodincaxdev=\hyperstressmodincaxdevsymskw=\hyperstressmodincaxhydro=0$
\begin{align}
	\begin{pmatrix}
		 a_{1}\\
		 a_{2}\\
		 a_{3}\\
		 a_{4}\\
		 a_{5}\\
		 a_{8}\\
		 a_{10}=a_{13}=a_{14}=a_{15}
%		 \smash{a_{10}}\\
%		 \smash{a_{11}}\\
%		 \smash{a_{13}}\\
%		 \smash{a_{14}}\\
%		 \smash{a_{15}}
	\end{pmatrix}
	=
	\frac{1}{12}
 \left( \begin {array}{ccccccc} 
	-1&\zz&-3&\zz&3&3&-3\\ 
	-1&\zz&-3&\zz&3&-3&3\\ 
	-1&6&3&\zz&-6&\zz&\zz\\ 
	-1&\zz&3&6&\zz&-6&9\\ 
	-1&\zz&3&-6&\zz&6&-3\\ 
	-1&\zz&3&6&\zz&-6&-3\\ 
	 2&\zz&\zz&\zz&\zz&\zz&\zz
	 %2&0&0&0&0&0&0\\ 
	 %2&0&0&0&0&0&0\\ 
	 %2&0&0&0&0&0&0\\ 
	 %2&0&0&0&0&0&0
\end {array} \right)
\cdot
  \begin{pmatrix}
		\hyperstressmodfulldevsym\\
		%\hyperstressmoddevskewax\\
		%\hyperstressmodincaxdev\\
		%\hyperstressmodincaxdevsymskw\\
		\hyperstressmodhydro\\
		\hyperstressmoddevhydro\\
		\hyperstressmodincaxskwax\\
		\hyperstressmodhydrodevhydro\\
		\hyperstressmodhydroincaxskwax\\
		\hyperstressmoddevhydroincaxskwax\\
		%\hyperstressmodincaxhydro
	\end{pmatrix}
	\label{eq:hypertressmodulimodaltoLameplane}
\end{align}
is obtained.
}

%---------------------------------------------------------------------------
\section{Solutions of Navier-Lamé equations in terms of spherical harmonics}
\label{sec:generalsolutions}

\citet[§172]{Love1944} provided three types of solutions to the homogeneous Navier-Lamé equations in terms of spherical harmonics $\sphericalharmonic$ of degree $\sphericalharmonicorder$. They have been transformed to tensor notation and been generalized here to incorporate the {plane} strain case $\ndim=2$ in addition to the spatial case $\ndim=3$:

\paragraph{Type $\omega$}

\begin{align}
	\displcomp_{i}&=r^2 \frac{\partial \sphericalharmonic\!\!\!\!\!\!}{\partial\locmicrocomp_i}+\spheresolomegacoeff \locmicrocomp_i \sphericalharmonic\\
	\tractioncomprad{i}&=\Lamesec(2\sphericalharmonicorder\!+\!\spheresolomegacoeff)r \frac{\partial \sphericalharmonic\!\!\!\!\!\!}{\partial\locmicrocomp_j}+\left[2\sphericalharmonicorder(\Lamefirst\!\!+\!\Lamesec)\!+\!((\sphericalharmonicorder\!+\!\ndim)\Lamefirst\!+\!(2\!+\!\sphericalharmonicorder)\Lamesec) \spheresolomegacoeff\right]  \frac{\locmicrocomp_j}{r} \sphericalharmonic
\end{align}
\begin{equation}
\text{with }
\spheresolomegacoeff=-2\frac{\sphericalharmonicorder\Lamefirst+\Lamesec(3\sphericalharmonicorder+\ndim-2)}{\Lamefirst(\ndim+\sphericalharmonicorder)+\Lamesec(\sphericalharmonicorder+\ndim+2)}
=-2\frac{3\sphericalharmonicorder + \ndim - 2+(4-4\sphericalharmonicorder - 2\ndim)\Poissrat}{\sphericalharmonicorder + \ndim - 4\Poissrat + 2}
\end{equation}

\paragraph{Type $\phi$ (Lamé potential)}

\begin{align}
	\displcomp_{i}&=\frac{\partial \sphericalharmonic}{\partial\locmicrocomp_i}, &
	\tractioncomprad{i}&=\frac{2\Lamesec}{r} (\sphericalharmonicorder-1) \frac{\partial \sphericalharmonic}{\partial\locmicrocomp_i}
\end{align}	

\paragraph{Type $\chi$}

\begin{align}
	\displcomp_{i}&=\ttpermutcomp_{ijk} \locmicrocomp_j \frac{\partial \sphericalharmonic}{\partial\locmicrocomp_k}, &
	\tractioncomprad{i}=\frac{2\Lamesec}{r} (\sphericalharmonicorder-1) \ttpermutcomp_{ijk} \locmicrocomp_j \frac{\partial \sphericalharmonic}{\partial\locmicrocomp_k}
\end{align}

%\paragraph{Spherical harmonics in tensor notation}

According to \citet{Applequist1989}, any spherical harmonic can be expressed via scalar products between a tensor and the location vector $\locmicrocomp_i$ and respective powers of its norm $r=\sqrt{\locmicrocomp_i\locmicrocomp_i}$. In particular,
\begin{equation}
	\sphericalharmonic=A_{i_1i_2i_3\dots i_{\sphericalharmonicorder}}\locmicrocomp_{i_1}\locmicrocomp_{i_2}\locmicrocomp_{i_3}\dots\locmicrocomp_{i_{\sphericalharmonicorder}}
\end{equation}
is an ordinary spherical harmonic of degree $\sphericalharmonicorder$ if the tensor $A_{i_1i_2i_3\dots i_{\sphericalharmonicorder}}$ is symmetric and traceless with respect to all of its indices $i_1$ to $i_{\sphericalharmonicorder}$. In this case, $\sphericalharmonic[2-\ndim-\sphericalharmonicorder]=\sphericalharmonic/r^{2\sphericalharmonicorder+\ndim-2}$ is a solid spherical harmonic of degree $2-\ndim-\sphericalharmonicorder$ with the same surface spherical harmonic part as $\sphericalharmonic$.

%=======================================================
\section[Elastic microscale solution for the hyperstress terms]{{Elastic microscale solution for the hyperstress terms}}
\label{sec:hyperstressdetails}

{The general solution procedure (which is actually not limited to the present problem), works as follows: From the general solution of the hollow-sphere problem (section~\ref{sec:spherical harmonics} and appendix~\ref{sec:generalsolutions}), an ansatz has to constructed which comprises all terms, whose underlying spherical harmonics can be expressed in terms of harmonic tensors of the same order. 
Thereby, it has to be taken into account whether the respective tensors are of polar ($\hyperstressfulldevsymcomp_{ijk}$, $\hyperstresshydrocomp_i$, $\hyperstressdevhydrocomp_i$) or axial ($\hyperstressdevskewaxmcomp_{ij}$) nature. In the former case, an ansatz incorporating the terms of types $\omega$ and $\phi$ is required (compare section~\ref{sec:microscaledevstrain}; also the dilatational solution in section~\ref{sec:microscaledilatation} can be represented this way) whereas the latter case requires an ansatz of type $\chi$ (compare section~\ref{sec:microscalerelrotation}).
The coefficients of the ansatz are obtained by comparison of polynomials with respective parts of boundary conditions. Finally the micro-macro relations 	
\eqref{eq:microdefgradgradmicrocomppartintdevsphere}--\eqref{eq:microdefgradgradmicrocomppartinthydsphere} can be evaluated.  Subsequently, a comparison with the general isotropic law \eqref{eq:hyperstressisotropicfulldevsym}--\eqref{eq:hyperstressisotropicinchydro} allows to extract the modal hyperstress moduli.}

%-------------------------------------------------------
\subsection{Totally symmetric part}
%-------------------------------------------------------

The boundary conditions of \eqref{eq:natBChyperstressspheresplit} for the totally symmetric and deviatoric part $\hyperstressfulldevsymcomp_{ijk}$ 
can be satisfied by an ansatz, which has also totally symmetric (non-axial) tensors of third order as coefficients. In particular, an ansatz of type $\omega$ is combined with an ansatz of type $\phi$, both with spherical harmonics of degree 3 and $-\ndim-1$ (see appendix~\ref{sec:generalsolutions}), respectively,
\begin{align}
	\displcomp_i=& 3r^2 A_{ijk} x_j x_k+\spheresolomegacoeff[3] x_i A_{jkl} x_j x_k x_l+\frac{3}{r^{\ndim+2}}  B_{ijk} x_j x_k+(\spheresolomegacoeff[-\ndim-1]-\ndim-4)\frac{1}{r^{\ndim+4}}  x_i B_{jkl} x_j x_k x_l \nonumber \\
	&+3C_{ijk} x_j x_k+\frac{3}{r^{\ndim+4}}  D_{ijk} x_j x_k-\frac{\ndim+4}{r^{\ndim+6}}  x_i D_{jkl} x_j x_k x_l \label{eq:displacementshyperstress_Mdds}\\
\intertext{corresponding to tractions}
	\tractioncomprad{i}=&\left[3r\Lamesec(6\!+\!\spheresolomegacoeff[3] ) A_{ijk}\!-\!(2\!+\!2\ndim\!-\!\spheresolomegacoeff[{\!-\!\ndim\!-\!1}] )  \frac{3\Lamesec}{r^{\ndim\!+\!3}}  B_{ijk}\right] x_j x_k \nonumber\\
	&+\left[6(\Lamefirst\!+\!\Lamesec)+((3\!+\!\ndim)\Lamefirst\!+\!5\Lamesec) \spheresolomegacoeff[3] \right]  \frac{1}{r}A_{jkl} x_i x_j x_k x_l \nonumber \\
&+
\left[2(\ndim+1)(\Lamesec(\ndim+3)+\Lamefirst)-(\Lamefirst+(2\ndim+3)\Lamesec) \spheresolomegacoeff[-\ndim-1] \right] 
B_{jkl} x_i x_j x_k x_l \nonumber\\
&+\frac{12\Lamesec}{r} C_{ijk} x_j x_k-\frac{2\Lamesec}{r}(\ndim+2)\left[\frac{3}{r^{\ndim+4}}  D_{ijk} x_j x_k-\frac{\ndim+4}{r^{\ndim+6}}  x_i D_{jkl} x_j x_k x_l \right]
\label{eq:tractionshyperstress_Mdds}
\end{align}	
Equating the coefficients of quadratic and quartic terms of the 
{respective part of boundary condition~\eqref{eq:natBChyperstressspheresplit}}
with ansatz~\eqref{eq:tractionshyperstress_Mdds} at $r=\radcell$ and the traction-free condition $\tractioncomprad[r=\radvoid]{i}=0$ at the pore surface yields four linear equations for the coefficient tensors $A_{ijk}$, $B_{ijk}$, $C_{ijk}$, $D_{ijk}$. These equations are decoupled for each set of indices $ijk$ so that the four coefficient tensors are coaxial to $\hyperstressfulldevsymcomp_{ijk}$. Finally, the kinematic micro-macro relation~\eqref{eq:microdefgradgradmicrosym} has to be evaluated for the respective displacement field~\eqref{eq:microdefgradgradmicrocomppartintdevsphere} yielding
\begin{equation}
	\microdefgradgradsymcomp_{ijk}=\microdefgradgradfulldevsymcomp_{ijk}=3 \frac{\ndim+5}{\ndim+4}\spheresolomegacoeff[3] \radcell^2 A_{ijk}+\frac{\spheresolomegacoeff[-\ndim-1]}{\ndim+4}  \frac{3}{\radcell^{\ndim+2}}  B_{ijk}+3C_{ijk}
  \label{eq:microdefgradgradsymdds}
\end{equation}
For obtaining this expression, surface averages over polynomials had to be computed which are listed in appendix~\ref{sec:surfintegrals}. Equation~\eqref{eq:microdefgradgradsymdds} shows that the resulting gradient of microdeformation $\microdefgradgradsymcomp_{ijk}$ is coaxial to $A_{ijk}$, $B_{ijk}$ and $C_{ijk}$, which are themselves coaxial to the totally symmetric and deviatioric part $\hyperstressfulldevsymcomp_{ijk}$ of the hyperstress. Thus, the resulting $\microdefgradgradsymcomp_{ijk}$ is coaxial to $\hyperstressfulldevsymcomp_{ijk}$, in accordance with the general isotropic law~\eqref{eq:hyperstressisotropicfulldevsym}.
Inserting the aforementioned solution for the coefficients $A_{ijk}$, $B_{ijk}$ and $C_{ijk}$ to Eq.~\eqref{eq:microdefgradgradsymdds} yields
\begin{align}
	\hyperstressfulldevsymcomp_{ijk}=\hyperstressmodfulldevsym \microdefgradgradfulldevsymcomp_{ijk}
\end{align}
with
{modulus $\hyperstressmodfulldevsym$ according to
Eqs.~\eqref{eq:hyperstressmodfulldevsymplane} and \eqref{eq:hyperstressmodfulldevsymspatial}.}

%-------------------------------------------------------
\subsection{Vectorial part of deviator}
%-------------------------------------------------------

For satisfying {the part of the boundary conditions \eqref{eq:natBChyperstressspheresplit} associated with the vectorial part $\hyperstressdevhydrocomp_k$ of the deviator}, an ansatz has to be found which has vectors as coefficients, i.e.\ which comprises surface spherical harmonics of degree one. This applies to ansatzes of type $\omega$ and $\phi$ with solid spherical harmonics of order $1$ and $1-\ndim$, respectively (see appendix~\ref{sec:generalsolutions}). The respective fields of displacements and the resulting tractions are
\begin{align}
	\displcomp_i=&r^2 A_i+\spheresolomegacoeff[1] x_i A_k x_k+\frac{1}{r^{\ndim-2}}  B_i+(\spheresolomegacoeff[1-\ndim]-\ndim) x_i  \frac{1}{r^{\ndim}}  B_k x_k
	+C_i+\frac{1}{r^{\ndim}}  D_i- \frac{\ndim}{r^{\ndim+2}}  x_i D_k x_k
	\label{eq:displacementshyperstress_Mdh}\\
	\tractioncomprad{i}=&\Lamesec(2+\spheresolomegacoeff[1] )r A_i+\Lamesec(2-2\ndim+\spheresolomegacoeff[1-\ndim] )  \frac{1}{r^{\ndim-1}}  B_i+\left[2(\Lamefirst+\Lamesec)+((1+\ndim)\Lamefirst+3\Lamesec) \spheresolomegacoeff[1] \right]\frac{1}{r}  x_i A_k x_k \nonumber\\
	&+\left[2(1\!-\!\ndim)(\Lamefirst\!+\!\Lamesec\!-\!\Lamesec \ndim)\!+\!(\Lamefirst\!+\!(3\!-\!2\ndim)\Lamesec) \spheresolomegacoeff[1\!-\!\ndim] \right]  \frac{1}{r^{\ndim+1}}  x_i B_k x_k %\nonumber \\
-2\ndim\Lamesec\left(\frac{1}{r^{\ndim\!+\!1}}  D_i\!-\!\frac{\ndim}{r^{\ndim+3}}  x_i D_k x_k \right)\,.
\label{eq:tractionshyperstress_Mdh}
\end{align}	
Obviously, the term $C_i$ corresponds to a pure rigid translation which remains undetermined (and irrelevant) for pure traction boundary conditions~\eqref{eq:natBChyperstressspheresplit}. Vice verse, the latter are self-equilibrating and so are the tractions~\eqref{eq:tractionshyperstress_Mdh}.
Inserting the tractions~\eqref{eq:tractionshyperstress_Mdh} into the boundary condition and the traction-free boundary condition at the pore $r=\radvoid$ and comparing the constant and quadratic terms under consideration of the self-equilibrating condition allows to determine the remaining coefficients as
\begin{align}
	A_j&=-\frac{1}{(2\!+\!\spheresolomegacoeff[1] )\Lamesec \radcell^2 }\,  \frac{1}{1\!-\!\VVF^{\frac{\ndim+2}{\ndim}}}  \frac{1}{\ndim-1} \hyperstressdevhydrocomp_j,&
	B_j&=0,&
	D_j&=-\frac{\radcell^{\ndim}}{2\ndim\Lamesec}  \frac{1}{\ndim-1}  \frac{\VVF^{\frac{\ndim+2}{\ndim}}}{1\!-\!\VVF^{\frac{\ndim+2}{\ndim}}} \hyperstressdevhydrocomp_j\,.
\end{align}
Finally, the kinematic relations \eqref{eq:microdefgradgradmicrocomppartintdevsphere} and \eqref{eq:microdefgradgradmicrocomppartinthydsphere} and the split of the gradient of microdeformations yield
\begin{align}
\microdefgradgradsymdevhydcomp_{i}
=&\underbrace{\frac{1}{8\Lamesec\radcell^2} \frac{(\ndim\!-\!1)(\ndim\!+\!2)^2}{\ndim^2}  \frac{\frac{1+\ndim(1-2\Poissrat)}{1+\Poissrat(\ndim-2)}+\VVF^{\frac{\ndim+2}{\ndim}}}{1-\VVF^{\frac{\ndim+2}{\ndim}} }}_{=:\hyperstressmoddevhydrocompl} \frac{2\ndim}{(\ndim\!-\!1)(\ndim\!+\!2)} \hyperstressdevhydrocomp_i
\label{eq:hyperstressmodcompldevhydroderiv} \\
\microdefgradgradsymhydcomp_{i}
=&-\underbrace{\frac{1}{8\Lamesec\radcell^2} \frac{(\ndim\!-\!1)(\ndim\!+\!2)^2}{\ndim} \frac{1}{1+\Poissrat(\ndim-2)}  \frac{1-\frac{\ndim}{\ndim+2}  \frac{1-\VVF^{\frac{\ndim+2}{\ndim}}}{1-\VVF}}{1-\VVF^{\frac{\ndim+2}{\ndim}}}}_{=:\hyperstressmodhydrodevhydrocompl} \frac{2\ndim}{(\ndim\!-\!1)(\ndim\!+\!2)} \hyperstressdevhydrocomp_i
\label{eq:hyperstressmodcompldevhydrohydroderiv}
\end{align}
as well as $\microdefgradgradfulldevsymcomp_{ijk}=0$ and $\microdefgradgradsymdevskewaxmcomp_{ij}=0$ in accordance with the general isotropic law \eqref{eq:hyperstressisotropicvectorpartsmatrix}. The compliance moduli $\hyperstressmoddevhydrocompl$ and $\hyperstressmodhydrodevhydrocompl$ for this mode are defined as indicated in the equation.

%-------------------------------------------------------
\subsection{Spherical Part}
%-------------------------------------------------------

The boundary term from the boundary condition~\eqref{eq:natBChyperstressspheresplit} which is associated with the spherical part $\hyperstresshydrocomp_j$ 
can be satisifed by an ansatz with vectorial coefficients \eqref{eq:displacementshyperstress_Mdh}--\eqref{eq:tractionshyperstress_Mdh} from the previous section.
{However, an additional particular integral $\displcomp_i=C \hyperstresshydrocomp_i r^2$ with $C=(1-2\Poissrat)/[(1-\VVF)(2\Lamesec \radcell^2 )(\ndim(1-2\Poissrat)+1)]$ is required for satisfying the local equilibrium condition~\eqref{eq:momentumbalancelocalminconstrsphere}.}
The tractions associated with this particular integral are $\tractioncomprad{j}=2C[\Lamesec \hyperstresshydrocomp_j r^2+(\Lamesec+\Lamefirst)\hyperstresshydrocomp_k x_k  x_j ]/r$.

Comparing the coefficients for the constant and quadratic terms at both, boundary and void surface, leads again to a system of equations for the coefficients $A_i$, $B_i$ and $D_i$.
Inserting the resulting displacement field into the kinematic relations \eqref{eq:microdefgradgradmicrocomppartintdevsphere} and \eqref{eq:microdefgradgradmicrocomppartinthydsphere} yields
\begin{align}
\microdefgradgradsymhydcomp_{i}=&\frac{1}{\ndim}\hyperstressmodhydrocompl \hyperstresshydrocomp_i\\
\microdefgradgradsymdevhydcomp_{i}=&-\underbrace{\frac{1}{8\Lamesec\radcell^2}  \frac{(\ndim+2)^2(\ndim-1)}{\ndim}  \frac{1}{1+\Poissrat(\ndim-2)}  \frac{1-\frac{\ndim}{\ndim+2}  \frac{1-\VVF^{\frac{\ndim+2}{\ndim}}}{1-\VVF}}{1-\VVF^{\frac{\ndim+2}{\ndim}}}}_{=\hyperstressmodhydrodevhydrocompl} \frac{1}{\ndim} \hyperstresshydrocomp_i
\label{eq:hyperstressmodcomplhydrodevhydroderiv}
\end{align}
with the compliance 
{$\hyperstressmodhydrocompl$ given in Eqs.~\eqref{eq:hyperstressmodcomplhydroplane} and \eqref{eq:hyperstressmodcomplhydrospatial}.}

Comparing equations~\eqref{eq:hyperstressmodcomplhydrodevhydroderiv} and \eqref{eq:hyperstressmodcompldevhydrohydroderiv} shows that the coupling term $\hyperstressmodhydrodevhydrocompl$ is identical in both relations in accordance with the general isotropic law~\eqref{eq:hyperstressisotropicvectorpartsmatrix}. 

%-------------------------------------------------------
\subsection{Skew part of deviator}
\label{sec:hyperstressskew}
%-------------------------------------------------------

The boundary conditions of \eqref{eq:natBChyperstressspheresplit} for the skew part of the deviator $\hyperstressdevskewaxmcomp_{ij}$ 
{(which appears only in the spatial case $\ndim=3$)}
requires
an ansatz of type $\chi$ with spherical harmonics of degree $2$ and $-\ndim$ (Appendix~\ref{sec:generalsolutions}):
\begin{align}
	\displcomp_i=&2 \left(A_{kj}+\frac{1}{r^{5}}  B_{kj} \right)\ttpermutcomp_{imk} x_m x_j  \label{eq:displacementshyperstress_Mdskew}\\
\intertext{with corresponding tractions}
	\tractioncomprad{j}=&\frac{2\Lamesec}{r}  \left(A_{im} -\frac{4}{r^{5}}  B_{im}  \right)\ttpermutcomp_{jkm} x_k x_i\,.
\label{eq:tractionshyperstress_Mdskew}
\end{align}	
Inserting the tractions~\eqref{eq:tractionshyperstress_Mdskew} to the 
%boundary condition~\eqref{eq:natBChyperstressspheresplitskew} 
{part of the respective part of the boundary condition~\eqref{eq:natBChyperstressspheresplit} }
and the traction-free boundary condition at the void surface yields two equations for the coefficient tensors $A_{im}$ and $B_{im}$, again decoupled for each set of indices $im$.
Solving this system and inserting the resulting displacements into the kinematic micro-macro relation~\eqref{eq:microdefgradgradmicrosym} yields 
\begin{align}
\microdefgradgradsymcomp_{ijk}=\microdefgradgradsymdevcomp_{ijk}=\frac{5\left(1+\frac{\VVF^{5/3}}{4}\right)}{6\Lamesec\radcell^2(1-\VVF^{5/3})} \left(\ttpermutcomp_{jkn} \hyperstressdevskewaxmcomp_{in}+\ttpermutcomp_{ikn} \hyperstressdevskewaxmcomp_{jn} \right)\,.
\end{align}
A comparison with the general harmonic decomposition~\eqref{eq:hyperstresssplit} shows that the resulting gradient of microdeformation $\microdefgradgradsymcomp_{ijk}$ has only a skew part as expected so that the respective modulus 
%from Eq.~\eqref{eq:hyperstressisotropicdevskew} can be identified 
{according to Eq.~\eqref{eq:hyperstressmoddevskewaxmicro}.}
%\begin{equation}
  %\hyperstressmoddevskewax=2(a_{10}-a_{11})=\Lamesec\radcell^2  \frac{8}{5}\,\frac{1-\VVF^{5/3}}{4+\VVF^{5/3}}\,.
	%\label{eq:hyperstressmoddevskewaxmicro}
%\end{equation}
%It is plotted against the porosity $\VVF$ in \figurename~\ref{fig:hyperstress_Mdskew}. Starting at a value of $\frac25\Lamesec\radcell^2$, $\hyperstressmoddevskewax$ decreases to zero as the porosity $\VVF$ tends to 1 (independent of $\Poissrat$ since the respective deformation field at the microscale is purely deviatoric).
{It can be remarked that fields of type $\chi$ are purely deviatoric. Consequently, the resulting modulus $\hyperstressmoddevskewax$ does depend only on the microscopic shear modulus $\Lamesec$, but not on Poisson's ratio $\Poissrat$.}

%---------------------------------------------------------------------------
\section{Surface integrals over polynomials}
\label{sec:surfintegrals}

The average of the quadratic term over a spherical surface amounts to $\averopsphereshell{x_j x_k}=r^2/\ndim\Kron{jk}$ as can be shown by using Gauss theorem together with the fact that location vector and normal of the surface are coaxial $x_j=r n_j$. In the same way, surface averages over higher order polynomials can be computed recursively, yielding
\begin{align}
 \averopsphereshell{x_i x_j x_k x_p}=&\frac{r^4}{\ndim(\ndim+2)}\left(\Kron{ip} \Kron{jk}+\Kron{ik} \Kron{jp}+\Kron{ij} \Kron{kp} \right)\\
 \averopsphereshell{x_i x_j x_k x_m x_n x_p}=&\frac{r^6}{\ndim(\ndim\!+\!2)(\ndim\!+\!4)}\left(\Kron{ij} \Kron{km} \Kron{np}\!+\!\Kron{ij} \Kron{kn} \Kron{mp}\!+\!\Kron{ij} \Kron{kp} \Kron{mn}\!+\!\Kron{ik} \Kron{jm} \Kron{np}\!+\!\Kron{ik} \Kron{jn} \Kron{mp}\! \right. \nonumber\\
&+\!\Kron{ik} \Kron{jp} \Kron{mn}\!+\!\Kron{im} \Kron{jk} \Kron{np}\!+\!\Kron{im} \Kron{jn} \Kron{kp}\!+\!\Kron{im} \Kron{jp} \Kron{kn}\!+\!\Kron{in} \Kron{jk} \Kron{mp}\!+\!\Kron{in} \Kron{jm} \Kron{kp} \nonumber\\
&+\left.\!\Kron{in} \Kron{jp} \Kron{km}\!+\!\Kron{ip} \Kron{jk} \Kron{mn}\!+\!\Kron{ip} \Kron{jm} \Kron{kn}\!+\!\Kron{ip} \Kron{jn} \Kron{km} \right)\,.
\end{align}
Averages over polynomials of odd degree vanish for symmetry reasons.

%---------------------------------------------------------------------------
\section{Torsion solution for full micromorphic theory}
\label{sec:torsionsolutionfullmicromorphic}

The ansatz \eqref{eq:torsionansatzdisplacement}, \eqref{eq:ansatzmicrodef} works also for the full isotropic micromorphic theory of Mindlin, i.e., incorporating the hyperstress constitutive law~\eqref{eq:hyperstressisotropic} including terms from the incompatible part of the gradient of microdeformation. 
In this case the solution~\eqref{eq:torsionsolutionfunc} for the radial functions $\torsionmicrostrainfunc(\locmacroradplane)$ and $\torsionmicrorotfunc(\locmacroradplane)$ remains valid if the characteristic length scales are defined as
\begin{align}
\intlengthtorsionsym&=\sqrt{\frac{a_{10}+a_{13}}{2\shearmodeffmicro}}, &
\intlengthtorsionskw&=\sqrt{\frac{a_{10}-a_{13}}{\CossParamSkew}}.
\label{eq:torsionintlengthsfullmicromorphic}
\end{align}
The boundary condition~\eqref{eq:BCtorsion} for the hyperstress becomes
\begin{align}
(2a_{10}\!-\!2 a_{11}\!+\!2 a_{13}\!-\!a_{14}\!-\!a_{15})\torsionmicrostrainfunc(\torsioncylinderrad) + 2(a_{10}\!+\!a_{13})\torsioncylinderrad \torsionmicrostrainfunc'(\torsioncylinderrad) + (a_{14}\!-\!a_{15})(\torsionmicrorotfunc(\torsioncylinderrad) + 2)=0 \\
(2a_{10}\!+\!2a_{11}\!-\!2a_{13}\!-\!a_{14}\!-\!a_{15})\torsionmicrorotfunc(\torsioncylinderrad)\!+\!2(a_{10}\!-\!a_{13})\torsioncylinderrad\torsionmicrorotfunc'(\torsioncylinderrad)\!+\!(a_{14}\!-\!a_{15})\torsionmicrostrainfunc(\torsioncylinderrad)\!-\!2(2a_{11}\!-\!a_{14}\!+\!a_{15})=0
	\label{eq:BCtorsionfullmicromorphic}
\end{align}
which allow to determine the coefficients $C_1$ and $C_3$.
Definition of the torque in Eq.~\eqref{eq:deftorque} remains valid. Evaluating the integral therein yields a normalized torsion stiffness of
\begin{equation}
\begin{split}
\torsionrelstiffness=&1+4{\frac {a_{10}\!-\!a_{11}\!-\!a_{13}\!+\!a_{14} 
}{ \shearmodeff \torsioncylinderrad^
{2}}}
+2\frac{\shearmodeffmixed}{\shearmodeffmicro} \frac {a_{14}\!-\!a_{15}}{
 \shearmodeff \torsioncylinderrad^2} \\
&+2C_1\left[ \frac { a_{14}\!-\!a_{15} }{ \shearmodeff\torsioncylinderrad^2 }\frac{I_{1}(\torsioncylinderrad/\intlengthtorsionsym)}{\torsioncylinderrad/\intlengthtorsionsym}-\frac{\shearmodeffmixed+\shearmodeffmicro }{\shearmodeffmicro}\frac {a_{10}\!+\!a_{13}}{\shearmodeff\torsioncylinderrad^2} I_{2}\left(\frac {\torsioncylinderrad}{\intlengthtorsionsym}\right)
\right]
\\&-2C_3\left[\frac{2a_{11}-a_{14}-a_{15}}{ \shearmodeff\torsioncylinderrad^2 } \frac{I_{1}(\torsioncylinderrad/\intlengthtorsionskw)}{\torsioncylinderrad/\intlengthtorsionskw}+\frac{a_{10}-a_{13}
 }{\shearmodeff\torsioncylinderrad^2 } I_{2}\left(\frac {\torsioncylinderrad}{\intlengthtorsionskw}\right)
\right]\,.
\end{split}
\end{equation}
Obviously, the resulting torsion stiffness does \emph{not} only depend on $a_{10}$ and $a_{13}$ as they appear in Eq.~\eqref{eq:torsionintlengthsfullmicromorphic}, but additionally on $a_{11}$, $a_{14}$ and $a_{15}$. Respective associated parameters of dimension length could be introduced similarly as in Eq.~\eqref{eq:torsionintlengthsfullmicromorphic}, if desired.
The solution from Section~\ref{sec:torsionsolution} for the micromorphic theory with compatible gradient of microdeformation is recovered, when the parameters $a_1=a_2$, $a_4=a_5=a_8$, $a_{10}=a_{14}$, $a_{11}=a_{13}=a_{15}$ are inserted. 

For completeness, the macroscopic stress tensor resulting from ansatz \eqref{eq:torsionansatzdisplacement}--\eqref{eq:ansatzmicrodef} is given here:
\begin{equation}
\stressmacrocomp_{ij}=\twist\left[(\shearmodeffmixed\!+\!\shearmodeffmicro)(\torsionmicrostrainfunc(\locmacroradplane)\!-\!1)\!-\!(\shearmodeffstrain+\shearmodeffmixed )\right](\ttpermutcomp_{ik3} \Kron{j3}\!+\!\ttpermutcomp_{jk3} \Kron{i3}) \locmacrocomp_k\!+\! \twist \CossParamSkew \frac12( \ttpermutcomp_{jk3} \Kron{i3}\!-\!\ttpermutcomp_{ik3} \Kron{j3}\!-\!\ttpermutcomp_{ijk} \torsionmicrorotfunc(\locmacroradplane)) \locmacrocompplane_k \,.
\end{equation}
The first term is symmetric whereas the second one is the skew contribution.

\end{document}